\documentclass{article}
\usepackage{amsfonts}
\usepackage{makeidx}
\usepackage{latexsym,amsmath,amssymb,amscd}
\usepackage{makecell}
\usepackage{xcolor}
\usepackage[all]{xy}
\usepackage{float}
\usepackage{longtable}
\usepackage{endnotes}

\let\footnote=\endnote

\allowdisplaybreaks
\newtheorem{theorem}{Theorem}

\newtheorem{proposition}[theorem]{Proposition}

\begin{document}

\title{Cubic first integrals of autonomous dynamical systems in $E^2$ by an algorithmic approach}
\author{Antonios Mitsopoulos$^{1,a)}$ and Michael Tsamparlis$^{2,3,b)}$ \\
{\ \ }\\
$^{1}${\textit{Faculty of Physics, Department of
Astronomy-Astrophysics-Mechanics,}}\\
{\ \textit{University of Athens, Panepistemiopolis, Athens 157 83, Greece}}
\\
$^{2}${\textit{NITheCS, National Institute for Theoretical and Computational Sciences,}}\\
{\textit{KwaZulu-Natal, South Africa}}\\
$^{3}${\textit{TCCMMP, Theoretical and Computational Condensed Matter
 and Materials Physics Group, }}\\
{\textit{School of Chemistry and Physics, }}\\
{\textit{University of KwaZulu-Natal, Pietermaritzburg, South Africa}}\\
\vspace{12pt} 
\\
$^{a)}$Author to whom correspondence should be addressed: antmits@phys.uoa.gr 
\\
$^{b)}$Email: mtsampa@phys.uoa.gr
\\
}
\date{}
\maketitle

\begin{abstract}
In a recent paper (A. Mitsopoulos and M. Tsamparlis, J. Geom. Phys. \textbf{170}, 104383, 2021), a general theorem is given which provides an algorithmic method for  the computation of first integrals (FIs) of autonomous dynamical systems in terms of the symmetries of the kinetic metric defined by the dynamical equations of the system. In the present work, we apply this theorem to compute the cubic FIs of autonomous conservative Newtonian dynamical systems with two degrees of freedom. We show that the known results on this topic, which have been obtained by means of various different methods, and additional ones derived in this work can be obtained by the single algorithmic method provided by this theorem. The results are collected in four Tables which can be used as an updated reference of this type of integrable and superintegrable potentials. The results we find are for special values of free parameters; therefore, using the methods developed here, other researchers by different suitable choice of the parameters will be able to find new integrable and superintegrable potentials.
\end{abstract}

\section{Introduction}

\label{sec.intro}

A Hamiltonian dynamical system with $n$ degrees of freedom is Liouville integrable\footnote{V.I. Arnold, \emph{`Mathematical Methods of Classical Mechanics'}, Springer (1989), proof in pp. 272-284. \label{Arnold 1989}} if it admits $n$ (functionally) independent first integrals (FIs) which are in involution. If the number of the independent FIs is $2n-1$, the system is called superintegrable. There can be at most $n$ independent FIs in involution. The maximum number of independent FIs is $2n-1$ only when the considered FIs are autonomous. If time-dependent FIs are used, this maximum limit can be exceeded. However, for superintegrability we always need $2n-1$ independent (autonomous or time-dependent) FIs.

In principle, the solution of the system of equations of a dynamical system which is Liouville integrable can be given in terms of quadratures either by constructing the action-angle coordinates from the $n$ FIs, or by using directly the FIs to find the general solution by means of direct integrations.

Most studies restrict their considerations to the integrability of autonomous conservative dynamical systems with two degrees of freedom in the Euclidean plane $E^{2}$. These systems already admit the Hamiltonian quadratic FI (QFI); therefore, in order to establish their integrability one more independent FI in involution is required. This FI must be autonomous of any order. It cannot be time-dependent because the Poisson bracket (PB) between $H$ and a time-dependent FI $J$ does not vanish; indeed, we have $\{H,J\}= \frac{\partial J}{\partial t} \neq 0$. However, if in addition to the autonomous FIs there exists one more independent time-dependent FI, then the system is superintegrable. We note that time-dependent FIs can be used for the integrability of a dynamical system provided they are in involution. In general, the time-dependent FIs of any order are used mainly for the establishment of the superintegrability where the non-vanishing PB with the Hamiltonian is not a problem.

From the above, it is apparent that a unified systematic, i.e. algorithmic, approach which would provide FIs, time-dependent or autonomous, of any order and for any space (flat or curved) is needed.

The methods which have been proposed so far are based on the following common strategy:\newline
{\emph{Transfer the determination of the FIs to an equivalent problem in another area of mathematics where there are available means to solve the problem.}}

This strategy has been realized in practice by the use of two major methods: (a) The method of the (point and generalized) Lie symmetries, and (b) The direct method.

A Lie symmetry of a differential equation is a point transformation in the solution space of the equation which preserves the set of solutions of the equation. Although a Lie symmetry is possible to lead to a FI\footnote{G.H. Katzin and J. Levine, \emph{`Related First Integral Theorem: A Method for Obtaining Conservation Laws of Dynamical Systems with Geodesic Trajectories in Riemannian Spaces Admitting Symmetries'}, J. Math. Phys. \textbf{9}(1), 8 (1968). \label{Katzin 1968}}$^{,}$\footnote{G.H. Katzin, \emph{`Related integral theorem II. A method for obtaining quadratic constants of the motion for conservative dynamical systems admitting symmetries'}, J. Math. Phys. \textbf{14}(9), 1213 (1973). \label{Katzin 1973}}, in general, it does not so and one has to use a special class of Lie symmetries called Noether symmetries; the latter being defined by the
additional requirement that they satisfy the Noether condition. Every Noether symmetry leads to a Noether FI. In general, the method of Noether symmetries is the major tool for the determination of FIs.

In the Lie symmetry method, as a rule, the integrability of autonomous conservative dynamical systems is studied, the Hamiltonian approach is used, and the linear FIs (LFIs), the QFIs, the cubic FIs (CFIs) and to a lesser extent the quartic FIs (QUFIs) have been considered\footnote{G.H. Katzin and J. Levine, \emph{`Dynamical symmetries and constants of motion for classical particle systems'}, J. Math. Phys. \textbf{15}(9), 1460 (1974). \label{Katzin 1974}}$^{,}$\footnote{A.S. Fokas, \emph{`Group Theoretical Aspects of Constants of Motion and Separable Solutions in Classical Mechanics'}, J. Math. Anal. Appl. \textbf{68}, 347 (1979). \label{Fokas 1979}}$^{,}$\footnote{A.S. Fokas and P.A. Lagerstrom, \emph{`Quadratic and Cubic Invariants in Classical Mechanics'}, J. Math. Anal. Appl. \textbf{74}, 325 (1980). \label{Fokas 1980}}$^{,}$\footnote{P.G.L. Leach, \emph{`Applications of the Lie theory of extended groups in Hamiltonian Mechanics: The oscillator and the Kepler problem'}, J. Austral. Math. Soc. B \textbf{23}, 173 (1981). \label{Leach 1981A}}$^{,}$\footnote{P.G.L. Leach, \emph{`A further note on the H\'{e}non-Heiles problem'}, J. Math. Phys. \textbf{22}(4), 679 (1981). \label{Leach 1981B}}$^{,}$\footnote{P.A. Damianou and C. Sophocleous, \emph{`Symmetries of Hamiltonian systems with two degrees of freedom'}, J. Math. Phys. \textbf{40}(1), 210 (1999). \label{Damianou 1999}}. The general approach of Refs. \ref{Fokas 1979} and \ref{Fokas 1980}, where the QFIs and CFIs are considered (among other items), is of special interest. In Ref. \ref{Fokas 1979}, an isomorphism between the autonomous FIs of Hamilton's equations of an autonomous conservative dynamical system and the admissible Lie-B\"{a}cklund symmetries of the Hamilton-Jacobi equation is established. This isomorphism is a relation expressing the derivative $\frac{dI}{dt}$ of an arbitrary FI $I$ in terms of the canonical coordinates used in the Hamilton-Jacobi formalism. The result is general, holds for all (finite) degrees of freedom and can be used to produce the (time-dependent and autonomous) FIs of any order.

In the direct method, instead of Lie/Noether symmetries, one
uses directly the condition $\frac{dI}{dt}=0$, where $I$ is assumed to be a FI which is polynomial in the velocities. Then, one works as in Ref. \ref{Fokas 1979} with the quantity $S_{,a}$, where $S$ is the Hamilton's principal function, replaced with $\dot{q}^{a}$ and the quantities $\ddot{q}^{a}$, whenever they appear, replaced from the dynamical equations. Again a system of equations is found, which involves the unknown coefficients defining $I$ as a polynomial of $\dot{q}^{a}$ together with the elements which characterize the dynamical system, that is, the potential $V$ and the non-conservative generalized forces $F^{a}$. The solution of this system provides the class of FIs defined by $I$. It appears that the direct method has been introduced for the first time by J. Bertrand\footnote{J. Bertrand, \emph{`M\'{e}moire sur les int\'{e}grales communes \`{a} plusieurs probl\`{e}mes de M\'{e}canique'}, Journal de Math\'{e}matiques pures et appliqu\'{e}es 1 s\'{e}rie \textbf{17}, 121 (1852). \label{Bertrand 1852}} in the study of integrable surfaces and, later on, used by Whittaker\footnote{E.T. Whittaker, \emph{`A Treatise on the Analytical Dynamics of Particles and Rigid Bodies'}, 2nd ed., Cambridge Univ. Press, (1917). \label{Whittaker}} in the determination of the integrable autonomous conservative Newtonian systems with two degrees of freedom. In the course of time, this method has been used by various authors$^{\text{\ref{Katzin 1973}},}$\footnote{J. Fris, V. Mandrosov, Ya. A. Smorodinsky, M. Uhlir and P. Winternitz, \emph{`On higher symmetries in quantum mechanics'}, Phys. Lett. \textbf{16}(3), 354 (1965). \label{Fris 1965}}$^{,}$\footnote{G.H. Katzin and J. Levine, \emph{`Geodesic first integrals with explicit path-parameter dependence in Riemannian space-times'}, J. Math. Phys. \textbf{22}(9), 1878 (1981). \label{Katzin 1981}}$^{,}$\footnote{L.S. Hall, \emph{`A Theory of Exact and Approximate Configurational Invariants'}, Physica D: Nonlin. Phen. \textbf{8}(1-2), 90 (1983). \label{Hall 1983}}$^{,}$\footnote{G. Thompson, \emph{`Polynomial constants of motion in flat space'}, J. Math. Phys. \textbf{25}(12), 3474 (1984). \label{Thompson 1984}}$^{,}$\footnote{T. Sen, \emph{`Integrable potentials with quadratic invariants'}, Phys. Lett. A \textbf{111}(3), 97 (1985). \label{Sen 1985}}$^{,}$\footnote{T. Sen, \emph{`Integrable potentials with cubic and quartic invariants'}, Phys. Lett. A \textbf{122}(2), 100 (1987). \label{Sen 1987B}}$^{,}$\footnote{S. Gravel and P. Winternitz, \emph{`Superintegrability with third-order integrals in quantum and classical mechanics'}, J. Math. Phys. \textbf{43}(12), 5902 (2002). \label{Gravel 2002}}$^{,}$\footnote{J.T. Horwood, \emph{`Higher order first integrals in classical mechanics'}, J. Math. Phys \textbf{48}, 102902 (2007). \label{Horwood 2007}}$^{,}$\footnote{I. Marquette and P. Winternitz, \emph{`Superintegrable systems with third-order integrals of motion'}, J. Phys. A: Math. Theor. \textbf{41}, 304031 (2008). \label{Marquette 2008}}$^{,}$\footnote{S. Post and P. Winternitz, \emph{`General Nth order integrals of motion in the Euclidean plane'}, J. Phys. A: Math. Gen. \textbf{48}, 405201 (2015). \label{Post 2015}}$^{,}$\footnote{M. Tsamparlis and A. Mitsopoulos, \emph{`Quadratic first integrals of autonomous conservative dynamical systems'}, J. Math. Phys. \textbf{61}, 072703 (2020). \label{Tsamparlis 2020}}$^{,}$\footnote{M. Tsamparlis and A. Mitsopoulos, \emph{`First integrals of holonomic systems without Noether symmetries'}, J. Math. Phys. \textbf{61}, 122701 (2020). \label{Tsamparlis 2020B}}$^{,}$\footnote{A. Mitsopoulos and M. Tsamparlis, \emph{`Higher order first integrals of autonomous dynamical systems'}, J. Geom. Phys. \textbf{170}, 104383 (2021). \label{Mits 2021}}.

Concerning the solution of the system of equations resulting from the requirement $\frac{dI}{dt}=0$, there have been employed two methods: the algebraic method and the geometric method.

In the algebraic method, the system of equations is solved using the standard approach, i.e direct integration and/or change of variables\footnote{C.R. Holt, \emph{`Construction of new integrable Hamiltonians in two degrees of freedom'}, J. Math. Phys. \textbf{23}(6), 1037 (1982). \label{Holt 1982}}$^{,}$\footnote{J. Hietarinta, \emph{`Direct methods for the search of the second invariant'}, Phys. Rep. \textbf{147}(2), 87 (1987). \label{Hietarinta 1987}}$^{,}$\footnote{M.F. Ra\~{n}ada, \emph{`Superintegrable $n=2$ systems, quadratic constants of motion and potentials of Drach'}, J. Math. Phys. \textbf{38}(8), 4165 (1997). \label{Ranada 1997}}$^{,}$\footnote{C. Daskaloyannis and K. Ypsilantis, \emph{`Unified treatment and classification of superintegrable systems with integrals quadratic in momenta on a two dimensional manifold'}, J. Math. Phys. \textbf{47} 042904 (2006). \label{Daskaloyannis 2006}}. As expected, the algebraic method becomes impossible even for small degrees of freedom or for FIs of higher order than QFIs even in the Euclidean space.

On the other hand, in the geometric method, one uses the results of Riemannian geometry concerning the collineations of the metric. Because these results are covariant, they make possible the systematic computation of FIs, autonomous or time-dependent, of any order and in a curved space. In the geometric method, two different approaches have been used depending on the Riemannian metric considered.

\subsubsection*{The Jacobi metric}

It is well-known that the geodesics of the Jacobi metric which is defined by the dynamical equations\footnote{O.C. Pin, \emph{`Curvature and Mechanics'}, Adv. Math. \textbf{15}, 269 (1975). \label{Pin 1975}}$^{,}$\footnote{C. Uggla, \emph{`Geometrizing the dynamics of Bianchi cosmology'}, K. Rosquist and R.T. Jantzen, Phys. Rev. D \textbf{42}(2), 404 (1990). \label{Uggla 1990}}$^{,}$\footnote{K. Rosquist and G. Pucacco, \emph{`Invariants at fixed and arbitrary energy. A unified geometric approach.'}, J. Phys. A: Math. Gen. \textbf{28}, 3235 (1995). \label{Rosquist 1995}} coincide with the paths/solutions of the dynamical system. Because the FIs of any order of the geodesic equations of a Riemannian space are computed in terms of the Killing tensors (KTs) of the metric (see e.g. Ref. \ref{Katzin 1981}), in order to compute the FIs of any order of a dynamical system it is enough to compute the KTs of the Jacobi metric.

With this approach many new CFIs and QUFIs have been found$^{\text{\ref{Rosquist 1995}},}$\footnote{M. Karlovini and K. Rosquist, \emph{`A unified treatment of cubic invariants at fixed and arbitrary energy'}, J. Math. Phys. \textbf{41}(1), 370 (2000). \label{Karlovini 2000}}$^{,}$\footnote{M. Karlovini, G. Pucacco, K. Rosquist and L. Samuelsson, \emph{`A unified treatment of quartic invariants at fixed and arbitrary energy'}, J. Math. Phys. \textbf{43}(8), 4041 (2002). \label{Karlovini 2002}}. However, certain drawbacks exist concerning the Jacobi metric: a. It is not a metric of constant curvature (where we know how to compute the KTs), and b. It assumes a reparametrization from the `physical time' of the system to the so-called `Jacobi time'. Both these facts make the computation of the KTs of the Jacobi metric, hence the computation of the FIs of higher order, a difficult task.

\subsubsection*{The kinetic metric}

The kinetic metric is defined by the kinetic energy of the dynamical equations and one `solves' the system of equations resulting from the requirement $\frac{dI}{dt}=0$ in terms of the symmetries (collineations) of this metric. This approach has been used extensively in the works of Katzin$^{\text{\ref{Katzin 1973}},\text{\ref{Katzin 1974}},\text{\ref{Katzin 1981}},}$\footnote{G.H. Katzin and J. Levine, \emph{`Time-dependent quadratic constants of motion, symmetries, and orbit equations for classical particle dynamical systems with independent Kepler potentials'}, J. Math. Phys. \textbf{23}(4), 552 (1982).\label{Katzin 1982}} and more recently by others$^{\text{\ref{Horwood 2007}},\text{\ref{Tsamparlis 2020}},\text{\ref{Tsamparlis 2020B}},\text{\ref{Mits 2021}}}$.

The difficulty in this approach is again the computation of the KTs of higher order. However, the situation is much easier than the case of the Jacobi metric. For example, for all holonomic Newtonian systems whose dynamical equations can be written in the form $\ddot{q}^{a} =F^{a}(t,q)$, the kinetic metric is the flat Euclidean metric whose KTs are known (they follow directly from the Killing vectors (KVs) of the metric). This approach can also be used directly in special relativistic problems without any change but the change of signature of the metric. For example, the kinetic metric has been used in scalar field cosmology in order to determine the FIs (Noether symmetries) of the mini superspace metric defined by the flat Friedmann-Robertson-Walker (FRW) metric\footnote{M. Tsamparlis and A. Paliathanasis, \emph{`Symmetries of Differential Equations in Cosmology'}, Symmetry \textbf{10}(7), 233 (2018). \label{Tsamparlis 2018}}.
\bigskip

It should be emphasized that all methods considered above, potentially, produce the autonomous and the time-dependent FIs of any order in any space provided certain quantities are possible to be computed.

Having reviewed the general methods and practices employed in the determination of the FIs of dynamical systems, we continue with the presentation of the existing results. The main source of the known integrable autonomous conservative dynamical systems with two degrees of freedom is the review article of Ref. \ref{Hietarinta 1987} together with recent additional results$^{\text{\ref{Karlovini 2000}},\text{\ref{Karlovini 2002}},}$\footnote{A. Mitsopoulos, M. Tsamparlis and A. Paliathanasis, \emph{`Integrable and Superintegrable Potentials of 2d Autonomous Conservative Dynamical Systems'}, Symmetry \textbf{12}(10), 1655 (2020). \label{MitsTsam sym}}. In the review paper of Ref. \ref{Hietarinta 1987}, as a rule, the integrability/superintegrability of the considered dynamical systems is established in terms of autonomous QFIs. The time-dependent FIs are absent, whereas there are occasional references mainly to CFIs and to a lesser extent to QUFIs. As it has been mentioned above, the time-dependent FIs are equally appropriate for establishing integrability\footnote{V.V. Kozlov, \emph{`Integrability and non-integrability in Hamiltonian mechanics'}, Russ. Math. Surv., Turpion, \textbf{38}(1), pp. 1-76 (1983), see p.17, Theorem 1, Chapter II, Paragraph 2. \label{Kozlov 1983}}$^{,}$\footnote{T.G. Vozmishcheva, \emph{`Integrable problems of celestial mechanics in spaces of constant curvature'}, J. Math. Sc. \textbf{125}(4), 419 (2005), see Theorem 3.4. \label{Vozmishcheva 2005}}; the same applies to higher order FIs. These facts emphasize the need for a systematic method to compute the autonomous and time-dependent FIs of any order.

In a recent paper (see Ref. \ref{Mits 2021}), using the direct method and the kinetic metric approach, a general theorem is provided (see Theorem 1 in Ref. \ref{Mits 2021}) which determines the autonomous and time-dependent FIs of any order. In the present work, we apply this theorem in order to determine third order FIs of two-dimensional (2d) Newtonian autonomous conservative systems. We find the following results: \newline
a. We recover (to our knowledge) the up to date known integrable and superintegrable potentials $V(x,y)$ admitting CFIs. \newline
b. We find that certain known CFIs are in fact special cases of other more general CFIs we determine, occurring for certain values of the parameters. \newline
c. We show that potentials which were believed to be integrable are in fact superintegrable. \newline
d. We find new time-dependent CFIs for known superintegrable potentials.

The structure of the paper is as follows. In section \ref{sec.theorem.CFIs}, we state Theorem \ref{theorem.CFIs} which determines the autonomous and time-dependent CFIs for autonomous holonomic dynamical systems of the general form (\ref{eq.hfi1}) in an arbitrary $n$-dimensional Riemannian space with metric the kinetic metric of the system. In section \ref{sec.KTs.notes}, we recall briefly some basic results concerning the KTs of spaces of constant curvature. We apply these results in section \ref{sec.E2.geometry} in order to compute the geometric quantities of $E^2$, which are needed in the computation of the CFIs $J^{(3,1)}_{\ell}$, $J^{(3,2)}_{\ell}$ and $I^{(3)}_{e}$ of Theorem \ref{theorem.CFIs}. In section \ref{sec.potentials.CFIs}, we give general guidelines on how Theorem \ref{theorem.CFIs} is to be used in order to determine integrable and superintegrable potentials $V(x,y)$ in $E^{2}$ that admit CFIs (the case of QFIs has already been considered extensively in the literature). In section \ref{sec.pot.cfi2}, we consider the CFI $J^{(3,2)}_{0}$ with $\ell=0$. The results of this section are collected in Tables \ref{Table.cfi1} and \ref{Table.cfi2}, where the integrable and superintegrable potentials that admit CFIs of the type $J^{(3,2)}_{0}$ are given. Subsequently, in section \ref{sec.pot.cfi1}, we do the same for the CFI $J^{(3,1)}_{1}$ with $\ell=1$. The superintegrable
potentials $V(x,y)$ that admit time-dependent CFIs of the type $J^{(3,1)}_{1}$ are collected in Table \ref{Table.time}. In section \ref{sec.pot.cfi3}, we consider the CFI $I^{(3)}_{e}$ and find one new superintegrable potential $V(x,y)$ which is given in Table \ref{Table.cfi3}. We note that in these Tables it is indicated which integrable/superintegrable potentials are new and which generalize already known ones. Finally, in section \ref{conclusions}, we draw our conclusions.
\bigskip

Notation: The Einstein summation convention is used, round (square) brackets indicate symmetrization (antisymmetrization) of the enclosed indices, indices enclosed between vertical lines are overlooked by  antisymmetrization or symmetrization symbols, a comma indicates partial derivative and a semicolon Riemannian covariant derivative.

\section{Algorithmic determination of the CFIs}

\label{sec.theorem.CFIs}

The general Theorem 1 in Ref. \ref{Mits 2021} for $m=3$ (CFIs) gives the following theorem.

\begin{theorem}
\label{theorem.CFIs} The autonomous $n$-dimensional holonomic dynamical systems\footnote{%
It is assumed that $\det \left[ \frac{\partial^{2}T}{\partial \dot{q}^{a}\partial \dot{q}^{b}} \right] \neq 0$, where $T=\frac{1}{2}\gamma_{ab}(q)\dot{q}^{a}\dot{q}^{b}$ is the kinetic energy of the system; therefore, the kinetic metric $\gamma_{ab}$ is nondegenerate. This condition also ensures that the dynamical equations can be solved in terms of $\ddot{q}^{a}$.} (not necessarily conservative)
\begin{equation}
\ddot{q}^{a} = -\Gamma^{a}_{bc}(q)\dot{q}^{b}\dot{q}^{c} - Q^{a}(q) \label{eq.hfi1}
\end{equation}
where $q^{a}$ are the coordinates of the configuration space, $\dot{q}^{a}= \frac{dq^{a}}{dt}$, $t$ is the time variable, $\Gamma^{a}_{bc}(q)$ are the Riemannian connection coefficients of the kinetic metric $\gamma_{ab}(q)$ defined by the kinetic energy of the system, and $-Q^{a}(q)$ are the generalized forces, admit the following three independent CFIs:
\bigskip

\textbf{CFI 1.}

\begin{eqnarray}
J^{(3,1)}_{\ell}&=& \left( -\frac{t^{2\ell-1}}{2\ell-1} L_{(2\ell-2)(i_{1}i_{2};i_{3})} - ... - \frac{t^{3}}{3} L_{(2)(i_{1}i_{2};i_{3})} -tL_{(0)(i_{1}i_{2};i_{3})} \right) \dot{q}^{i_{1}} \dot{q}^{i_{2}}\dot{q}^{i_{3}} + \left( t^{2\ell} L_{(2\ell)i_{1}i_{2}} + \right. \notag \\
&& \left. + ... + t^{2}L_{(2)i_{1}i_{2}} +L_{(0)i_{1}i_{2}} \right) \dot{q}^{i_{1}} \dot{q}^{i_{2}} + \left( t^{2\ell-1} L_{(2\ell-1)i_{1}} + ... + t^{3}L_{(3)i_{1}} +tL_{(1)i_{1}} \right) \dot{q}^{i_{1}}+ \notag \\
&& + \frac{t^{2\ell}}{2\ell}L_{(2\ell-1)c}Q^{c} + ... + \frac{t^{2}}{2}L_{(1)c}Q^{c} + G(q) \label{eq.cfi1}
\end{eqnarray}
where $\ell=1,2,3,...$, $L_{(N)(i_{1}i_{2};i_{3})}(q)$ for $N=0,2,...,2\ell-2$ are third order KTs, $L_{(2\ell)i_{1}i_{2}}(q)$ is a second order KT, while the symmetric tensors $L_{(N)i_{1}i_{2}}(q)$, the vectors $L_{(A)i_{1}}(q)$, $A=1,3,...2\ell-1$, and the function $G(q)$ satisfy the conditions:
\begin{eqnarray}
L_{(k-1)(i_{1};i_{2})}&=& -\frac{3}{k-1} L_{(k-2)(i_{1}i_{2};i_{3})} Q^{i_{3}} -kL_{(k)i_{1}i_{2}}, \enskip k=2,4,...,2\ell \label{eq.cfi1a} \\
\left( L_{(2\ell-1)c}Q^{c} \right)_{,i_{1}} &=& 4\ell L_{(2\ell)i_{1}i_{2}}Q^{i_{2}} \label{eq.cfi1b} \\
\left( L_{(k-2)c}Q^{c} \right)_{,i_{1}} &=& 2(k-1)L_{(k-1)i_{1}i_{2}}Q^{i_{2}} -k(k-1)L_{(k)i_{1}}, \enskip k=3,5,...,2\ell-1 \label{eq.cfi1c} \\
G_{,i_{1}}&=& 2L_{(0)i_{1}i_{2}}Q^{i_{2}} -L_{(1)i_{1}}. \label{eq.cfi1d}
\end{eqnarray}

\textbf{CFI 2.}

\begin{eqnarray}
J^{(3,2)}_{\ell}&=&\left( -\frac{t^{2\ell}}{2\ell} L_{(2\ell-1)(i_{1}i_{2};i_{3})} - ... -\frac{t^{2}}{2}L_{(1)(i_{1}i_{2};i_{3})} +L_{(0)i_{1}i_{2}i_{3}} \right) \dot{q}^{i_{1}} \dot{q}^{i_{2}} \dot{q}^{i_{3}}+ \left( t^{2\ell+1} L_{(2\ell+1)i_{1}i_{2}} +\right. \notag \\
&& \left.+ ... + t^{3}L_{(3)i_{1}i_{2}} +tL_{(1)i_{1}i_{2}} \right) \dot{q}^{i_{1}} \dot{q}^{i_{2}} + \left( t^{2\ell} L_{(2\ell)i_{1}} + ... + t^{2}L_{(2)i_{1}} +L_{(0)i_{1}} \right) \dot{q}^{i_{1}} + \notag \\
&& + \frac{t^{2\ell+1}}{2\ell+1}L_{(2\ell)c}Q^{c} + ... + \frac{t^{3}}{3}L_{(2)c}Q^{c} + tL_{(0)c}Q^{c} \label{eq.cfi2}
\end{eqnarray}
where $\ell=0,1,2,...$, $L_{(0)i_{1}i_{2}i_{3}}(q)$ and $L_{(N)(i_{1}i_{2};i_{3})}(q)$ for $N=1,3...,2\ell-1$ are third order KTs, $L_{(2\ell+1)i_{1}i_{2}}(q)$ is a second order KT, while the symmetric tensors $L_{(N)i_{1}i_{2}}(q)$ and the vectors $L_{(A)i_{1}}(q)$, $A=0,2,...,2\ell$, satisfy the conditions:
\begin{eqnarray}
L_{(0)(i_{1};i_{2})}&=& 3L_{(0)i_{1}i_{2}i_{3}} Q^{i_{3}} -L_{(1)i_{1}i_{2}} \label{eq.cfi2a} \\
L_{(k-1)(i_{1};i_{2})}&=& -\frac{3}{k-1}L_{(k-2)(i_{1}i_{2};i_{3})} Q^{i_{3}} -kL_{(k)i_{1}i_{2}}, \enskip k=3,5,...,2\ell+1 \label{eq.cfi2b} \\
\left( L_{(2\ell)c}Q^{c} \right)_{,i_{1}} &=& 2(2\ell+1)L_{(2\ell+1)i_{1}i_{2}}Q^{i_{2}} \label{eq.cfi2c} \\
\left( L_{(k-2)c}Q^{c} \right)_{,i_{1}} &=& 2(k-1)L_{(k-1)i_{1}i_{2}}Q^{i_{2}} -k(k-1)L_{(k)i_{1}}, \enskip k=2,4,...,2\ell. \label{eq.cfi2d}
\end{eqnarray}

\textbf{CFI 3.}

\begin{equation}
I^{(3)}_{e}= e^{\lambda t} \left( -L_{(i_{1}i_{2};i_{3})} \dot{q}^{i_{1}} \dot{q}^{i_{2}} \dot{q}^{i_{3}} + \lambda L_{i_{1}i_{2}} \dot{q}^{i_{1}} \dot{q}^{i_{2}} + \lambda L_{i_{1}} \dot{q}^{i_{1}} + L_{i_{1}}Q^{i_{1}} \right) \label{eq.cfi3}
\end{equation}
where $\lambda\neq0$, $L_{(i_{1}i_{2};i_{3})}(q)$ is a third order KT, while the symmetric tensor $L_{i_{1}i_{2}}(q)$ and the vector $L_{i_{1}}(q)$ satisfy the conditions:
\begin{eqnarray}
L_{(i_{1};i_{2})}&=& -\frac{3}{\lambda} L_{(i_{1}i_{2};i_{3})} Q^{i_{3}} -\lambda L_{i_{1}i_{2}} \label{eq.cfi3a} \\ \left(L_{c}Q^{c}\right)_{,i_{1}}&=& 2\lambda L_{i_{1}i_{2}} Q^{i_{2}} -\lambda^{2}L_{i_{1}}. \label{eq.cfi3b}
\end{eqnarray}
\end{theorem}

Theorem 3 in Ref. \ref{Tsamparlis 2020B} concerning QFIs is derived as a subcase from Theorem \ref{theorem.CFIs} when the third order KT $L_{(0)i_{1}i_{2}i_{3}}$ vanish and all the symmetric tensors $L_{(N)i_{1}i_{2}}$ are assumed to be second order KTs.

The application of Theorem \ref{theorem.CFIs} requires the knowledge of second and third order KTs of the kinetic metric. These tensors of any order and in covariant form for a flat space can be found in Table 1 of Ref. \ref{Horwood 2007}. However, in the present study, beyond the KTs we shall need the symmetric tensors mentioned in Theorem \ref{theorem.CFIs}; therefore, the covariant form is not useful. In the next section \ref{sec.KTs.notes}, we recall some general results on KTs and, in the subsequent section \ref{sec.E2.geometry}, we give the coordinate form of the required geometric tensors relevant to the Euclidean space $E^{2}$.

\section{Killing tensors (KTs) of spaces of constant curvature}

\label{sec.KTs.notes}

A KT of order $m$ in an $n$-dimensional Riemannian space $(V^{n}, g_{ab})$ is a totally symmetric tensor\footnote{The independent components of a totally symmetric tensor of rank $m$ in an $n$-dimensional manifold are $\frac{(n+m-1)!}{m!(n-1)!}$. For $n=2$ we have $m+1$; and for $n=3$ we have $\frac{(m+1)(m+2)}{2}$.} of type $(0,m)$ defined by the requirement
\begin{equation}
C_{(i_{1}i_{2}...i_{m};k)}=0. \label{eq.syKT1}
\end{equation}
We have the following result$^{\text{\ref{Thompson 1984}},}$\footnote{G. Thompson, \emph{`Killing tensors in spaces of constant curvature'}, J. Math. Phys. \textbf{27}(11), 2693 (1986). \label{Thompson 1986}}$^{,}$\footnote{J.T. Horwood, \emph{`On the theory of algebraic invariants of vector spaces of Killing tensors'}, J. Geom. Phys. \textbf{58}, 487 (2008). \label{Horwood 2008}}$^{,}$\footnote{M. Takeuchi, \emph{`Killing tensor fields on spaces of constant curvature'}, Tsukuba J. Math. \textbf{7}(2), 233 (1983). \label{Takeuchi 1983}}$^{,}$\footnote{M. Eastwood, \emph{`Higher symmetries of the Laplacian'}, Ann. Math. \textbf{161}, 1645 (2005). \label{Eastwood}}$^{,}$\footnote{A.G. Nikitin and O.I. Prylypko, \emph{`Generalized Killing tensors and symmetry of Klein-Gordon-Fock equations'}, arXiv:math-ph/0506002v1. \label{Nikiting}}.

\begin{proposition}
\label{pro.KT} On a general $n$-dimensional (pseudo-Riemannian) manifold $V^{n}$, the (vector) space $\mathcal{K}^{(m,n)}$ of KTs of order $m$ has dimension
\begin{equation*}
\dim\left(\mathcal{K}^{(m,n)}\right) \leq \frac{(n+m-1)!(n+m)!}{(n-1)!n!m!(m+1)!}
\end{equation*}%
and the equality is attained if and only if $V^{n}$ is of constant curvature. Moreover, in the case of spaces of constant curvature, since there are not proper affine collineations (ACs) and projective collineations (PCs), all KTs can be expressed as a sum of symmetrized tensor products of KVs.
\end{proposition}

We note that in a space of dimension $n$ the number $N_{m}$ of KTs of order $m=2,3,4$ is
\begin{equation*}
N_{2} \leq \frac{n(n+1)^{2}(n+2)}{12}, \enskip N_{3} \leq \frac{n(n+1)^{2}(n+2)^{2}(n+3)}{3!4!}, \enskip N_{4} \leq \frac{n(n+1)^{2}(n+2)^{2}(n+3)^{2}(n+4)}{4!5!}.
\end{equation*}
In the case of $E^{2}$, we have $N_{2}=6$, $N_{3}=10$, $N_{4}=15$; and for $E^{3}$ we find $N_{2}=20$, $N_{3}=50$, $N_{4}=105$.

If a Riemannian manifold admits $n_{0}$ KVs (gradient and non-gradient) $X_{Ia}$ where $I=1,2,...,n_{0}$, then one constructs the generic KT of order $m$ as follows:
\begin{equation}
K_{i_{1}...i_{m}}= \alpha^{I_{1}...I_{m}} X_{I_{1}(i_{1}} X_{|I_{2}|i_{2}}... X_{|I_{m}|i_{m})} \label{eq.syKT2}
\end{equation}
where the Einstein summation convention is used and $1\leq I_{1} \leq I_{2} \leq ... \leq I_{m}\leq n_{0}$. According to proposition \ref{pro.KT}, in spaces of constant curvature, all KTs of order $m$ are of the form (\ref{eq.syKT2}). We note that in general not all the symmetrized products are linearly independent. This means that the parameters $\alpha^{I_{1}...I_{m}}$ are not all independent. Moreover, the number of these parameters is always larger or equal ($n=2$) to the dimension of the associated KT space computed in the proposition \ref{pro.KT} (see also a useful remark below eq. (2.12) in Ref. \ref{Horwood 2008}).

In the following sections, we apply these results in the case of $E^{2}$ where we compute the CFIs.

\section{The geometric quantities of $E^{2}$}

\label{sec.E2.geometry}

$E^{2}$ admits two gradient KVs $\partial_{x}, \partial _{y}$ whose generating functions are $x, y$ respectively and one non-gradient KV (the rotation) $y\partial _{x}-x\partial _{y}$. These vectors are written collectively as
\begin{equation}
L_{a}=\left(
\begin{array}{c}
b_{1}+b_{3}y \\
b_{2}-b_{3}x%
\end{array}%
\right)  \label{FL.15}
\end{equation}%
where $b_{1},b_{2}$, and $b_{3}$ are arbitrary constants, possibly zero.

The symmetrized tensor products of the KVs produce the KTs of various orders in $E^{2}$.

\subsection{KTs of order 2 in $E^{2}$}

- The general KT of order 2 in $E^{2}$ is\footnote{C. Chanu, L. Degiovanni and R.G. McLenaghan, \emph{`Geometrical classification of Killing tensors on bidimensional flat manifolds'}, J. Math. Phys. \textbf{47}, 073506 (2006). \label{Chanu 2006}}$^{,}$\footnote{C.M. Adlam, R.G. McLenaghan and R.G. Smirnov, \emph{`On geometric properties of joint invariants of Killing tensors'}, Symmetries and overdetermined systems of partial differential equations, IMA Vol. Math. Appl., Vol. 144, pp 205-221, Springer, New York (2008). \label{Adlam}}
\begin{equation}
C_{ab}=\left(
\begin{array}{cc}
\gamma y^{2}+2\alpha y+A & -\gamma xy-\alpha x-\beta y+C \\
-\gamma xy-\alpha x-\beta y+C & \gamma x^{2}+2\beta x+B%
\end{array}%
\right)  \label{FL.14b}
\end{equation}
where $\alpha, \beta, \gamma, A, B$, and $C$ are arbitrary constants.

- The vectors $L_{a}$ generating KTs of $E^{2}$ of the form $%
C_{ab}=L_{(a;b)}$ are\footnote{%
We note that $L_{a}$ in (\ref{FL.14}) is the sum of the non-proper ACs of $E^{2}$ and not of its KVs which give $C_{ab}=0.$}
\begin{equation}
L_{a}=\left(
\begin{array}{c}
-2\beta y^{2}+2\alpha xy+Ax+(2C-a_{1})y+a_{2} \\
-2\alpha x^{2}+2\beta xy+a_{1}x+By+a_{3}
\end{array}%
\right)  \label{FL.14}
\end{equation}
where $a_{1}, a_{2}$, and $a_{3}$ are also arbitrary constants.

- The KTs $C_{ab}=L_{(a;b)}$ in $E^{2}$ generated from the vector (\ref{FL.14}) are
\begin{equation}
C_{ab}=L_{(a;b)}=\left(
\begin{array}{cc}
L_{x,x} & \frac{1}{2}(L_{x,y}+L_{y,x}) \\
\frac{1}{2}(L_{x,y}+L_{y,x}) & L_{y,y}%
\end{array}%
\right) =\left(
\begin{array}{cc}
2\alpha y+A & -\alpha x-\beta y+C \\
-\alpha x-\beta y+C & 2\beta x+B%
\end{array}%
\right).  \label{FL.14.1}
\end{equation}
These KTs are special cases of the general KT (\ref{FL.14b}) for $\gamma =0$.

We note that the vector $L_{a}$ given by (\ref{FL.14}) depends on eight parameters and the generated KT $L_{(a;b)}$ depends on only five of them, i.e. the $\alpha, \beta, A, B,$ and $C$. We also note that the remaining $8-5=3$ parameters $a_{1}, a_{2}, a_{3}$ of the vector $L_{a}$ generate the KVs in $E^{2}$, which generate the zero KT.

\subsection{KTs of order 3 in $E^{2}$}

- The general KT $C_{abc}$ of order 3 in $E^{2}$ has independent components\footnote{R.G. McLenaghan, R.G. Smirnov and D. The, \emph{`Towards a classification of cubic integrals of motion'}, p. 199, Superintegrability in classical and quantum systems, CRM Proc. Lecture Notes, Vol. 37, Amer. Math. Soc., Providence, RI (2004). arXiv:nlin/0305048v1. \label{McLenaghan2004}}$^{,}$\footnote{J.T. Horwood, R.G. McLenaghan, R.G. Smirnov and D. The, \emph{`Fundamental covariants in the invariant theory of Killing tensors'}, SPT 2004, Symmetry and Perturbation Theory, pp. 124-131, World Sci. Publ., Hackensack, NJ (2005). \label{HorwoodMc}}
\begin{eqnarray}
C_{111}&=& a_{1}y^{3} +3a_{2}y^{2} + 3a_{3}y +a_{4} \notag \\
C_{112}&=& -a_{1}xy^{2} -2a_{2}xy +a_{5}y^{2} -a_{3}x +a_{8}y +a_{9} \notag \\
C_{221}&=& a_{1}x^{2}y +a_{2}x^{2} -2a_{5}xy -a_{8}x -a_{6}y +a_{10} \label{eq.KT1} \\
C_{222}&=&-a_{1}x^{3} +3a_{5}x^{2} +3a_{6}x +a_{7} \notag
\end{eqnarray}
where $a_{K}$ with $K=1,2,...,10$ are arbitrary constants\footnote{These are the eqs. (3.3.11) - (3.3.14) found in Ref. \ref{Hietarinta 1987}. In the notation of Ref. \ref{Hietarinta 1987}: $A=C_{111}$, $B=3C_{112}$, $C=3C_{221}$ and $D=C_{222}$. The extra factor $3$ arises from the fact that the author in Ref. \ref{Hietarinta 1987} uses algebraic methods and not the techniques of differential geometry.}.

- The reducible KT $C_{abc}= L_{(ab;c)}$ in $E^{2}$ is generated by the symmetric tensor
\begin{eqnarray}
L_{11}&=& 3b_{2}xy^{2} +3b_{5}y^{3} +3b_{3}xy +3(b_{10}+b_{8})y^{2} +b_{4}x +3b_{15}y +b_{12} \notag \\
L_{12}&=& -3b_{2}x^{2}y -3b_{5}xy^{2} -\frac{3}{2}b_{3}x^{2} -\frac{3}{2}(2b_{10}+b_{8})xy -\frac{3}{2}b_{6}y^2 +\frac{3}{2}(b_{9} -b_{15})x -\frac{3}{2}b_{11}y +b_{13} \label{eq.KT2} \\
L_{22}&=& 3b_{2}x^{3} +3b_{5}x^{2}y +3b_{10}x^{2} +3b_{6}xy +3(b_{1} +b_{11})x +b_{7}y +b_{14} \notag
\end{eqnarray}
where $b_{1}, b_{2}, ..., b_{15}$ are arbitrary constants. The independent components of the generated KT are
\begin{eqnarray}
L_{(11;1)}&=& 3b_{2}y^{2} +3b_{3}y +b_{4} \notag \\
L_{(11;2)}&=& -2b_{2}xy +b_{5}y^{2} -b_{3}x +b_{8}y +b_{9} \notag \\
L_{(22;1)}&=& b_{2}x^{2}-2b_{5}xy -b_{8}x -b_{6}y +b_{1} \label{eq.KT3} \\
L_{(22;2)}&=& 3b_{5}x^{2} +3b_{6}x +b_{7}. \notag
\end{eqnarray}
We note that the KT (\ref{eq.KT3}) is just a subcase of the general KT (\ref{eq.KT1}) for $a_{1}=0$.

Furthermore, we see that from the fifteen parameters of the symmetric tensor $L_{ab}$ given by (\ref{eq.KT2}), the nine first parameters $b_{1}, b_{2}, ..., b_{9}$ generate the reducible KT $L_{(ab;c)}$, while the remaining six parameters $b_{10}, b_{11}, ..., b_{15}$ generate all the second order KTs of the general form (\ref{FL.14b}) where $\alpha=\frac{3}{2}b_{15}$, $\beta= \frac{3}{2}b_{11}$, $\gamma=3b_{10}$, $A=b_{12}$, $B=b_{14}$, and $C=b_{13}$.

\section{Potentials $V(x,y)$ in $E^{2}$ that admit CFIs}

\label{sec.potentials.CFIs}

In Ref. \ref{Hietarinta 1987}, the known integrable and superintegrable potentials $V(x,y)$ in $E^{2}$ in terms of autonomous QFIs, CFIs and QUFIs are given. Since then, using different techniques, many new results have been produced in the field. Therefore, it is reasonable one to update and enrich the existing reviews focusing on CFIs (the case of QFIs has been extensively considered) via a single systematic approach. This is what is done in the following sections by means of Theorem \ref{theorem.CFIs}.

\subsection{Method of work}

We apply Theorem \ref{theorem.CFIs} to 2d Newtonian autonomous conservative systems of the form (\ref{eq.hfi1}). In this case, we have $q^{a}=(x,y)$, $\Gamma^{a}_{bc}=0$ and $Q^{a}=V^{,a}$, where $V(x,y)$ denotes the potential. The kinetic metric $\gamma_{ab}= \delta_{ab}= diag(1,1)$. For such systems one QFI is the Hamiltonian; therefore, one needs one more independent autonomous FI in involution in order to prove that the system is (Liouville) integrable. If the system is integrable, then one more independent autonomous or time-dependent FI is required in order to establish its superintegrability.

The algorithm we follow consists of the following actions: \newline
1) Substitute the quantities $q^{a}=(x,y)$, $Q^{a}=V^{,a}$ and $\gamma_{ab}= \delta_{ab}$ in Theorem \ref{theorem.CFIs}. \newline
2) Compute for each of the three cases, i.e. CFI 1, CFI 2 and CFI 3, of Theorem \ref{theorem.CFIs} the general expression of the CFI and the associated system of partial differential equations (PDEs). \newline
3) Use the geometric quantities of $E^{2}$, as determined in section \ref{sec.E2.geometry}, in order to fix the involved tensor quantities in the resulting systems of PDEs. Specifically, the involved symmetric tensors $L_{(N)i_{1}i_{2}}$ are given by either (\ref{FL.14b}) or (\ref{eq.KT2}), while the third order KT $L_{(0)i_{1}i_{2}i_{3}}$ is given by (\ref{eq.KT1}). Then, the system of PDEs has as unknowns the potential $V$, the vectors $L_{(N)i_{1}}$, the function $G$, and the free parameters defining the geometric quantities. \newline
4) Because the general solution of the system of conditions cannot be found, fix the degree $\ell$ of time-dependence of the CFI and solve the resulting system of PDEs for special values of the free parameters. \newline
5) Compare with the existing results and draw the appropriate conclusions.

In the following sections, we show how this algorithm works.

\section{The CFI $J^{(3,2)}_{0}$ ($\ell=0$)}

\label{sec.pot.cfi2}

We set $L_{(0)abc}=L_{abc}$, $L_{(1)ab}=C_{ab}$, and $L_{(0)a}=B_{a}$. Then, the CFI (\ref{eq.cfi2}) for $\ell=0$ becomes
\begin{equation}
J^{(3,2)}_{0}= L_{abc} \dot{q}^{a} \dot{q}^{b} \dot{q}^{c} +tC_{ab} \dot{q}^{a} \dot{q}^{b} +B_{a} \dot{q}^{a} + tB_{a}V^{,a} \label{eq.pcf2}
\end{equation}
where $L_{abc}$ is a third order KT given by (\ref{eq.KT1}), $C_{ab}$ is a second order KT given by (\ref{FL.14b}), and the vector $B_{a}$ satisfies the conditions:
\begin{eqnarray}
B_{(a;b)}&=& 3L_{abc} V^{,c} -C_{ab} \label{eq.pcf2.1} \\
\left( B_{c}V^{,c} \right)_{,a} &=& 2C_{ab}V^{,b} \label{eq.pcf2.2}.
\end{eqnarray}

From section \ref{sec.E2.geometry}, we have:
\begin{equation}
C_{ab}=\left(
\begin{array}{cc}
\gamma y^{2}+2\alpha y+A & -\gamma xy-\alpha x-\beta y+C \\
-\gamma xy-\alpha x-\beta y+C & \gamma x^{2}+2\beta x+B%
\end{array}%
\right)  \label{eq.pcf2.KT1}
\end{equation}
and
\begin{eqnarray}
L_{111}&=& a_{1}y^{3} +3a_{2}y^{2} + 3a_{3}y +a_{4} \notag \\
L_{112}&=& -a_{1}xy^{2} -2a_{2}xy +a_{5}y^{2} -a_{3}x +a_{8}y +a_{9} \notag \\
L_{221}&=& a_{1}x^{2}y +a_{2}x^{2} -2a_{5}xy -a_{8}x -a_{6}y +a_{10} \label{eq.pcf2.KT2} \\
L_{222}&=&-a_{1}x^{3} +3a_{5}x^{2} +3a_{6}x +a_{7}. \notag
\end{eqnarray}

Substituting the quantities (\ref{eq.pcf2.KT1}) and (\ref{eq.pcf2.KT2}) in the CFI conditions (\ref{eq.pcf2.1}) and (\ref{eq.pcf2.2}), we obtain a system of six PDEs (including the integrability condition of (\ref{eq.pcf2.2}) - see eq. (\ref{eq.pcf1h}) ) with three unknown functions $B_{a}(x,y)$ and $V(x,y)$. This system of equations cannot be solved in full generality; therefore, we consider special cases.

\subsection{Case $C_{ab}=0$}

\label{sec.pot.cfi2.1}

In this case, the quadratic term in the FI (\ref{eq.pcf2}) vanishes; therefore, we expect to find new 2d integrable potentials $V(x,y)$ that admit additional CFIs and not QFIs.

Equation (\ref{eq.pcf2}) becomes
\begin{equation}
J^{(3,2)}_{0}(C_{ab}=0)= L_{abc} \dot{q}^{a} \dot{q}^{b} \dot{q}^{c} +B_{a} \dot{q}^{a} + st \label{eq.pcf2.3}
\end{equation}
where $L_{abc}$ is a third order KT given by (\ref{eq.pcf2.KT2}), $s$ is an arbitrary constant, and the vector $B_{a}$ satisfies the conditions:
\begin{eqnarray}
B_{(a;b)}&=& 3L_{abc} V^{,c} \label{eq.pcf2.4} \\
B_{a}V^{,a} &=& s \label{eq.pcf2.5}.
\end{eqnarray}
We note that for $s=0$ the CFI (\ref{eq.pcf2.3}) is the standard autonomous CFI (3.3.1) discussed in Ref. \ref{Hietarinta 1987}, while conditions (\ref{eq.pcf2.4}) and (\ref{eq.pcf2.5}) coincide with eqs. (3.3.7) - (3.3.10) of Ref. \ref{Hietarinta 1987}. In the notation of Ref. \ref{Hietarinta 1987}, we have $A=L_{111}$, $B=3L_{112}$, $C=3L_{221}$, $D=L_{222}$, $F=B_{1}$, and $G=B_{2}$.

The system of equations (\ref{eq.pcf2.4}) - (\ref{eq.pcf2.5}) consists of four PDEs with three unknown functions $B_{1}(x,y)$, $B_{2}(x,y)$, $V(x,y)$ and eleven free constants, i.e. $s, a_{1}, a_{2}, ..., a_{10}$. Even this system cannot be solved in full generality and we look for special solutions.

\subsection{Case $C_{ab}=0$, $B^{a}=Z(V^{,y}, -V^{,x})$ - Holt's method}

\label{sec.pot.cfi2.2Holt}

This case was first introduced by Holt in Ref. \ref{Holt 1982} in order to discuss the integrability of 2d potentials that admit CFIs. It has as follows.

Because $C_{ab}=0$, the CFI is of the form (\ref{eq.pcf2.3}) with associated conditions the equations (\ref{eq.pcf2.4}) and (\ref{eq.pcf2.5}).

We assume that the vector\footnote{There is a misprint in eq. (3.3.15) of Ref. \ref{Hietarinta 1987}. The correct equation is the (\ref{eq.pcf2.6}).}
\begin{equation}
B_{a}= Z
\left(
  \begin{array}{c}
    V_{,y} \\
    -V_{,x} \\
  \end{array}
\right) \label{eq.pcf2.6}
\end{equation}
where $Z(x,y)$ is an arbitrary function.

Replacing (\ref{eq.pcf2.6}) in (\ref{eq.pcf2.5}), we find $s=0$ and the remaining condition (\ref{eq.pcf2.4}) gives:
\begin{eqnarray}
Z_{,x}V_{,y} +ZV_{,xy} -3L_{111}V_{,x} -3L_{112}V_{,y} &=& 0 \label{eq.pcf2.7a}\\
Z(V_{,yy}-V_{,xx}) +Z_{,y}V_{,y} -Z_{,x}V_{,x} -6L_{112}V_{,x} -6L_{221}V_{,y} &=&0 \label{eq.pcf2.7b} \\
-Z_{,y}V_{,x} -ZV_{,xy} -3L_{221}V_{,x} -3L_{222}V_{,y} &=& 0. \label{eq.pcf2.7c}
\end{eqnarray}
These are eqs. (3.3.16) - (3.3.18) in Ref. \ref{Hietarinta 1987}. We have three PDEs (\ref{eq.pcf2.7a}) - (\ref{eq.pcf2.7c}) with two unknown functions $Z(x,y)$, $V(x,y)$ and ten free parameters $a_{1}$, $a_{2}$, ..., $a_{10}$.

If we add equations (\ref{eq.pcf2.7a}) and (\ref{eq.pcf2.7c}), the following equation is obtained\footnote{We note that there is a misprint (a plus sign $+$ is missing between $C$ and $Z_{y}$) in eq. (3.3.19) of Ref. \ref{Hietarinta 1987}. The correct equation is the (\ref{eq.pcf2.8}).}:
\begin{equation}
\left( 3L_{111} +3L_{221} +Z_{,y} \right)V_{,x} +\left( 3L_{222} +3L_{112} -Z_{,x} \right)V_{,y} =0 \label{eq.pcf2.8}
\end{equation}
and, then, we have to solve the three PDEs (\ref{eq.pcf2.7a}), (\ref{eq.pcf2.7b}), and (\ref{eq.pcf2.8}).

The PDE (\ref{eq.pcf2.8}) is solved for
\begin{equation}
Z(x,y)= F(V) + Y(x,y) \label{eq.pcf2.9}
\end{equation}
where $F(V)$ is an arbitrary (smooth) function of the potential $V(x,y)$, and the function $Y(x,y)$ satisfies the system of equations:
\begin{eqnarray}
Y_{,y} &=& -3(L_{111} +L_{221}) \label{eq.pcf2.10a} \\
Y_{,x} &=& 3(L_{222} +L_{112}). \label{eq.pcf2.10b}
\end{eqnarray}
Replacing $L_{abc}$ from (\ref{eq.pcf2.KT2}), the system of equations (\ref{eq.pcf2.10a}) - (\ref{eq.pcf2.10b}) is integrated and yields the function
\begin{eqnarray}
Y(x,y) &=& -\frac{3}{4}a_{1}(x^{2}+y^{2})^{2} +3(a_{5}x -a_{2}y)(x^{2}+y^{2}) +\frac{3}{2} (3a_{6}-a_{3})x^{2} -\frac{3}{2}(3a_{3}-a_{6})y^{2} + \notag \\
&& +3a_{8}xy +3(a_{7}+a_{9})x -3(a_{4}+a_{10})y +c \label{eq.pcf2.11}
\end{eqnarray}
where $c$ is an arbitrary constant.

Substituting the solution (\ref{eq.pcf2.9}) in the remaining PDEs (\ref{eq.pcf2.7a}) and (\ref{eq.pcf2.7b}), we obtain the system of equations\footnote{Use also equations (\ref{eq.pcf2.10a}) and (\ref{eq.pcf2.10b}).}:
\begin{eqnarray}
YV_{,xy} +3L_{222}V_{,y} -3L_{111}V_{,x} &=& -(FV_{,y})_{,x} = -F'V_{,x}V_{,y} -FV_{,xy} \label{eq.pcf2.12a} \\
Y(V_{,yy}-V_{,xx}) -3\left( L_{111} +3L_{221} \right)V_{,y} -3\left( L_{222} +3L_{112} \right)V_{,x} &=& (FV_{,x})_{,x} -(FV_{,y})_{,y} \notag \\
&=& -F'\left[ (V_{,y})^{2} - (V_{,x})^{2} \right] - F(V_{,yy} -V_{,xx}) \notag \\
&& \label{eq.pcf2.12b} 
\end{eqnarray}
where $F' \equiv \frac{dF}{dV}$. By introducing the function $N(V)$ from
\begin{equation}
N' \equiv \frac{dN}{dV}= F(V) \label{eq.pcf2.13}
\end{equation}
the system of equations (\ref{eq.pcf2.12a}) - (\ref{eq.pcf2.12b}) becomes\footnote{We note that
\[
dN =FdV \implies dN= FV_{,x}dx +FV_{,y}dy \implies N_{,x}=FV_{,x}, \enskip N_{,y}=FV_{,y}.
\]
}:
\begin{eqnarray}
YV_{,xy} +3L_{222}V_{,y} -3L_{111}V_{,x} &=& -N(V)_{,xy} \label{eq.pcf2.13a} \\
Y(V_{,yy}-V_{,xx}) -3\left( L_{111} +3L_{221} \right)V_{,y} -3\left( L_{222} +3L_{112} \right)V_{,x} &=& N(V)_{,xx} -N(V)_{,yy}. \label{eq.pcf2.13b}
\end{eqnarray}
This is the system of equations\footnote{These are eqs. (170) and (171) found by Holt in Ref. \ref{Holt 1982}. We note that the left-hand sides of eqs. (3.3.22) and (3.3.23) of Ref. \ref{Hietarinta 1987} need minor corrections.} we have to solve in order to compute potentials $V(x,y)$ that admit CFIs of the form (\ref{eq.pcf2.3}). Since we do not have the general solution $V(x,y)$ of the system of equations (\ref{eq.pcf2.13a}) - (\ref{eq.pcf2.13b}), we consider several cases concerning the eleven free parameters $a_{1}$, $a_{2}, ..., a_{10}$, $c$ and the function $N(V)$.

\subsubsection{Parameters $a_{4}, a_{7}, a_{9}, a_{10}$: The components $L_{abc}$ given by (\ref{eq.pcf2.KT2}) are constant}

\label{sec.Holt1}

We have the following cases:
\bigskip

1) $N(V)=0 \implies F(V)=0$, $a_{10}=-a_{4}$ and $c\neq0$ (the remaining parameters are fixed to zero).

We find $L_{111}=a_{4}$, $L_{221}=-a_{4}$, $L_{112}=L_{222}=0$ and $Z=Y=c$.

The system of equations (\ref{eq.pcf2.13a}) - (\ref{eq.pcf2.13b}) becomes:
\begin{eqnarray}
cV_{,xy} -3a_{4}V_{,x} &=& 0 \label{eq.pcf2.15a} \\
c(V_{,yy}-V_{,xx}) +6a_{4}V_{,y} &=& 0. \label{eq.pcf2.15b}
\end{eqnarray}
The solution of this system is the potential
\begin{equation}
V(x,y)= c_{+} e^{\frac{3a_{4}}{c}(y+\sqrt{3}x)} +c_{-} e^{\frac{3a_{4}}{c}(y-\sqrt{3}x)} + c_{0} e^{-\frac{6a_{4}}{c}y} \label{eq.pcf2.16}
\end{equation}
where $c_{+}, c_{-}$, and $c_{0}$ are arbitrary constants. This is a family of 2d integrable potentials which for $a_{4}=1$ and $c=3$ reduces to the well-known \emph{Toda lattice potential}\footnote{M. Toda, \emph{`Wave Propagation in Anharmonic Lattices'}, J. Phys. Soc. Jap. \textbf{23}(3), 501 (1967). \label{Toda 1967}}$^{,}$\footnote{J. Ford, S.D. Stoddard and J.S. Turner, \emph{`On the Integrability of the Toda Lattice'}, Progr. Theor. Phys. \textbf{50}(5), 1547 (1973). \label{Ford 1973}}.

The CFI (\ref{eq.pcf2.3}) is
\begin{eqnarray}
J^{(3,2)}_{0}&=& a_{4}\dot{x}^{3} -3a_{4}\dot{x}\dot{y}^{2} + cV_{,y}\dot{x} -cV_{,x}\dot{y} \notag \\
&=& a_{4}\dot{x}^{3} -3a_{4}\dot{x}\dot{y}^{2} + 3a_{4} \left[c_{+} e^{\frac{3a_{4}}{c}(y+\sqrt{3}x)} +c_{-} e^{\frac{3a_{4}}{c}(y-\sqrt{3}x)}-2c_{0} e^{-\frac{6a_{4}}{c}y}\right] \dot{x} - \notag \\
&& -3\sqrt{3}a_{4} \left[ c_{+} e^{\frac{3a_{4}}{c}(y+\sqrt{3}x)} -c_{-} e^{\frac{3a_{4}}{c}(y-\sqrt{3}x)} \right]\dot{y}. \label{eq.pcf2.17}
\end{eqnarray}
\bigskip

2) $N(V)=0 \implies F(V)=0$, $a_{4}=1$ and $a_{10}=\frac{1}{2}$ (the remaining parameters are fixed to zero).

In this case, $L_{111}=1$, $L_{221}=\frac{1}{2}$, $L_{112}=L_{222}=0$ and $Z=Y=-\frac{9}{2}y$.

The system of equations (\ref{eq.pcf2.13a}) - (\ref{eq.pcf2.13b}) becomes:
\begin{eqnarray}
3y V_{,xy} +2V_{,x} &=& 0 \label{eq.pcf2.17a} \\
3y(V_{,yy}-V_{,xx}) +5V_{,y} &=& 0. \label{eq.pcf2.17b}
\end{eqnarray}
The solution of this system is the potential
\begin{equation}
V(x,y)= \left( \frac{4}{3}c_{1}x^{2} + c_{2}x + c_{3} \right) y^{-2/3} + c_{1}y^{4/3} \label{eq.pcf2.18}
\end{equation}
where $c_{1}, c_{2}$, and $c_{3}$ are arbitrary constants. For $c_{1}=\frac{3}{4}$ and $c_{2}=0$ the potential (\ref{eq.pcf2.18}) reduces to the potential (173) found in Ref. \ref{Holt 1982}. In the case that\footnote{S. Post and P. Winternitz, \emph{`A nonseparable quantum superintegrable system in 2D real Euclidean space'}, J. Phys. A: Math. Theor. \textbf{44}, 162001 (2011). \label{Post 2011}} $c_{1}=c_{3}=0$ and $x \leftrightarrow y$, we find the non-separable potential defined by the Hamiltonian (25) in Ref. \ref{Post 2011}. It is proved that $V=c_{2}yx^{-2/3}$ is actually a fourth order superintegrable potential due to the additional quartic FI given in eq. (27) of Ref. \ref{Post 2011}. We couldn't find this fourth order FI because we consider only FIs up to third order.

The CFI (\ref{eq.pcf2.3}) is
\begin{eqnarray}
J^{(3,2)}_{0} &=& \dot{x}^{3} +\frac{3}{2}\dot{x}\dot{y}^{2} -\frac{9}{2}y V_{,y} \dot{x} +\frac{9}{2}y V_{,x} \dot{y} \notag \\
&=& \dot{x}^{3} +\frac{3}{2}\dot{x}\dot{y}^{2} + \left[ \left( 4c_{1}x^{2} + 3c_{2}x + 3c_{3} \right) y^{-2/3} - 6c_{1}y^{4/3} \right] \dot{x} +\left( 12c_{1}x + \frac{9}{2}c_{2} \right) y^{1/3}\dot{y}. \label{eq.pcf2.19}
\end{eqnarray}
We note that for $c_{1}=\frac{3}{4}\epsilon$, $c_{2}=0$, and $c_{3}=\epsilon \delta$, where $\epsilon$ and $\delta$ are arbitrary constants, the CFI (\ref{eq.pcf2.19}) coincides with eq. (174) of Ref. \ref{Holt 1982} divided by 2.
\bigskip

3) $N(V)=0 \implies F(V)=0$ and $a_{9}\neq0$ is the only non-vanishing parameter\footnote{If instead of $a_{9}\neq0$ we take $a_{10}\neq0$, then we find symmetric results transformed to each other by the interchange $x \leftrightarrow y$.}.

The only quantities that survive are the $L_{112}=a_{9}$ and $Z=Y= 3a_{9}x$.

The system of equations (\ref{eq.pcf2.13a}) - (\ref{eq.pcf2.13b}) becomes:
\begin{eqnarray}
V_{,xy} &=& 0 \label{eq.pcf2.19a} \\
x(V_{,yy}-V_{,xx}) -3V_{,x} &=& 0. \label{eq.pcf2.19b}
\end{eqnarray}
The solution of this system is the potential
\begin{equation}
V(x,y)= c_{1}x^{2} +\frac{c_{2}}{x^{2}} +4c_{1}y^{2} +c_{3}y \label{eq.pcf2.20}
\end{equation}
where $c_{1}, c_{2}$, and $c_{3}$ are arbitrary constants. For $c_{1}=1$ and $c_{3}=0$ the potential (\ref{eq.pcf2.20}) reduces to the potential (175) found in Ref. \ref{Holt 1982}.

The CFI (\ref{eq.pcf2.3}) is
\begin{equation}
J^{(3,2)}_{0} =\dot{x}^{2}\dot{y} +x V_{,y} \dot{x} -xV_{,x} \dot{y} =\dot{x}^{2}\dot{y} +x (8c_{1}y +c_{3})\dot{x} -2 \left(c_{1}x^{2} -\frac{c_{2}}{x^{2}}\right)\dot{y}. \label{eq.pcf2.21}
\end{equation}
This is the CFI (176) of Ref. \ref{Holt 1982}.

The potential (\ref{eq.pcf2.20}) is of the separable form $V(x,y)= F_{1}(x) +F_{2}(y)$. It is well-known (see Tables in Ref. \ref{MitsTsam sym}) that such potentials are integrable and they admit the additional QFIs
\begin{equation}
I_{71a}= \frac{1}{2}\dot{x}^{2} +F_{1}(x), \enskip I_{71b}= \frac{1}{2}\dot{y}^{2} + F_{2}(y). \label{eq.pcf2.21.1}
\end{equation}
Therefore, using the CFI (\ref{eq.pcf2.21}), it is shown that the potential (\ref{eq.pcf2.20}) is also superintegrable because the FIs (\ref{eq.pcf2.21}), $I_{71a}$ and $I_{71b}$ are functionally independent. However, directly from the last Table of Ref. \ref{MitsTsam sym} (see also eq. (67) in Ref. \ref{MitsTsam sym}), we see that the potential (\ref{eq.pcf2.20}) is superintegrable because it also admits the additional QFI
\begin{equation}
I_{s2a}=\dot{x}(y\dot{x}-x\dot{y})-2c_{1}x^{2}y +\frac{2c_{2}y}{x^{2}}-\frac{c_{3}}{2}x^{2}. \label{eq.pcf2.22}
\end{equation}
Hence the CFI (\ref{eq.pcf2.21}) can be expressed as a function of $I_{71a}, I_{71b}$, and $I_{s2a}$.
\bigskip

4) $N(V)=0 \implies F(V)=0$, $a_{7}=-a_{4}$ and $c$ are the only non-vanishing parameters.

We find $L_{111}=-L_{222}=a_{4}$ and $Z=Y=-3a_{4}(x+y)+c$.

Solving the system of equations (\ref{eq.pcf2.13a}) - (\ref{eq.pcf2.13b}), we find the integrable potential
\begin{equation}
V= \frac{k}{-3a_{4}(x+y)+c} \label{eq.pcf2.23}
\end{equation}
where $k$ is an arbitrary constant.

The CFI (\ref{eq.pcf2.3}) is
\begin{equation}
J^{(3,2)}_{0}= \dot{x}^{3} -\dot{y}^{3} +\frac{3k}{-3a_{4}(x+y)+c}(\dot{x} -\dot{y}). \label{eq.pcf2.23a}
\end{equation}

From Ref. \ref{MitsTsam sym}, the potential (\ref{eq.pcf2.23}) is of the form $V=F(x+y)$; therefore, it admits also the following LFIs/QFIs:
\[
I_{1}= \dot{x}-\dot{y}, \enskip I_{2}=t(\dot{x}-\dot{y})-x+y, \enskip I_{3}=\dot{x}\dot{y} +F(x+y).
\]
This means that the potential (\ref{eq.pcf2.23}) is superintegrable without using higher order FIs. The CFI (\ref{eq.pcf2.23a}) is expressed as follows:
\[
J^{(3,2)}_{0}= I_{1}^{3} +3I_{1}I_{3}= \dot{x}^{3} -\dot{y}^{3} +3F(x+y)(\dot{x} -\dot{y}).
\]
\bigskip

5)$F(V)=\lambda V$, $F'=\lambda\neq0$ and $a_{4}\neq0$ (the remaining parameters are fixed to zero).

We find $L_{111}=a_{4}$, $Y=-3a_{4}y$ and $Z=\lambda V -3a_{4}y$.

The system of equations (\ref{eq.pcf2.13a}) - (\ref{eq.pcf2.13b}) becomes:
\begin{eqnarray}
\left[ F(V) -3a_{4}y \right]V_{,xy} +F'V_{,x}V_{,y} -3a_{4}V_{,x} &=& 0 \label{eq.pcf2.44a} \\
\left[ F(V) -3a_{4}y \right](V_{,yy} - V_{,xx}) +F' \left[(V_{,y})^{2} -(V_{,x})^{2}\right] -3a_{4}V_{,y} &=& 0. \label{eq.pcf2.44b}
\end{eqnarray}
The structure of the above system indicates that $F(V)= x^{\nu} +3a_{4}y$, where $\nu$ is an arbitrary constant. Replacing $F(V)=\lambda V$, we find
\[
V(x,y)= \frac{x^{\nu}}{\lambda} +\frac{3a_{4}}{\lambda}y
\]
which when replaced into the system of equations (\ref{eq.pcf2.44a}) - (\ref{eq.pcf2.44b}) results to $\nu= \frac{1}{2}$. Therefore, we obtain the separable superintegrable potential
\begin{equation}
V(x,y)= c_{1}\sqrt{x} +c_{2}y \label{eq.pcf2.45}
\end{equation}
where $c_{1}\equiv\frac{1}{\lambda}\neq0$ and $c_{2}\equiv \frac{3a_{4}}{\lambda}=3c_{1}a_{4}$. The potential (\ref{eq.pcf2.45}) for $c_{1}=1$ reduces to the one found in Table I of Ref. \ref{Karlovini 2000}.

The function $Z= \sqrt{x}$ and the associated CFI (\ref{eq.pcf2.3}) is
\begin{equation}
J^{(3,2)}_{0}= c_{2}\dot{x}^{3} +3c_{1}c_{2}\sqrt{x}\dot{x} -\frac{3}{2}c_{1}^{2}\dot{y} \label{eq.pcf2.46}
\end{equation}
where $c_{1}$ and $c_{2}$ are arbitrary constants.
\bigskip

6) $F(V)=\lambda V^{2} \implies F'=2\lambda V$, $\lambda\neq0$ and $a_{7}=-a_{4}\neq0$ (the remaining parameters are fixed to zero).

We find $L_{111}=a_{4}$, $L_{222}=-a_{4}$, $Y=-3a_{4}(x+y)$ and $Z=\lambda V^{2} -3a_{4}(x+y)$.

The system of equations (\ref{eq.pcf2.13a}) - (\ref{eq.pcf2.13b}) becomes:
\begin{eqnarray}
\left[ F(V) -3a_{4}(x+y) \right]V_{,xy} +2\lambda VV_{,x}V_{,y} -3a_{4}(V_{,x}+V_{,y}) &=& 0 \label{eq.pcf2.47a} \\
\left[ F(V) -3a_{4}(x+y) \right](V_{,yy} - V_{,xx}) +2\lambda V \left[(V_{,y})^{2} -(V_{,x})^{2}\right] +3a_{4}(V_{,x}-V_{,y}) &=& 0. \label{eq.pcf2.47b}
\end{eqnarray}
From the structure of the above system and the complete square assumption $F(V)= \lambda V^{2}$, we deduce that the function $F(V)$ must have the following form:
\[
F(V)= 3a_{4}(x+y) \pm6a_{4}\sqrt{xy} \implies \lambda V^{2}= 3a_{4}\left( \sqrt{x} \pm \sqrt{y} \right)^{2} \implies
\]
\begin{equation}
V(x,y)= k \left( \sqrt{x} \pm \sqrt{y} \right) \label{eq.pcf2.48}
\end{equation}
where $k^{2}\equiv \frac{3a_{4}}{\lambda}$. It can be easily checked that the potential (\ref{eq.pcf2.48}) satisfies trivially the system of equations (\ref{eq.pcf2.47a}) - (\ref{eq.pcf2.47b}). This potential has been found also by Karlovini in Table I of Ref. \ref{Karlovini 2000} using the Jacobi metric of the system.

The function $Z=\pm 6a_{4}\sqrt{xy}$ and the associated CFI (\ref{eq.pcf2.3}) is
\begin{equation}
J^{(3,2)}_{0}= \dot{x}^{3} -\dot{y}^{3} + 3k\left( \sqrt{x}\dot{x} \mp \sqrt{y}\dot{y} \right). \label{eq.pcf2.49}
\end{equation}
Since the potential (\ref{eq.pcf2.48}) is separable, the additional CFI (\ref{eq.pcf2.49}) makes it superintegrable.

\subsubsection{Parameters $a_{3}, a_{6}, a_{8}$: The components $L_{abc}$ given by (\ref{eq.pcf2.KT2}) are linearly dependent on $x,y$}

\label{sec.Holt2}

We have the following cases:
\bigskip

1) $F(V)\neq0$ and $a_{8}=\frac{1}{3}$ (the remaining parameters are set to zero).

We find $L_{111}=L_{222}=0$, $L_{112}=\frac{y}{3}$, $L_{221}=-\frac{x}{3}$, $Y= xy$ and $Z= F(V)+xy$.

The system of equations (\ref{eq.pcf2.12a}) - (\ref{eq.pcf2.12b}) becomes:
\begin{eqnarray}
\left[ xy + F(V) \right]V_{,xy} +F'V_{,x}V_{,y} &=& 0 \label{eq.pcf2.24a} \\
\left[ xy +F(V) \right](V_{,yy} -V_{,xx}) +F'\left[(V_{,y})^{2} -(V_{,x})^{2}\right] +3(xV_{,y} -yV_{,x})&=&0. \label{eq.pcf2.24b}
\end{eqnarray}
This system has been solved previously by Inozemtsev\footnote{V.I. Inozemtsev, \emph{`New integrable classical system with two degrees of freedom'}, Phys. Lett. A \textbf{96}(9), 447 (1983). \label{Inozemtsev 1983}} as follows.

From the structure of the system (\ref{eq.pcf2.24a}) - (\ref{eq.pcf2.24b}), we assume $V= f(w)$ where $w=xy$ and $f(w)$ is an arbitrary smooth function. Then, the above system of equations becomes\footnote{
We compute:
\[
V_{,x}= y\frac{df}{dw}, \enskip V_{,y}=x\frac{df}{dw}, \enskip V_{,xy}= w\frac{d^{2}f}{dw^{2}} +\frac{df}{dw}, \enskip V_{,xx}= y^{2}\frac{d^{2}f}{dw^{2}}, \enskip V_{,yy}= x^{2}\frac{d^{2}f}{dw^{2}}.
\]
}:
\begin{eqnarray}
\left[ w + F(f) \right]\left(w \frac{d^{2}f}{dw^{2}} +\frac{df}{dw} \right) +wF'\left( \frac{df}{dw} \right)^{2} &=& 0 \label{eq.pcf2.25a} \\
\left[ w +F(f) \right]\frac{d^{2}f}{dw^{2}} +F'\left( \frac{df}{dw} \right)^{2} +3\frac{df}{dw} &=&0. \label{eq.pcf2.25b}
\end{eqnarray}
Substituting equation (\ref{eq.pcf2.25a}) in (\ref{eq.pcf2.25b}), we find \Big(assume $\frac{df}{dw}\neq0$\Big)
\[
F(f(w))=2w \implies F' \frac{df}{dw}=2
\]
and the remaining equation gives
\[
3w\frac{d^{2}f}{dw^{2}} +5\frac{df}{dw} =0 \implies f(w)=kw^{-2/3}
\]
where $k$ is an arbitrary non-zero constant.

Therefore, we get the integrable potential
\begin{equation}
V(x,y)= k(xy)^{-2/3} \label{eq.pcf2.26}
\end{equation}
and the function $F(V)=2\left(\frac{V}{k}\right)^{-3/2} \implies Z=3xy$.

The CFI (\ref{eq.pcf2.3}) is
\begin{equation}
J^{(3,2)}_{0}=(y\dot{x} -x\dot{y})\dot{x}\dot{y} -2k(xy)^{-2/3} \left(x\dot{x} -y\dot{y}\right)= (\dot{x}^{2}-2H)x\dot{x} -(\dot{y}^{2}-2H)y\dot{y} \label{eq.pcf2.27}
\end{equation}
where the Hamiltonian $H= \frac{1}{2}(\dot{x}^{2}+\dot{y}^{2}) +k(xy)^{-2/3}$.
\bigskip

2) $N(V)=0 \implies F(V)=0$ and $a_{8}=\frac{1}{3}$ (the remaining parameters are set to zero).

We find $L_{221}=-\frac{x}{3}$, $L_{112}=\frac{y}{3}$ and $Z=Y=xy$.

The solution of the system of equations (\ref{eq.pcf2.13a}) - (\ref{eq.pcf2.13b}) is the potential (see Table II of Ref. \ref{Karlovini 2000})
\begin{equation}
V= c_{1}(x^{2}+y^{2}) +\frac{c_{2}}{x^{2}} +\frac{c_{3}}{y^{2}} \label{eq.pcf2.28}
\end{equation}
and the associated CFI (\ref{eq.pcf2.3}) is
\begin{eqnarray}
J^{(3,2)}_{0}&=& y\dot{x}^{2}\dot{y}-x\dot{x}\dot{y}^{2} +xy( 2c_{1}y -2c_{3}y^{-3})\dot{x} -xy(2c_{1}x -2c_{2}x^{-3})\dot{y} \notag \\
&=& (y\dot{x}-x\dot{y})(\dot{x}\dot{y} +2c_{1}xy) +2c_{2}x^{-2}y\dot{y} -2c_{3}xy^{-2}\dot{x} \label{eq.pcf2.29}
\end{eqnarray}
where $c_{1}, c_{2}$, and $c_{3}$ are arbitrary constants.

The potential (\ref{eq.pcf2.28}) is superintegrable because it is of the separable form $V=F_{1}(x) +F_{2}(y)$; therefore, it also admits the QFIs (\ref{eq.pcf2.21.1}).

It is important to note that directly from the last Table of Ref. \ref{MitsTsam sym} -without using CFIs- the potential (\ref{eq.pcf2.28}) is superintegrable because it is of the form
\begin{equation}
V_{274}=-\frac{\lambda ^{2}}{8}(x^{2}+y^{2})-\frac{\lambda
^{2}}{4}\left( d_{1}x+ d_{2}y\right) -\frac{k_{1}}{(x+d_{1})^{2}} -\frac{k_{2}}{(y+d_{2})^{2}} \label{eq.pcf2.30}
\end{equation}
where $\lambda\neq0$, $k_{1}, k_{2}, d_{1}$, and $d_{2}$ are arbitrary constants. It is proved in Ref. \ref{MitsTsam sym} that the potential (\ref{eq.pcf2.30}) is superintegrable because besides the QFIs (\ref{eq.pcf2.21.1}) it admits also the time-dependent QFIs
\begin{eqnarray}
I_{73a}&=& e^{\lambda t}\left[ -\dot{x}^{2}+\lambda (x+d_{1})\dot{x}-\frac{\lambda^{2}}{4}(x+d_{1})^{2} +\frac{2k_{1}}{(x+d_{1})^{2}}\right] \label{eq.pcf2.31a} \\
I_{73b}&=& e^{\lambda t}\left[ -\dot{y}^{2}+\lambda
(y+d_{2})\dot{y}-\frac{\lambda
^{2}}{4}(y+d_{2})^{2}+\frac{2k_{2}}{(y+d_{2})^{2}}\right]. \label{eq.pcf2.31b}
\end{eqnarray}
It follows that \emph{it is possible to use time-dependent FIs of lower order in order to show that a potential is superintegrable. This is a helpful conclusion because the computation of higher order FIs is in general a major task.}
\bigskip

3) $F(V)=\lambda V$, $F'=\lambda\neq0$ and $a_{6}=\frac{1}{3}$ (the remaining parameters are fixed to zero).

We find $L_{221}=-\frac{y}{3}$, $L_{222}=x$, $Y=\frac{3}{2}x^{2} +\frac{y^{2}}{2}$ and $Z=\lambda V +\frac{3}{2}x^{2} +\frac{y^{2}}{2}$.

Substituting in the system of equations (\ref{eq.pcf2.12a}) - (\ref{eq.pcf2.12b}), we obtain:
\begin{eqnarray}
\left( \frac{3}{2}x^{2} +\frac{y^{2}}{2} +\lambda V \right)V_{,xy} +3xV_{y} +\lambda V_{,x}V_{,y} &=& 0 \label{eq.pcf2.32a} \\
\left[ \frac{3}{2}x^{2} +\frac{y^{2}}{2} +\lambda V \right]( V_{,yy} -V_{,xx}) +3yV_{,y} -3xV_{,x} + \lambda\left[ (V_{,y})^{2} -(V_{,x})^{2} \right] &=&0. \label{eq.pcf2.32b}
\end{eqnarray}
In order to solve the above system, we assume\footnote{We make this assumption in order $Z(x,y)$ to be a function of $y^{2}$ alone.} that $F(V)= -\frac{3}{2}x^{2} +ky^{2}$ where $k$ is an arbitrary constant. Then, we find that
\[
V(x,y)= \frac{1}{\lambda}\left(-\frac{3}{2}x^{2} +ky^{2}\right).
\]
Replacing the above potential in the system of equations (\ref{eq.pcf2.32a}) - (\ref{eq.pcf2.32b}), we find that the only non-trivial solution is for $k=-\frac{1}{6}$.

Therefore, we find the potential (contained in Table II of Ref. \ref{Karlovini 2000} for $c_{0}=1$)
\begin{equation}
V(x,y)= c_{0}\left(9x^{2} +y^{2}\right) \label{eq.pcf2.33}
\end{equation}
where $c_{0}\equiv -\frac{1}{6\lambda}$ and hence $Z=\frac{y^{2}}{3}$. This potential includes as subcases the potentials (3.15a) and (3.15b) found by Fokas in Ref. \ref{Fokas 1980} for $c_{0}=\frac{1}{2}$ (with $x \leftrightarrow y$) and $c_{0}=\frac{1}{18}$, respectively.

The associated CFI (\ref{eq.pcf2.3}) is
\begin{equation}
J^{(2,3)}_{0}= (x\dot{y} - y\dot{x})\dot{y}^{2} +\frac{2c_{0}}{3}y^{3}\dot{x} -6c_{0}xy^{2}\dot{y}. \label{eq.pcf2.34}
\end{equation}
Therefore, the potential (\ref{eq.pcf2.33}) is superintegrable because it is of the separable form $V(x,y)= F_{1}(x) +F_{2}(y)$.
\bigskip

4) $N(V)=0 \implies F(V)=0$ and $a_{3}=a_{6}\neq0$ (the remaining parameters are fixed to zero).

We find $L_{111}=3a_{3}y$, $L_{112}=-a_{3}x$, $L_{221}=-a_{3}y$, $L_{222}=3a_{3}x$ and $Z=Y=3a_{3}(x^{2}-y^{2})$.

The system of equations (\ref{eq.pcf2.13a}) - (\ref{eq.pcf2.13b}) becomes:
\begin{eqnarray}
(x^{2}-y^{2})V_{,xy} +3xV_{,y} -3yV_{,x} &=& 0 \label{eq.pcf2.36a} \\
V_{,yy} - V_{,xx} &=& 0. \label{eq.pcf2.36b}
\end{eqnarray}
Solving the above system, we find the potential (see Table II of Ref. \ref{Karlovini 2000})
\begin{equation}
V(x,y)= c_{1}(x^{2}+y^{2}) + \frac{c_{2}xy +c_{3}(x^{2}+y^{2})}{(x^{2}-y^{2})^{2}} \label{eq.pcf2.37}
\end{equation}
where $c_{1}, c_{2}$, and $c_{3}$ are arbitrary constants.

The associated CFI (\ref{eq.pcf2.3}) is
\begin{eqnarray}
J^{(3,2)}_{0}&=& \left( x\dot{y} -y\dot{x} \right) \left( \dot{y}^{2} -\dot{x}^{2} \right) + (x^{2}-y^{2})(V_{,y}\dot{x} -V_{,x}\dot{y}) \notag \\
&=& \left( x\dot{y} -y\dot{x} \right) \left( \dot{y}^{2} -\dot{x}^{2} \right) -2c_{1}(x\dot{y} -y\dot{x})(x^{2}-y^{2}) + \frac{x(x^{2}+3y^{2})(c_{2}\dot{x} +2c_{3}\dot{y})}{(x^{2}-y^{2})^{2}}+ \notag \\
&& +\frac{y(3x^{2}+y^{2})(2c_{3}\dot{x} +c_{2} \dot{y})}{(x^{2} -y^{2})^{2}}. \label{eq.pcf2.38}
\end{eqnarray}

Since the PDE (\ref{eq.pcf2.36b}) is a wave equation, it admits a solution of the form $V= F_{3}(y+x) +F_{4}(y-x)$ where $F_{3}$ and $F_{4}$ are arbitrary smooth functions of their arguments. Indeed, the potential (\ref{eq.pcf2.37}) is of this form with
\[
F_{3}(y+x)= \frac{c_{1}}{2}(y+x)^{2} +\frac{2c_{3}-c_{2}}{4(y+x)^{2}} , \enskip F_{4}(y-x)= \frac{c_{1}}{2}(y-x)^{2} +\frac{2c_{3}+c_{2}}{4(y-x)^{2}}.
\]
Therefore, the potential (\ref{eq.pcf2.37}) is superintegrable because besides the Hamiltonian $H$ and the CFI (\ref{eq.pcf2.38}) it admits also the QFI (see the fourth Table in Ref. \ref{MitsTsam sym})
\begin{equation}
I_{8}= \dot{x}\dot{y} + F_{3}(y+x) -F_{4}(y-x). \label{eq.pcf2.39}
\end{equation}
\bigskip

5) $F(V)\neq0$ and $a_{3}=a_{6}\neq0$ (the remaining parameters are fixed to zero).

The difference with the previous case is that now we assume $F(V)\neq0$.

We find $L_{111}=3a_{3}y$, $L_{112}=-a_{3}x$, $L_{221}=-a_{3}y$, $L_{222}=3a_{3}x$, $Y=3a_{3}(x^{2}-y^{2})$ and $Z=F(V) +Y$.

The system of equations (\ref{eq.pcf2.13a}) - (\ref{eq.pcf2.13b}) becomes:
\begin{eqnarray}
\left[ 3a_{3}(x^{2}-y^{2}) +F(V) \right]V_{,xy} +F'V_{,x}V_{,y} +9a_{3}xV_{,y} -9a_{3}yV_{,x} &=& 0 \label{eq.pcf2.40a} \\
\left[ 3a_{3}(x^{2}-y^{2}) +F(V) \right](V_{,yy} - V_{,xx}) +F' \left[(V_{,y})^{2} -(V_{,x})^{2}\right] &=& 0. \label{eq.pcf2.40b}
\end{eqnarray}
The structure of the above system indicates that
\begin{equation}
V(x,y)= \phi(w) \label{eq.pcf2.41}
\end{equation}
where $\phi$ is an arbitrary smooth function of $w\equiv x^{2} -y^{2}$.

The system of equations (\ref{eq.pcf2.40a}) - (\ref{eq.pcf2.40b}) becomes\footnote{
We compute:
\[
V_{,x}= 2x\frac{d\phi}{dw}, \enskip V_{,y}= -2y\frac{d\phi}{dw}, \enskip V_{,xy}= -4xy\frac{d^{2}\phi}{dw^{2}}, \enskip V_{,xx}= 2\left( 2x^{2}\frac{d^{2}\phi}{dw^{2}} +\frac{d\phi}{dw} \right), \enskip V_{,yy}= 2\left( 2y^{2}\frac{d^{2}\phi}{dw^{2}} -\frac{d\phi}{dw} \right).
\]
}:
\begin{eqnarray}
\left[ 3a_{3}w +F(\phi) \right]\frac{d^{2}\phi}{dw^{2}} +F'\left(\frac{d\phi}{dw}\right)^{2} +9a_{3}\frac{d\phi}{dw} &=& 0 \label{eq.pcf2.41a} \\
\left[ 3a_{3}w +F(\phi) \right] \left( w\frac{d^{2}\phi}{dw^{2}} +\frac{d\phi}{dw} \right) +wF'\left( \frac{d\phi}{dw} \right)^{2} &=& 0. \label{eq.pcf2.41b}
\end{eqnarray}
Multiplying equation (\ref{eq.pcf2.41a}) with $-w$ and then adding the resulting equation to (\ref{eq.pcf2.41b}), we find that
\begin{equation}
F= 6a_{3}w \implies F'\frac{d\phi}{dw}=6a_{3} \label{eq.pcf2.41c}
\end{equation}
where $\frac{d\phi}{dw} \neq0$.

Substituting equation (\ref{eq.pcf2.41c}) in the remaining equation (\ref{eq.pcf2.41a}), we get
\[
3w\frac{d^{2}\phi}{dw^{2}} +5\frac{d\phi}{dw}=0 \implies \phi= kw^{-2/3}
\]
where $k$ is an arbitrary constant.

Returning to the original coordinates, we obtain the integrable potential
\begin{equation}
V(x,y)= k(x^{2} -y^{2})^{-2/3}. \label{eq.pcf2.42}
\end{equation}
The function $Z= 9a_{3}(x^{2}-y^{2})$ and the associated CFI (\ref{eq.pcf2.3}) is (see Table II in Ref. \ref{Karlovini 2000})
\begin{eqnarray}
J^{(3,2)}_{0}&=& \left( x\dot{y} -y\dot{x} \right) \left( \dot{y}^{2} -\dot{x}^{2} \right) +4k(x^{2} -y^{2})^{-2/3}(y\dot{x} +x\dot{y}) \notag \\
&=& \left( x\dot{y} -y\dot{x} \right) \left( \dot{y}^{2} -\dot{x}^{2} \right) +4V(y\dot{x} +x\dot{y}). \label{eq.pcf2.43}
\end{eqnarray}
The above results coincide with eq. (3.19) of Ref. \ref{Fokas 1980}. In Ref. \ref{Fokas 1980}, the authors derive this result by applying directly the CFI conditions to a potential of type $V=f(x^{2}+\nu y^{2})$, where $f$ is an arbitrary smooth function and $\nu$ an arbitrary constant.
\bigskip

6) $N(V)=0 \implies F(V)=0$ and $a_{3}=-a_{6}\neq0$ (the remaining parameters are fixed to zero).

We find $L_{111}=3a_{3}y$, $L_{112}=-a_{3}x$, $L_{221}=a_{3}y$, $L_{222}=-3a_{3}x$, and $Z=Y=-6a_{3}r^{2}$ where $r=\sqrt{x^{2} +y^{2}}$.

The system of equations (\ref{eq.pcf2.13a}) - (\ref{eq.pcf2.13b}) becomes:
\begin{eqnarray}
2r^{2}V_{,xy} +3xV_{,y} +3yV_{,x} &=& 0 \label{eq.pcf2.56a} \\
r^{2}(V_{,yy} - V_{,xx}) +3yV_{,y} -3xV_{,x} &=& 0. \label{eq.pcf2.56b}
\end{eqnarray}
Solving the above system, we find the potential (see Table V of Ref. \ref{Karlovini 2000})
\begin{equation}
V(x,y)= \frac{c_{1}}{r} +\frac{c_{2}\sqrt{r+x}}{r} +\frac{c_{3} \sqrt{r-x}}{r} \label{eq.pcf2.57}
\end{equation}
where $c_{1}, c_{2}$, and $c_{3}$ are arbitrary constants.

The associated CFI (\ref{eq.pcf2.3}) is
\begin{eqnarray}
J^{(3,2)}_{0}&=& \frac{1}{2}(x\dot{y} -y\dot{x})(\dot{x}^{2} +\dot{y}^{2}) +r^{2}(V_{,y}\dot{x} -V_{,x}\dot{y}) \notag \\
&=& (x\dot{y} -y\dot{x})H +(r^{2}V_{,y} +yV)\dot{x} -(r^{2}V_{,x} +xV)\dot{y} \notag \\
&=& (x\dot{y} -y\dot{x})H + \frac{1}{2} \left( c_{2}\sqrt{r-x} +c_{3}\sqrt{r+x} \right)\dot{x} -\frac{1}{2} \left( c_{2}\sqrt{r+x} -c_{3}\sqrt{r-x} \right) \dot{y} \label{eq.pcf2.58}
\end{eqnarray}
where the Hamiltonian $H= \frac{1}{2}(\dot{x}^{2} +\dot{y}^{2}) +V$.

We note that in the last Table of Ref. \ref{MitsTsam sym} it is proved by using only autonomous QFIs that the potential (\ref{eq.pcf2.57}) is superintegrable because it admits the functionally independent QFIs:
\begin{eqnarray}
I_{s4a} &=&(x\dot{y}-y\dot{x})\dot{y} + \frac{c_{1}x}{r} + \frac{c_{3}(r+x)\sqrt{r-x} -c_{2}(r-x)\sqrt{r+x}}{r}  \label{eq.pcf2.59a} \\
I_{s4b} &=&-(x\dot{y}-y\dot{x})\dot{x} + G(x,y)  \label{eq.pcf2.59b}
\end{eqnarray}
where the smooth function $G(x,y)$ satisfies the conditions $G_{,y} + xV_{,x}=0$ and $G_{,x} + xV_{,y} -2yV_{,x}=0$.

\subsubsection{Parameters $a_{2}, a_{5}$: The components $L_{abc}$ given by (\ref{eq.pcf2.KT2}) depend on $x^{2}, xy, y^{2}$}

\label{sec.Holt3}

We have the following cases:
\bigskip

1) $N(V)=0 \implies F(V)=0$ and $a_{2}\neq0$ (the remaining parameters are fixed to zero).

We find $L_{111}=3a_{2}y^{2}$, $L_{112}=-2a_{2}xy$, $L_{221}= a_{2}x^{2}$, and $Z=Y=-3a_{2}yr^{2}$ where $r= \sqrt{x^{2}+y^{2}}$.

The system of equations (\ref{eq.pcf2.13a}) - (\ref{eq.pcf2.13b}) becomes:
\begin{eqnarray}
r^{2}V_{,xy} +3yV_{,x} &=& 0 \label{eq.pcf2.50a} \\
yr^{2}(V_{,yy} - V_{,xx}) +3r^{2}V_{,y} -6xyV_{,x} &=& 0. \label{eq.pcf2.50b}
\end{eqnarray}
Solving this system, we find the integrable potential (see Table III in Ref. \ref{Karlovini 2000})
\begin{equation}
V(x,y)= \frac{k_{1}}{y^{2}} +\frac{k_{2}}{r} +\frac{k_{3}x}{ry^{2}} \label{eq.pcf2.51}
\end{equation}
where $k_{1}, k_{2}$, and $k_{3}$ are arbitrary constants.

In the last Table of Ref. \ref{MitsTsam sym}, it has been proved that the potential\footnote{To compare with the potential $V_{s3}$ of Ref. \ref{MitsTsam sym}, interchange $x\leftrightarrow y$. This happens because the analysis for $a_{2}\neq0$ is symmetric with the analysis for $a_{5}\neq0$.} (\ref{eq.pcf2.51}) is superintegrable because it admits the functionally independent QFIs:
\begin{eqnarray}
I_{s3a}&=& (x\dot{y} - y\dot{x})^{2} +
2k_{1}\frac{x^{2}}{y^{2}} + 2k_{3}\frac{rx}{y^{2}} \label{eq.pcf2.52a} \\
I_{s3b}&=&(x\dot{y} - y\dot{x})\dot{y} + 2k_{1} \frac{x}{y^{2}} + k_{2}\frac{x}{r} +k_{3}\frac{2x^{2}+y^{2}}{ry^{2}}. \label{eq.pcf2.52b}
\end{eqnarray}
Therefore, the higher order FIs are not necessary. The same applies to the potential (see Table III of Ref. \ref{Karlovini 2000})
\begin{equation}
V=\frac{A}{r} + \frac{B}{r(r+x)} +\frac{C}{r(r-x)} \label{eq.pcf2.53}
\end{equation}
which is of the type (\ref{eq.pcf2.51}) for $k_{1}=B+C, k_{2}=A$, and $k_{3}=C-B$.
\bigskip

2) $N(V)=0 \implies F(V)=0$ and $a_{2},a_{5}$ are the only non-vanishing parameters.

We find $L_{111}=3a_{2}y^{2}$, $L_{112}=-2a_{2}xy +a_{5}y^{2}$, $L_{221}= a_{2}x^{2} -2a_{5}xy$, $L_{222}= 3a_{5}x^{2}$, and $Z= Y=3(a_{5}x-a_{2}y)r^{2}$ where $r= \sqrt{x^{2}+y^{2}}$.

The system of equations (\ref{eq.pcf2.13a}) - (\ref{eq.pcf2.13b}) becomes:
\begin{eqnarray}
3(a_{5}x-a_{2}y)r^{2}V_{,xy} +9a_{5}x^{2}V_{,y} -9a_{2}y^{2} V_{,x} &=& 0 \label{eq.pcf2.60a} \\
3(a_{5}x-a_{2}y)r^{2}(V_{,yy} - V_{,xx}) -9(a_{2}r^{2} -2a_{5}xy)V_{,y} -9(a_{5}r^{2} -2a_{2}xy)V_{,x} &=& 0. \label{eq.pcf2.60b}
\end{eqnarray}
Solving this system, we find the integrable potential
\begin{equation}
V(x,y)= \frac{k_{1}}{(a_{2}y -a_{5}x)^{2}} +\frac{k_{2}}{r} +\frac{k_{3}(a_{2}x +a_{5}y)}{r(a_{2}y -a_{5}x)^{2}} \label{eq.pcf2.61}
\end{equation}
where $k_{1}, k_{2}$, and $k_{3}$ are arbitrary constants. This is a new potential whose integrability is proved by means of CFIs and not by QFIs. We note that for $a_{5}=0$ we obtain the potential (\ref{eq.pcf2.51}). Moreover, the potential (\ref{eq.pcf2.61}) holds for arbitrary values of the constants $a_{2}$ and $a_{5}$ since the system of PDEs (\ref{eq.pcf2.60a}) - (\ref{eq.pcf2.60b}) has been solved under this assumption.

The associated CFI (\ref{eq.pcf2.3}) is
\begin{eqnarray}
J^{(3,2)}_{0} &=& (x\dot{y} -y\dot{x})^{2}(a_{2}\dot{x} +a_{5}\dot{y}) +r^{2}(a_{5}x-a_{2}y)(V_{,y}\dot{x} -V_{,x}\dot{y}) \notag \\
&=&  (x\dot{y} -y\dot{x})^{2}(a_{2}\dot{x} +a_{5}\dot{y}) + \frac{2k_{1}r^{2}}{(a_{2}y -a_{5}x)^{2}} (a_{2}\dot{x} +a_{5}\dot{y}) -\frac{k_{2}(a_{2}y -a_{5}x)}{r} (x\dot{y} -y\dot{x}) + \notag \\
&& + \frac{k_{3}r}{a_{2}y -a_{5}x}(a_{2}\dot{y} -a_{5}\dot{x}) - \frac{k_{3}(a_{2}x +a_{5}y)}{r(a_{2}y -a_{5}x)} (x\dot{y} -y\dot{x}) +\frac{2k_{3}(a_{2}x+a_{5}y)r}{(a_{2}y -a_{5}x)^{2}} (a_{2}\dot{x} +a_{5}\dot{y}). \label{eq.pcf2.62}
\end{eqnarray}

\subsubsection{Parameter $a_{1}$: The components $L_{abc}$ given by (\ref{eq.pcf2.KT2}) depend on $x^{3}, xy^{2}, x^{2}y, y^{3}$}

\label{sec.Holt4}

We have the following cases:
\bigskip

1) $N(V)=0 \implies F(V)=0$ and $a_{1}\neq0$ (the remaining parameters are fixed to zero).

We find $L_{111}=a_{1}y^{3}$, $L_{112}=-a_{1}xy^{2}$, $L_{221}= a_{1}x^{2}y$, $L_{222}= -a_{1}x^{3}$, and $Z=Y=-\frac{3}{4}a_{1}r^{4}$ where $r= \sqrt{x^{2}+y^{2}}$.

The system of equations (\ref{eq.pcf2.13a}) - (\ref{eq.pcf2.13b}) becomes:
\begin{eqnarray}
\frac{r^{4}}{4}V_{,xy} +x^{3}V_{,y} +y^{3}V_{,x} &=& 0 \label{eq.pcf2.54a} \\
\frac{r^{4}}{4}(V_{,yy} - V_{,xx}) +(y^{3}+3x^{2}y)V_{,y} -(x^{3} +3xy^{2})V_{,x} &=& 0. \label{eq.pcf2.54b}
\end{eqnarray}
Solving this system, we find the integrable potential (see Table IV of Ref. \ref{Karlovini 2000} and eq. (3.3.44) of Ref. \ref{Hietarinta 1987})
\begin{equation}
V(x,y)= \frac{k_{1}}{r^{2}} + \frac{k_{2}e^{\sqrt{3}\theta} +k_{3}e^{-\sqrt{3}\theta}}{r^{3}} \label{eq.pcf2.54c}
\end{equation}
where $k_{1}, k_{2}, k_{3}$ are arbitrary constants and $\theta= \tan^{-1}\left(\frac{y}{x}\right)$.

The associated CFI (\ref{eq.pcf2.3}) is
\begin{eqnarray}
J^{(3,2)}_{0}&=& \left(x\dot{y} - y\dot{x} \right)^{3} +\frac{3}{4}r^{4}(V_{,y}\dot{x} -V_{,x}\dot{y}) \notag \\
&=& p_{\theta}^{3} +\frac{3}{4} \left[ 2k_{1} +\frac{3}{r}\left( k_{2}e^{\sqrt{3}\theta} +k_{3}e^{-\sqrt{3}\theta} \right) \right]p_{\theta} +\frac{3\sqrt{3}}{4}\left( k_{2}e^{\sqrt{3}\theta} -k_{3}e^{-\sqrt{3}\theta} \right)p_{r} \label{eq.pcf2.55}
\end{eqnarray}
where $r\dot{r}=x\dot{x} +y\dot{y}$, $p_{\theta}= x\dot{y} - y\dot{x}$ and $p_{r}=\dot{r}$.
\bigskip

2) $N(V)=0 \implies F(V)=0$ and $a_{1}, c\neq0$ are the only non-vanishing parameters.

This case gives nothing new.
\bigskip

3) $N(V)=0 \implies F(V)=0$ and $a_{1}=-\frac{4}{3}$.

This case gives nothing new.

\subsubsection{Mixed choice of the ten parameters $a_{1}, ..., a_{10}$: The components $L_{abc}$ given by (\ref{eq.pcf2.KT2}) depend on products of $x, y$ of mixed degree}

\label{sec.Holt5}

We have the following cases:
\bigskip

1) $F(V)=\lambda V \implies F'=\lambda\neq0$ and $a_{3}, a_{4}$ are the only non-vanishing parameters.

We find $L_{111}=3a_{3}y +a_{4}$, $L_{112}= -a_{3}x$, $Y= -\frac{3}{2}a_{3}(x^{2} +3y^{2}) -3a_{4}y$ and $Z= \lambda V +Y$.

The system of equations (\ref{eq.pcf2.13a}) - (\ref{eq.pcf2.13b}) becomes:
\begin{align}
& \left[ \lambda V -\frac{3}{2}a_{3}(x^{2} +3y^{2}) -3a_{4}y \right] V_{,xy} +\lambda V_{,x}V_{,y} -3(3a_{3}y +a_{4})V_{,x} = 0 \label{eq.pcf2.67a} \\
& \left[ \lambda V -\frac{3}{2}a_{3}(x^{2} +3y^{2}) -3a_{4}y  \right] (V_{,yy} - V_{,xx}) +\lambda \left[ (V_{,y})^{2} -(V_{,x})^{2} \right] -3(3a_{3}y+a_{4})V_{,y} +9a_{3}xV_{,x} = 0. \label{eq.pcf2.67b}
\end{align}
From the structure of the system of equations (\ref{eq.pcf2.67a}) - (\ref{eq.pcf2.67b}), we assume that
\[
F(V)=\lambda V = x^{\nu} + \frac{3}{2}a_{3}(x^{2} +3y^{2}) +3a_{4}y \implies V= \frac{1}{\lambda} \left[ x^{\nu} + \frac{3}{2}a_{3}(x^{2} +3y^{2}) +3a_{4}y \right].
\]
Then, the function $Z= x^{\nu}$ and, by substituting the potential $V$ in the system of equations (\ref{eq.pcf2.67a}) - (\ref{eq.pcf2.67b}), we find that $\nu=2$ and $a_{3}=\frac{1}{3}, -1$. We consider the following subcases.

- Subcase $a_{3}=\frac{1}{3}$.

The potential is of the type
\[
V(x,y)= k_{0}(x^{2} +y^{2}) +k_{1}y
\]
where $k_{0}$ and $k_{1}$ are arbitrary constants. It has been proved in Ref. \ref{MitsTsam sym}, using only QFIs, that this potential is superintegrable.

- Subcase $a_{3}=-1$.

The potential is
\begin{equation}
V(x,y)= \frac{1}{\lambda} \left[ -\frac{1}{2}(x^{2} +9y^{2}) +3a_{4}y \right] = c_{0}(x^{2} +9y^{2}) +c_{1}y \label{eq.pcf2.68}
\end{equation}
where $c_{0}\equiv -\frac{1}{2\lambda}\neq0$ and $c_{1}\equiv \frac{3a_{4}}{\lambda}$. We find that $a_{4}= -\frac{c_{1}}{6c_{0}}$. This is a new integrable potential.

The associated CFI (\ref{eq.pcf2.3}) is (divided by three)
\begin{equation}
J^{(3,2)}_{0}= (x\dot{y} -y\dot{x})\dot{x}^{2} -\frac{c_{1}}{18c_{0}}\dot{x}^{3}+\frac{c_{1}}{3}x^{2}\dot{x} +6c_{0}x^{2}y\dot{x} - \frac{2c_{0}}{3}x^{3}\dot{y}. \label{eq.pcf2.69}
\end{equation}
We note that if we take $c_{1}=0$ and we interchange $x\leftrightarrow y$, then we recover the results of the case 3) of section \ref{sec.Holt2}.

The potential (\ref{eq.pcf2.68}) is also superintegrable because it is of the separable form $V= F_{1}(x) + F_{2}(y)$.
\bigskip

2) $N(V)=0 \implies F(V)=0$, $a_{9}=ka_{8}$ where $k$ is an arbitrary constant, and the remaining parameters are fixed to zero.

This case gives nothing new.

\subsection{Case $C_{ab}=0$ and $V= F_{1}(y+\sqrt{3}x) +F_{2}(y-\sqrt{3}x) +F_{3}(-2y)$}

\label{sec.Toda.potential}

The potential
\begin{equation}
V(x,y)= F_{1}(y+\sqrt{3}x) +F_{2}(y-\sqrt{3}x) +F_{3}(-2y) \label{eq.Tod1}
\end{equation}
where $F_{1}, F_{2}$, and $F_{3}$ are arbitrary smooth functions, is the 2d equivalent system of a three-particle pairwise interacting system whose internal interactions depend on the distances between the particles. The Toda type potential  (\ref{eq.pcf2.16}) is a special case of the potential (\ref{eq.Tod1}).

For the potential (\ref{eq.Tod1}) the conditions (\ref{eq.pcf2.4}) and (\ref{eq.pcf2.5}) become:
\begin{eqnarray}
B_{1,x} -3L_{111} V_{,x} -3L_{112}V_{,y} &=& 0 \label{eq.Tod1a} \\
B_{1,y} + B_{2,x} -6L_{112}V_{,x} -6L_{221}V_{,y} &=& 0 \label{eq.Tod1b} \\
B_{2,y} -3L_{221}V_{,x} -3L_{222}V_{,y} &=& 0 \label{eq.Tod1c} \\
B_{1}V_{,x} + B_{2}V_{,y} &=& s. \label{eq.Tod1d}
\end{eqnarray}

In section \ref{sec.pot.cfi2.2Holt}, using Holt's method, we proved that the Toda type potential admits the CFI (\ref{eq.pcf2.17}). This CFI has a leading part of the type $(\dot{x}^{2} - 3\dot{y}^{2})\dot{x}$. Therefore, in order to solve the system of equations (\ref{eq.Tod1a}) - (\ref{eq.Tod1c}) for the vector $B_{a}(x,y)$, it is proper to assume that $a_{4}=-a_{10}=1$ are the only non-vanishing parameters of the third order KT $L_{abc}$ given by (\ref{eq.pcf2.KT2}), that is, $L_{111}=-L_{221}= 1$ and $L_{112}=L_{222}=0$. Then, the system of equations (\ref{eq.Tod1a}) - (\ref{eq.Tod1c}) becomes:
\begin{eqnarray}
B_{1,x} &=& 3V_{,x} \label{eq.Tod2a} \\
B_{1,y} + B_{2,x} &=& -6V_{,y} \label{eq.Tod2b} \\
B_{2,y} &=& -3V_{,x}. \label{eq.Tod2c}
\end{eqnarray}
The system of equations (\ref{eq.Tod2a}) - (\ref{eq.Tod2c}) has the solution (see eq. (3.3.35) of Ref. \ref{Hietarinta 1987})
\begin{equation}
B_{a}=
\left(
  \begin{array}{c}
    3\left[ F_{1}(y+\sqrt{3}x) +F_{2}(y-\sqrt{3}x) -2F_{3}(-2y) \right] \\
    -3\sqrt{3} \left[ F_{1}(y+\sqrt{3}x) -F_{2}(y-\sqrt{3}x) \right] \\
  \end{array}
\right). \label{eq.Tod3}
\end{equation}
Substituting the vector $B_{a}$ in the remaining equation (\ref{eq.Tod1d}), we find $s=0$ and the cyclic condition (see eq. (51) of Ref. \ref{Karlovini 2000})
\begin{equation}
\frac{dF_{1}}{dw_{1}}(F_{2}-F_{3}) +\frac{dF_{2}}{dw_{2}}(F_{3}-F_{1}) +\frac{dF_{3}}{dw_{3}}(F_{1}-F_{2}) =0 \label{eq.Tod4}
\end{equation}
where $w_{1}\equiv y+\sqrt{3}x$, $w_{2}\equiv y-\sqrt{3}x$, and $w_{3}\equiv -2y$. Condition (\ref{eq.Tod4}) can also be derived using the Lax-pair approach\footnote{M.A. Olshanetsky and A.M. Perelomov, \emph{`Classical integrable finite-dimensional systems related to Lie algebras'}, Phys. Rep. \textbf{71}(5), 313 (1981). \label{Olshanetsky 1981}} (see e.g. Refs. \ref{Hietarinta 1987}, \ref{Olshanetsky 1981} and references therein).

The associated CFI (\ref{eq.pcf2.3}) is
\begin{eqnarray}
J^{(3,2)}_{0}&=& \dot{x}^{3} -3\dot{x}\dot{y}^{2} + 3\left[ F_{1}(y+\sqrt{3}x) +F_{2}(y-\sqrt{3}x) -2F_{3}(-2y) \right]\dot{x} - \notag \\
&& -3\sqrt{3} \left[ F_{1}(y+\sqrt{3}x) -F_{2}(y-\sqrt{3}x) \right]\dot{y} \label{eq.Tod5}
\end{eqnarray}
provided that the functions $F_{1}, F_{2}$, and $F_{3}$ satisfy the condition (\ref{eq.Tod4}). It can be easily proved that the Toda type potential (\ref{eq.pcf2.16}) satisfies this condition and the CFI (\ref{eq.Tod5}) reduces to the CFI (\ref{eq.pcf2.17}).

\subsection{Case $C_{ab}=0$ and $s=0$ - The integrability conditions method}

\label{sec.integr.con}

In this case, the CFI (\ref{eq.pcf2.3}) becomes the autonomous CFI\footnote{We have replaced the third order KT (\ref{eq.pcf2.KT2}) in the CFI (\ref{eq.pcf2.3}).}
\begin{eqnarray}
J^{(3,2)}_{0}&=& -a_{1}L^{3} +3\left( a_{2}\dot{x} +a_{5}\dot{y} \right) L^{2} -3\left( a_{3}\dot{x}^{2} +a_{8}\dot{x}\dot{y} -a_{6}\dot{y}^{2} \right) L +a_{4}\dot{x}^{3} +3a_{9}\dot{x}^{2}\dot{y} + \notag \\
&& +3a_{10}\dot{x}\dot{y}^{2} +a_{7}\dot{y}^{3} +B_{1}(x,y)\dot{x} +B_{2}(x,y)\dot{y} \label{eq.intc.0}
\end{eqnarray}
where $L= x\dot{y} -y\dot{x}$ is the angular momentum, $a_{1}, ..., a_{10}$ are the parameters of the third order KT (\ref{eq.pcf2.KT2}), and the vector components $B_{1}(x,y)$ and $B_{2}(x,y)$ satisfy the four PDEs (see eqs. (\ref{eq.pcf2.4}) and (\ref{eq.pcf2.5})):
\begin{eqnarray}
B_{1}V_{,x} +B_{2}V_{,y} &=& 0 \label{eq.intc.0.1} \\
B_{1,x} &=& 3 \left( L_{111}V_{,x} +L_{112}V_{,y} \right) \label{eq.intc.0.2} \\
B_{2,y}&=& 3\left( L_{221}V_{,x} +L_{222}V_{,y} \right) \label{eq.intc.0.3} \\
B_{1,y} +B_{2,x} &=& 6 \left( L_{112}V_{,x} +L_{221}V_{,y} \right). \label{eq.intc.0.4}
\end{eqnarray}
The tensor $L_{abc}$ is the third order KT (\ref{eq.pcf2.KT2}) and $V(x,y)$ is the potential. The above PDEs are the conditions (2.6) - (2.9) of Ref. \ref{Gravel 2002}. In the notation of Ref. \ref{Gravel 2002}, we have: $g_{a}= B_{a}$, $f_{1}= L_{111}$, $f_{2}= 3L_{112}$, $f_{3}= 3L_{221}$, $f_{4}= L_{222}$, $A_{300}= -a_{1}$, $A_{210}= 3a_{2}$, $A_{201}= 3a_{5}$, $A_{120}= -3a_{3}$, $A_{111}= -3a_{8}$, $A_{102}= 3a_{6}$, $A_{030}= a_{4}$, $A_{021}= 3a_{9}$, $A_{012}= 3a_{10}$, and $A_{003}= a_{7}$.

In sections \ref{sec.pot.cfi2.2Holt} and \ref{sec.Toda.potential}, we solved directly the system of PDEs (\ref{eq.intc.0.1}) - (\ref{eq.intc.0.4}) by assuming either a specific form for the vector $B_{a}$ such that the quantity $B_{a}V^{,a}$ vanishes identically (Holt's method), or a specific functional form for the potential $V$. These assumptions resulted: \newline
a. In the first case (see sec. \ref{sec.pot.cfi2.2Holt}), to the second order non-linear system of PDEs (\ref{eq.pcf2.13a}) - (\ref{eq.pcf2.13b}). For certain choices of the involved parameters $a_{1}, ..., a_{10}, c$ and the function $N(V)$, we solved this system and determined third order integrable and superintegrable potentials $V(x,y)$. \newline
b. In the second case (see sec. \ref{sec.Toda.potential}), to a linear first order system of PDEs, which for certain choices of the third order KT (\ref{eq.pcf2.KT2}), gave compatible solutions for the vector $B_{a}(x,y)$.

In this section, we shall treat the system of PDEs (\ref{eq.intc.0.1}) - (\ref{eq.intc.0.4}) differently. In particular, we will derive the integrability conditions of  (\ref{eq.pcf2.4}), i.e. equations (\ref{eq.intc.0.2}) - (\ref{eq.intc.0.4}), from which we will find for specific choices of the KT $L_{abc}$ compatible potentials $V(x,y)$. These potentials will be replaced in the original system (\ref{eq.intc.0.1}) - (\ref{eq.intc.0.4}) in order to compute the vector $B_{a}$ and, finally, the associated CFI (\ref{eq.intc.0}).

\subsubsection{The integrability conditions of (\ref{eq.pcf2.4}) in a flat space}

\label{sec.integr.con.1}

In a general $n$-dimensional flat space $\{q^{a}\}$, the condition (\ref{eq.pcf2.4}) reads
\begin{equation}
B_{a,b} +B_{b,a}= M_{ab} \label{eq.intc.1}
\end{equation}
where $M_{ab}\equiv 6L_{abc} V^{,c}$ is a second order symmetric tensor and $L_{abc}$ is given by (\ref{eq.pcf2.KT2}).

Taking the partial derivative of (\ref{eq.intc.1}) with cyclic permutation of the indices, we obtain:
\begin{eqnarray}
B_{a,bc} +B_{b,ac} &=& M_{ab,c} \label{eq.intc.2} \\
B_{a,cb} +B_{c,ab} &=& M_{ac,b} \label{eq.intc.3} \\
-B_{b,ca} -B_{c,ba} &=& -M_{bc,a}. \label{eq.intc.4}
\end{eqnarray}
Adding by parts equations (\ref{eq.intc.2}) - (\ref{eq.intc.4}), we find
\begin{equation}
B_{a,bc}= \frac{1}{2} \left( M_{ab,c} +M_{ac,b} -M_{bc,a} \right). \label{eq.intc.5}
\end{equation}
Since $B_{a,bc}= B_{a,cb}$, the pair of indices $b, c$ may be thought of as a collective index $A\equiv bc$. Then, we have the integrability conditions:
\begin{equation}
B_{a,[Ad]}=0 \implies B_{a,bcd}= B_{a,dbc} \implies B_{a,bcd}= B_{a,bdc} \implies B_{a,b[cd]}=0. \label{eq.intc.6}
\end{equation}
Replacing (\ref{eq.intc.5}) in (\ref{eq.intc.6}), we find the condition (see eq. (2.23) of Ref. \ref{Horwood 2007})
\begin{equation}
M_{[a|[c,d]|b]} = 0. \label{eq.intc.7}
\end{equation}

In $E^{2}$, only the conditions with $a\neq b$ and $c\neq d$ survive, and equation (\ref{eq.intc.7}) gives
\[
M_{11,22} +M_{22,11} -2M_{12,12} =0 \implies
\]
\[
\left( L_{111} V_{,x} +L_{112}V_{,y} \right)_{,yy} +\left( L_{221}V_{,x} +L_{222}V_{,y} \right)_{,xx} -2\left( L_{112}V_{,x} +L_{221}V_{,y} \right)_{,xy} =0 \implies
\]
\begin{eqnarray}
0&=& L_{221}V_{,xxx} +\left( L_{222} -2L_{112} \right)V_{,xxy} +\left(L_{111} -2L_{221}\right)V_{,xyy} +L_{112}V_{,yyy} + \notag \\
&& +2\left( L_{221,x} -L_{112,y} \right)V_{,xx} +2\left( L_{111,y} +L_{222,x} -L_{112,x} -L_{221,y} \right) V_{,xy} + \notag \\
&& +2\left( L_{112,y} -L_{221,x} \right) V_{,yy} +\left( L_{221,xx} -2L_{112,xy} +L_{111,yy} \right)V_{,x} + \notag \\
&& +\left( L_{222,xx} -2L_{221,xy} +L_{112,yy} \right)V_{,y}. \label{eq.intc.8}
\end{eqnarray}
This is a third order linear PDE, which coincides with eq. (2.10) of Ref. \ref{Gravel 2002}.

Therefore, in order to find integrable potentials $V(x,y)$ that admit CFIs of the form (\ref{eq.intc.0}), we have to solve the system of PDEs (\ref{eq.intc.0.1}) - (\ref{eq.intc.0.4}) together with the integrability condition (\ref{eq.intc.8}). The general solution for the above system is not possible; therefore, we consider several cases concerning the parameters of the KT $L_{abc}$ (\ref{eq.pcf2.KT2}) and the functional form of $V$.

\subsubsection{Separable potentials $V(x,y)= F_{1}(x) +F_{2}(y)$}

\label{sec.intc.1}

We consider separable potentials of the general form\footnote{S. Gravel, \emph{`Hamiltonians separable in Cartesian coordinates and third-order integrals of motion'}, J. Math. Phys. \textbf{45}(3), 1003 (2004). \label{Gravel 2004}}
\begin{equation}
V(x,y)= F_{1}(x) +F_{2}(y) \label{eq.sep.1}
\end{equation}
where $F_{1}$ and $F_{2}$ are arbitrary smooth functions of their arguments. Due to the form (\ref{eq.sep.1}), such potentials already admit the independent autonomous QFIs
\begin{equation}
I_{1}= \frac{1}{2}\dot{x}^{2} +F_{1}(x), \enskip I_{2}= \frac{1}{2}\dot{y}^{2} +F_{2}(y). \label{eq.sep.2}
\end{equation}
Therefore, if we find for special choices of $F_{1}$ and $F_{2}$ one additional CFI of the form (\ref{eq.intc.0}), the resulting potentials (\ref{eq.sep.1}) become superintegrable. This way of research has been followed in Refs \ref{Marquette 2008} and \ref{Gravel 2004}, and eight superintegrable potentials have been found (see potentials (C.1) - (C.8) in the appendix of Ref. \ref{Gravel 2004} and sec. 4 of Ref. \ref{Marquette 2008}). However, from these potentials, only five are purely third order superintegrable because the potentials (C.1) - (C.3) of Ref. \ref{Gravel 2004} can be proved to be superintegrable by using only QFIs.

Comparing the results of Ref. \ref{Gravel 2004} with those found in the previous sections by using Holt's method, we note that the potentials (C.1) and (C.2) are subcases of (\ref{eq.pcf2.30}); (C.3) is the (\ref{eq.pcf2.20}) for $x \leftrightarrow y$; (C.4) is a subcase of (\ref{eq.pcf2.68}) for $c_{1}=0$; (C.7) is the (\ref{eq.pcf2.45}); the potential (\ref{eq.pcf2.48}) is a special case of (C.5); and the potentials (C.6) and (C.8) could not be found due to the restrictive assumption (\ref{eq.pcf2.6}) for the vector $B_{a}$. However, in what follows, we shall derive these two potentials as subcases of more general ones. In addition, we shall find some new ones.

Replacing (\ref{eq.sep.1}) in the PDEs (\ref{eq.intc.0.1}) - (\ref{eq.intc.0.4}) and (\ref{eq.intc.8}), we obtain\footnote{The vector components (\ref{eq.sep3.1}) and (\ref{eq.sep3.2}) are derived from the PDEs (\ref{eq.intc.0.2}) and (\ref{eq.intc.0.3}).}:
\begin{eqnarray}
B_{1}(x,y)&=& 3\left( a_{1}y^{3} +3a_{2}y^{2} +3a_{3}y +a_{4} \right) F_{1} -\frac{3}{2}x^{2}\left( a_{1}y^{2} +2a_{2}y +a_{3} \right) F_{2,y} + \notag \\
&& + 3x \left( a_{5}y^{2} +a_{8}y +a_{9} \right) F_{2,y} +f_{1}(y) \label{eq.sep3.1} \\
B_{2}(x,y)&=& -3\left( a_{1}x^{3} -3a_{5}x^{2} -3a_{6}x -a_{7} \right) F_{2} +\frac{3}{2}y^{2}\left( a_{1}x^{2} -2a_{5}x -a_{6} \right) F_{1,x}+ \notag \\
&& +3y \left( a_{2}x^{2} -a_{8}x +a_{10} \right) F_{1,x} +f_{2}(x) \label{eq.sep3.2} \\
B_{1,y} +B_{2,x}&=& 6 \left( L_{112} F_{1,x} +L_{221} F_{2,y} \right) \label{eq.sep3.3} \\
B_{1}F_{1,x} +B_{2}F_{2,y}&=& 0 \label{eq.sep3.4} \\
0&=& L_{221} F_{1,xxx} +L_{112}F_{2,yyy} +2\left( L_{221,x} -L_{112,y} \right)F_{1,xx} +2\left( L_{112,y} -L_{221,x} \right) F_{2,yy} +\notag \\
&& +\left( L_{221,xx} -2L_{112,xy} +L_{111,yy} \right)F_{1,x} + +\left( L_{222,xx} -2L_{221,xy} +L_{112,yy} \right)F_{2,y} \label{eq.sep3.5}
\end{eqnarray}
where $f_{1}(y)$ and $f_{2}(x)$ are arbitrary smooth functions.

The components of the vector $B_{a}$ are given in (\ref{eq.sep3.1}) - (\ref{eq.sep3.2}); hence, we have to solve the system of ODEs (\ref{eq.sep3.3}) - (\ref{eq.sep3.5}) with four unknown functions $F_{1}(x), F_{2}(y), f_{1}(y)$, and $f_{2}(x)$. We consider several cases concerning the parameters of the KT (\ref{eq.pcf2.KT2}).
\bigskip

1) The only non-vanishing parameters are the $a_{4}a_{9}\neq 0$.

The KT (\ref{eq.pcf2.KT2}) is $L_{111}= a_{4}$, $L_{112}= a_{9}$ and $L_{221}=L_{222}=0$.

From (\ref{eq.sep3.1}) - (\ref{eq.sep3.2}), the vector $B_{a}$ is
\begin{equation}
B_{a}=
\left(
  \begin{array}{c}
    3a_{4}F_{1} +3a_{9}xF_{2,y} +f_{1} \\
    f_{2} \\
  \end{array}
\right). \label{eq.sep4}
\end{equation}

The system of ODEs (\ref{eq.sep3.3}) - (\ref{eq.sep3.5}) becomes:
\begin{eqnarray}
3a_{9} \left( xF_{2,yy} -2F_{1,x} \right) +f_{1,y} +f_{2,x} &=& 0 \label{eq.sep5.1} \\
F_{2,yyy}&=& 0 \label{eq.sep5.2} \\
3a_{4}F_{1}F_{1,x} +3a_{9}xF_{1,x}F_{2,y} +F_{1,x}f_{1} +F_{2,y}f_{2}&=& 0. \label{eq.sep5.3}
\end{eqnarray}

From (\ref{eq.sep5.2}), we find that $F_{2}(y)= c_{1}y^{2} +c_{2}y$ where $c_{1}$ and $c_{2}$ are arbitrary constants. Replacing this function in the above equations, we find that
\begin{equation}
B_{a}=
\left(
  \begin{array}{c}
    3a_{4}F_{1}(x) +3a_{9}x\left( 2c_{1}y +c_{2}\right) +f_{1}(y) \\
    f_{2}(x) \\
  \end{array}
\right) \label{eq.sep6.1}
\end{equation}
and the conditions:
\begin{eqnarray}
6a_{9} \left( c_{1}x -F_{1,x} \right) +f_{1,y} +f_{2,x} &=& 0 \label{eq.sep6.2} \\
3a_{4}F_{1}(x)F_{1,x} +3a_{9}xF_{1,x}\left( 2c_{1}y +c_{2}\right) +F_{1,x}f_{1}(y) +\left( 2c_{1}y +c_{2}\right)f_{2}(x)&=& 0. \label{eq.sep6.3}
\end{eqnarray}
Moreover, the potential (\ref{eq.sep.1}) becomes
\begin{equation}
V(x,y)= c_{1}y^{2} +c_{2}y +F_{1}(x). \label{eq.sep6.4}
\end{equation}

Integrating (\ref{eq.sep6.2}), we find
\[
f_{1}(y)= c_{3}y +c_{4}, \enskip f_{2}(x)= 6a_{9}F_{1}(x) -3a_{9}c_{1}x^{2} -c_{3}x +c_{5}
\]
where $c_{3}, c_{4}$, and $c_{5}$ are arbitrary constants. Substituting these functions in (\ref{eq.sep6.1}) and (\ref{eq.sep6.3}), we get
\begin{equation}
B_{a}=
\left(
  \begin{array}{c}
    3a_{4}F_{1}(x) +3a_{9}c_{2}x +6a_{9}c_{1}xy +c_{3}y +c_{4} \\
    6a_{9}F_{1}(x) -3a_{9}c_{1}x^{2} -c_{3}x +c_{5} \\
  \end{array}
\right) \label{eq.sep7.1}
\end{equation}
where the remaining unknown function $F_{1}(x)$ satisfies the condition
\begin{equation}
3a_{4}F_{1}F_{1}' + \left( 3a_{9}xF_{1}' +6a_{9}F_{1} -3a_{9}c_{1}x^{2} -c_{3}x +c_{5} \right) \left( 2c_{1}y +c_{2}\right) +F_{1}' \left( c_{3}y +c_{4} \right)= 0 \label{eq.sep7.2}
\end{equation}
where $F_{1}'\equiv \frac{dF_{1}}{dx}= F_{1,x}$.

Condition (\ref{eq.sep7.2}) is written equivalently as
\begin{eqnarray}
0&=& \left( 3a_{4}F_{1}F_{1}' +3a_{9}c_{2}xF_{1}' +c_{4}F_{1}' +6a_{9}c_{2}F_{1} -3a_{9}c_{1}c_{2}x^{2} -c_{2}c_{3}x +c_{2}c_{5} \right) + \notag \\
&& + 2y \left( 3a_{9}c_{1}xF_{1}' +F_{1}'\frac{c_{3}}{2} +6a_{9}c_{1}F_{1} -3a_{9}c_{1}^{2}x^{2} -c_{1}c_{3}x +c_{1}c_{5} \right). \label{eq.sep8}
\end{eqnarray}
This equation is of the form $A(x) +yB(x) =0$; therefore, it follows that $A(x)=B(x)=0$, that is:
\begin{eqnarray}
3a_{4}F_{1}F_{1}' +3a_{9}c_{2}xF_{1}' +c_{4}F_{1}' +6a_{9}c_{2}F_{1} -3a_{9}c_{1}c_{2}x^{2} -c_{2}c_{3}x +c_{2}c_{5} &=& 0 \label{eq.sep9.1} \\
3a_{9}c_{1}xF_{1}' +F_{1}'\frac{c_{3}}{2} +6a_{9}c_{1}F_{1} -3a_{9}c_{1}^{2}x^{2} -c_{1}c_{3}x +c_{1}c_{5} &=& 0. \label{eq.sep9.2}
\end{eqnarray}

The only non-trivial case is for \underline{$c_{1}=c_{3}=0$.}

In this case, potential (\ref{eq.sep6.4}) is $V= c_{2}y +F_{1}(x)$ and
\begin{equation}
B_{a}=
\left(
  \begin{array}{c}
    3a_{4}F_{1}(x) +3a_{9}c_{2}x +c_{4} \\
    6a_{9}F_{1}(x) +c_{5} \\
  \end{array}
\right). \label{eq.sep10}
\end{equation}

Condition (\ref{eq.sep9.2}) vanishes identically, while condition (\ref{eq.sep9.1}) becomes
\begin{equation}
3a_{4}F_{1}F_{1}' +3a_{9}c_{2}xF_{1}' +c_{4}F_{1}' +6a_{9}c_{2}F_{1} +c_{2}c_{5}= 0. \label{eq.sep11}
\end{equation}
This first order non-linear ODE admits the following FI
\begin{equation}
\left( F_{1} +\frac{c_{5}}{6a_{9}} \right) \left( F_{1} +\frac{3a_{9}c_{2}}{a_{4}}x +\frac{c_{4}}{a_{4}} -\frac{c_{5}}{3a_{9}} \right)^{2}= k \label{eq.sep12}
\end{equation}
where $k$ is an arbitrary constant along solutions $F_{1}(x)$ of (\ref{eq.sep11}). For $c_{4}=c_{5}=0$ and by setting $b\equiv -\frac{3a_{9}c_{2}}{a_{4}}$, the FI (\ref{eq.sep12}) reduces to the well-known condition\footnote{The condition (\ref{eq.sep13}) is written $F_{1}^{3} -2bxF_{1}^{2} +b^{2}x^{2}F_{1} -k =0$. Therefore, in last lines of page 6 in Ref. \ref{Marquette 2008}, the term $x^{4}$ should be corrected to $x^{2}$.
} (see eqs. (32) and (34) of Ref. \ref{Gravel 2004})
\begin{equation}
F_{1} \left( F_{1} -bx \right)^{2}= k. \label{eq.sep13}
\end{equation}
We conclude that the superintegrable potential (C.8) of Ref. \ref{Gravel 2004} is a subcase of $V= c_{2}y +F_{1}(x)$ for $c_{4}=c_{5}=0$. Recall that the function $F_{1}(x)$ is determined now from the more general cubic algebraic equation (\ref{eq.sep12}).

The associated CFI (\ref{eq.intc.0}) is
\begin{equation}
J^{(3,2)}_{0}= a_{4}\dot{x}^{3} +3a_{9}\dot{x}^{2}\dot{y} +\left(3a_{4}F_{1} +3a_{9}c_{2}x +c_{4}\right) \dot{x} +\left( 6a_{9}F_{1} +c_{5} \right) \dot{y} \label{eq.sep14}
\end{equation}
where the function $F_{1}(x)$ is a solution of (\ref{eq.sep12}). We note that for $c_{4}=c_{5}=0$, $b\equiv -3a_{9}c_{2}$ and by multiplying with $\frac{2c_{2}}{a_{4}}$ the CFI (\ref{eq.sep14}) coincides with eq. (41) of Ref. \ref{Gravel 2004}.

The number of the constants is reduced by one if we rename them as follows: $k_{1}\equiv \frac{a_{4}}{3a_{9}}$, $k_{2}\equiv \frac{c_{4}}{3a_{9}}$, $k_{3}\equiv \frac{c_{5}}{3a_{9}}$, $k_{4}\equiv k$, and $c\equiv c_{2}$. Then, the potential
\begin{equation}
V(x,y)= cy +F(x) \label{eq.sep15}
\end{equation}
admits the CFI
\begin{equation}
I= k_{1}\dot{x}^{3} +\dot{x}^{2}\dot{y} +\left( 3k_{1}F +cx +k_{2} \right) \dot{x} +\left( 2F +k_{3}\right)\dot{y} \label{eq.sep16}
\end{equation}
where the function $F(x)$ is a solution of the cubic algebraic equation
\begin{equation}
\left( F +\frac{k_{3}}{2} \right) \left( F +\frac{c}{k_{1}}x +\frac{k_{2}}{k_{1}} -k_{3} \right)^{2} =k_{4}. \label{eq.sep17}
\end{equation}

We note that in Ref. \ref{MitsTsam sym} it has been shown that all the potentials of the form (\ref{eq.sep15}) are superintegrable because of the additional time-dependent LFI $I= \dot{y} +ct$. Therefore, there is no need for looking for CFIs.
\bigskip

2) The only non-vanishing parameters are the $a_{4}a_{7}\neq0$.

The KT (\ref{eq.pcf2.KT2}) is $L_{111}=a_{4}$, $L_{222}=a_{7}$ and $L_{112}= L_{221}=0$.

From equations (\ref{eq.sep3.1}) - (\ref{eq.sep3.2}), the vector
\begin{equation}
B_{a}=
\left(
  \begin{array}{c}
    3a_{4}F_{1}(x) +f_{1}(y) \\
    3a_{7}F_{2}(y) +f_{2}(x) \\
  \end{array}
\right). \label{eq.sep18}
\end{equation}

Equations (\ref{eq.sep3.3}) - (\ref{eq.sep3.4}) become:
\begin{eqnarray}
\frac{df_{1}(y)}{dy} +\frac{df_{2}(x)}{dx} &=& 0 \label{eq.sep19.1} \\
3a_{4}F_{1}(x)\frac{dF_{1}}{dx} +f_{1}(y)\frac{dF_{1}}{dx} +3a_{7}F_{2}(y)\frac{dF_{2}}{dy} +f_{2}(x)\frac{dF_{2}}{dy}&=& 0 \label{eq.sep19.2}
\end{eqnarray}
and the integrability condition (\ref{eq.sep3.5}) vanishes identically.

From (\ref{eq.sep19.1}), we find that $f_{1}(y)= c_{1}y +c_{2}$ and $f_{2}(x)= -c_{1}x +c_{3}$, where $c_{1}, c_{2}$, and $c_{3}$ are arbitrary constants. Replacing these functions in (\ref{eq.sep19.2}), we see that non-trivial potentials exist only when $c_{1}=0$; therefore, the vector (\ref{eq.sep18}) is
\begin{equation}
B_{a}=
\left(
  \begin{array}{c}
    3a_{4}F_{1}(x) +c_{2} \\
    3a_{7}F_{2}(y) +c_{3} \\
  \end{array}
\right) \label{eq.sep20}
\end{equation}
and the functions $F_{1}, F_{2}$ satisfy the condition
\begin{equation}
\underbrace{3a_{4}F_{1}(x)\frac{dF_{1}}{dx} +c_{2}\frac{dF_{1}}{dx}}_{\equiv A(x)} \underbrace{+3a_{7}F_{2}(y)\frac{dF_{2}}{dy} +c_{3}\frac{dF_{2}}{dy}}_{B(y)}=0. \label{eq.sep21}
\end{equation}
The later is of the form $A(x) +B(y) =0$; therefore, $A(x)=\frac{3c_{4}}{2}$ and $B(y)=-\frac{3c_{4}}{2}$ where $c_{4}$ is an arbitrary constant. Solving these ODEs, we find that
\begin{equation}
F_{1}(x)= \pm \sqrt{-\frac{c_{4}}{a_{4}}x +\frac{c_{2}^{2}}{9a_{4}^{2}}} -\frac{c_{2}}{3a_{4}}, \enskip F_{2}(y)= \pm \sqrt{\frac{c_{4}}{a_{7}}y +\frac{c_{3}^{2}}{9a_{7}^{2}}} -\frac{c_{3}}{3a_{7}}. \label{eq.sep22}
\end{equation}

The associated CFI (\ref{eq.intc.0}) is
\begin{equation}
J^{(3,2)}_{0}= a_{4}\dot{x}^{3} +a_{7}\dot{y}^{3} +\left( 3a_{4}F_{1} +c_{2} \right)\dot{x} +\left( 3a_{7}F_{2} +c_{3} \right)\dot{y}. \label{eq.sep23}
\end{equation}

The above expressions are simplified by renaming the constants as follows:
\[
k_{1}\equiv -\frac{c_{4}}{a_{4}}, \enskip k_{2}\equiv \frac{c_{4}}{a_{7}}, \enskip k_{3}= \frac{c_{2}}{3a_{4}}, \enskip k_{4}\equiv \frac{c_{3}}{3a_{7}}.
\]

Then, we have the potentials $V= F_{1}(x) +F_{2}(y)$ with
\begin{equation}
F_{1}(x)= \pm \sqrt{k_{1}x +k_{3}^{2}} -k_{3}, \enskip F_{2}(y)= \pm \sqrt{k_{2}y +k_{4}^{2}} -k_{4} \label{eq.sep24}
\end{equation}
which admit the CFIs
\begin{equation}
J^{(3,2)}_{0}= k_{2}\dot{x}^{3} -k_{1}\dot{y}^{3} +3k_{2}\left( F_{1} +k_{3} \right)\dot{x} -3k_{1}\left( F_{2} +k_{4} \right)\dot{y}. \label{eq.sep25}
\end{equation}
If we introduce the new constants $d_{1}= \frac{k_{3}^{2}}{k_{1}}$ and $d_{2}= \frac{k_{4}^{2}}{k_{2}}$, then the constants $k_{3}$ and $k_{4}$ can be removed due to the translations $x \to x -d_{1}$ and $y \to y -d_{2}$. Therefore, we can set $k_{3}=k_{4}=0$.

We observe that the superintegrable potential (C.5) of Ref. \ref{Gravel 2004} coincides with the potentials defined by (\ref{eq.sep24}). Moreover, the potential (\ref{eq.pcf2.48}) is also a subcase of the above potentials for $k_{1}=k_{2}=k^{2}$, $k_{3}=k_{4}=0$ and $F_{1}(x)= \sqrt{k_{1}x}$.
\bigskip

3) The only non-vanishing parameter is the $a_{3}\neq0$.

In this case, $L_{111}=3a_{3}y$, $L_{112}= -a_{3}x$ and $L_{221}= L_{222}= 0$.

The vector given by (\ref{eq.sep3.1}) - (\ref{eq.sep3.2}) is
\begin{equation}
B_{a}=
\left(
  \begin{array}{c}
    9a_{3}yF_{1}(x) -\frac{3}{2}a_{3}x^{2}\frac{dF_{2}}{dy} +f_{1}(y) \\
    f_{2}(x) \\
  \end{array}
\right). \label{eq.sep26}
\end{equation}

The system of ODEs (\ref{eq.sep3.3}) - (\ref{eq.sep3.5}) becomes:
\begin{eqnarray}
6a_{3}x \frac{dF_{1}}{dx} +9a_{3}F_{1} +\frac{df_{2}}{dx} +\frac{df_{1}}{dy} -\frac{3}{2}a_{3}x^{2} \frac{d^{2}F_{2}}{dy^{2}} &=& 0 \label{eq.sep27.1} \\
f_{2}\frac{dF_{2}}{dy} -\frac{3}{2}a_{3}x^{2} \frac{dF_{1}}{dx} \frac{dF_{2}}{dy} +f_{1}\frac{dF_{1}}{dx} +9a_{3}yF_{1} \frac{dF_{1}}{dx} &=& 0 \label{eq.sep27.2} \\
\frac{d^{3}F_{2}}{dy^{3}} &=& 0. \label{eq.sep27.3}
\end{eqnarray}
From (\ref{eq.sep27.3}), we find that $F_{2}(y)= c_{1}y^{2} +c_{2}y$ where $c_{1}$ and $c_{2}$ are arbitrary constants. Replacing this function in the remaining equations, we obtain:
\begin{eqnarray}
6a_{3}x \frac{dF_{1}}{dx} -3c_{1}a_{3}x^{2} +9a_{3}F_{1} +\frac{df_{2}}{dx} +\frac{df_{1}}{dy} &=& 0 \label{eq.sep28.1} \\
c_{2}f_{2} +2c_{1}yf_{2} -\frac{3}{2}a_{3}c_{2}x^{2} \frac{dF_{1}}{dx} -3a_{3}c_{1}x^{2}y \frac{dF_{1}}{dx} +f_{1}\frac{dF_{1}}{dx} +9a_{3}yF_{1} \frac{dF_{1}}{dx} &=& 0. \label{eq.sep28.2}
\end{eqnarray}
Solving (\ref{eq.sep28.1}), we find that $f_{1}(y)= c_{3}y +c_{4}$ where $c_{3}$ and $c_{4}$ are arbitrary constants; therefore, the system of ODEs (\ref{eq.sep28.1}) - (\ref{eq.sep28.2}) becomes:
\begin{eqnarray}
6a_{3}x \frac{dF_{1}}{dx} -3c_{1}a_{3}x^{2} +9a_{3}F_{1} +\frac{df_{2}}{dx} +c_{3} &=& 0 \label{eq.sep29.1} \\
c_{2}f_{2} -\frac{3}{2}a_{3}c_{2}x^{2} \frac{dF_{1}}{dx} +c_{4}\frac{dF_{1}}{dx} &=& 0. \label{eq.sep29.2} \\
2c_{1}f_{2} -3a_{3}c_{1}x^{2} \frac{dF_{1}}{dx} +c_{3}\frac{dF_{1}}{dx} +9a_{3}F_{1} \frac{dF_{1}}{dx} &=& 0. \label{eq.sep29.3}
\end{eqnarray}

We consider the following non-trivial cases:
\bigskip

3.1. Case $c_{2}=0$ and $c_{1}\neq0$.

From equations (\ref{eq.sep29.1}) - (\ref{eq.sep29.3}), we find that $c_{4}=0$ and $f_{2}(x)= -\frac{1}{2} F_{1}' \left( \frac{9a_{3}}{c_{1}}F_{1} -3a_{3}x^{2} +\frac{c_{3}}{c_{1}} \right)$, where the function $F_{1}(x)$ satisfies the condition
\begin{equation}
D \equiv 3a_{3}\left[ \frac{c_{1}}{2}x^{2}F_{1}'' +3c_{1}\left( xF_{1} \right)' -\frac{3}{4}\left(F_{1}^{2}\right)'' -c_{1}^{2}x^{2} \right] +c_{3}\left( c_{1}-\frac{F_{1}''}{2}\right) =0. \label{eq.sep30}
\end{equation}
If we multiply with $8$ and set $a_{3}= -\frac{A_{120}}{3}$, $c_{1}=a$ and $c_{3}=\eta$, we get condition (16) of Ref. \ref{Gravel 2004} for $\hbar=0$ and $A_{021}=0$.

ODE (\ref{eq.sep30}) admits a LFI of the form $A(x)F_{1}' +B(x)= k_{1}$ where $k_{1}$ is an arbitrary constant iff it satisfies the condition
\begin{equation}
A'F_{1}' +AF_{1}'' +B' =0 \implies A'F_{1}' +AF_{1}'' +B'= g(x) D \label{eq.sep31}
\end{equation}
along solutions of (\ref{eq.sep30}). By setting the function $g=\frac{x}{3a_{3}}$, we find:
\begin{eqnarray*}
A&=& c_{1}x^{3} -3xF_{1} -\frac{c_{3}}{3a_{3}}x \\
B&=& \frac{3}{2}F_{1}^{2} +3c_{1}x^{2}F_{1} -\frac{c_{1}^{2}}{2}x^{4} +\frac{c_{3}}{3a_{3}} \left( F_{1} +c_{1}x^{2} \right).
\end{eqnarray*}
Therefore, ODE (\ref{eq.sep30}) admits the LFI
\begin{equation}
x \left( c_{1}x^{2} -3F_{1} \right)F_{1}' +\frac{3}{2}F_{1}^{2} +3c_{1}x^{2}F_{1} -\frac{c_{1}^{2}}{2}x^{4} +\frac{c_{3}}{3a_{3}} \left( F_{1} -xF_{1}' +c_{1}x^{2} \right) =k_{1}. \label{eq.sep32}
\end{equation}
This is eq. (18) of Ref. \ref{Gravel 2004} for $\hbar=0$, $k=4k_{1}$ if we set $c_{3}=0$.

Solving (\ref{eq.sep32}), we find that $F_{1}(x)$ must satisfy the quartic algebraic equation
\begin{eqnarray}
0&=& k_{2}x^{2} +4k_{1}^{2} +\left( 9F_{1} -c_{1}x^{2} \right) \left( F_{1} -c_{1}x^{2} \right)^{3} -4k_{1} \left( F_{1} -c_{1}x^{2} \right) \left( 3F_{1} +c_{1}x^{2} \right) + \notag \\
&& +4k_{3} \left( 3F_{1} -c_{1}x^{2} \right) \left( F_{1} -c_{1}x^{2} \right)^{2} +4k_{3}^{2} \left( F_{1} -c_{1}x^{2} \right)^{2} -\frac{8k_{1}k_{3}}{3} \left( 3F_{1} -c_{1}x^{2} \right) \label{eq.sep33}
\end{eqnarray}
where $k_{3}\equiv \frac{c_{3}}{3a_{3}}$ and $k_{2}$ is an arbitrary constant. For $k_{3}=0$ we obtain eq. (24) of Ref. \ref{Gravel 2004}.

Moreover, the vector (\ref{eq.sep26}) becomes
\begin{equation}
B_{a}=
\left(
  \begin{array}{c}
    9a_{3}yF_{1}(x) -3a_{3}c_{1}x^{2}y +c_{3}y \\
    -\frac{1}{2} F_{1}' \left( \frac{9a_{3}}{c_{1}}F_{1} -3a_{3}x^{2} +\frac{c_{3}}{c_{1}} \right) \\
  \end{array}
\right). \label{eq.sep34}
\end{equation}

We conclude that the potential
\begin{equation}
V(x,y)= c_{1}y^{2} +F_{1}(x) \label{eq.sep35}
\end{equation}
admits the CFI
\begin{equation}
J^{(3,2)}_{0}= L\dot{x}^{2} -\left( 3yF_{1} -c_{1}x^{2}y +k_{3}y \right) \dot{x} +\frac{1}{2c_{1}} F_{1}' \left( 3F_{1} -c_{1}x^{2} +k_{3} \right)\dot{y} \label{eq.sep36}
\end{equation}
where $F_{1}(x)$ satisfies the quartic algebraic equation (\ref{eq.sep33}). We note that the superintegrable potential (C.6) of Ref. \ref{Gravel 2004} is a subcase of the above result for $k_{3}=0$.

\subsubsection{Parameters $a_{4}, a_{7}, a_{9}, a_{10}$: The components $L_{abc}$ given by (\ref{eq.pcf2.KT2}) are constant}

\label{sec.int.1}

1) The only non-vanishing parameters are the $a_{4}=a$ and $a_{9}=b$, where $a$ and $b$ are arbitrary constants.

The KT (\ref{eq.pcf2.KT2}) is $L_{221}=L_{222}=0$, $L_{111}=a$ and $L_{112}=b$.

The PDEs (\ref{eq.intc.0.1}) - (\ref{eq.intc.0.4}) and (\ref{eq.intc.8}) become:
\begin{eqnarray}
B_{1}V_{,x} +B_{2}V_{,y} &=& 0 \label{eq.intc.9.1} \\
B_{1,x} &=& 3 \left( a V_{,x} +b V_{,y} \right) \label{eq.intc.9.2} \\
B_{2,y} &=& 0 \label{eq.intc.9.3} \\
B_{1,y} +B_{2,x} &=& 6bV_{,x} \label{eq.intc.9.4} \\
b V_{,yyy} +a V_{,xyy} -2b V_{,xxy} &=& 0. \label{eq.intc.9.5}
\end{eqnarray}
No new potentials are found.
\bigskip

2) $a_{10}=- a_{4}$ (the remaining parameters are fixed to zero); without loss of generality, we set $a_{4}=1 \implies a_{10}= -1$.

The associated CFIs (\ref{eq.intc.0}) are
\begin{equation}
J^{(3,2)}_{0}= \dot{x}^{3} -3\dot{x}\dot{y}^{2} +B_{1}\dot{x} +B_{2}\dot{y}. \label{eq.intc.9.5.1}
\end{equation}

Solving the integrability condition (\ref{eq.intc.8}), we find potentials of the general form
\begin{equation}
V= F_{1}(y) +F_{2}(y +\sqrt{3}x) +F_{3}(y -\sqrt{3}x). \label{eq.intc.9.5.2}
\end{equation}
These are the integrable potentials discussed thoroughly in section \ref{sec.Toda.potential}, where it has been shown that the Toda type potentials are just special cases. We note that all CFIs of the form (\ref{eq.intc.9.5.1}) are allowed only by potentials of the form (\ref{eq.intc.9.5.2}).

The major difference with the method followed in section \ref{sec.Toda.potential} is that here we determined the functional form (\ref{eq.intc.9.5.2}) by solving the integrability condition (\ref{eq.intc.8}) for KT parameters $a_{10}=- a_{4}$; whereas, in section \ref{sec.Toda.potential}, we took (\ref{eq.intc.9.5.2}) as an ansatz and we replaced it into the CFI conditions (\ref{eq.pcf2.4}) and (\ref{eq.pcf2.5}) without using the integrability condition.

The potential\footnote{F. Tremblay and P. Winternitz, \emph{`Third-order superintegrable systems separating in polar coordinates'}, J. Phys. A: Math. Theor. \textbf{43}, 175206 (2010). \label{Tremblay 2010}} given in eq. (4.21) of Ref. \ref{Tremblay 2010} is a subcase of (\ref{eq.intc.9.5.2}) for
\[
F_{1}=\frac{k}{9y^{2}}, \enskip F_{2}= \frac{4k}{9(y +\sqrt{3}x)^{2}}, \enskip F_{3}= \frac{4k}{9(y-\sqrt{3}x)^{2}}
\]
where $k$ is an arbitrary constant. Indeed, we have
\begin{eqnarray*}
V&=& \frac{k}{9y^{2}} + \frac{4k}{9(y +\sqrt{3}x)^{2}} +\frac{4k}{9(y-\sqrt{3}x)^{2}} \\
&=& k \frac{\frac{1}{9}(y^{2} -3x^{2})^{2} +\frac{4}{9}y^{2}(y -\sqrt{3}x)^{2} +\frac{4}{9}y^{2} (y +\sqrt{3}x)^{2}}{y^{2}(y +\sqrt{3}x)^{2}(y -\sqrt{3}x)^{2}} \\
&=& \frac{kr^{4}}{y^{2}(y +\sqrt{3}x)^{2}(y -\sqrt{3}x)^{2}}= \frac{k}{y^{2}} \left( \frac{r^{2}}{y^{2} -3x^{2}} \right)^{2} \\
&=& \frac{k}{r^{2}\sin^{2}\theta (3 -4\sin^{2}\theta)^{2}}= \frac{k}{r^{2} \sin^{2}3\theta}
\end{eqnarray*}
where we used the transformation $x=r\cos\theta$ and $y= r\sin\theta$, and the trigonometric identity $\sin(3\theta)= 3\sin\theta -4\sin^{3}\theta$.

\subsubsection{Parameters $a_{3}, a_{6}, a_{8}$: The components $L_{abc}$ given by (\ref{eq.pcf2.KT2}) are linearly dependent on $x,y$}

\label{sec.int.2}

We find potentials already found in sections (\ref{sec.Holt2}) and (\ref{sec.intc.1}).

\subsubsection{Parameters $a_{2}, a_{5}$: The components $L_{abc}$ given by (\ref{eq.pcf2.KT2}) depend on $x^{2}, xy, y^{2}$}

\label{sec.int.3}

1) The only non-vanishing parameter is the $a_{2}$; without loss of generality, we set $a_{2}= \frac{1}{3}$.

The KT (\ref{eq.pcf2.KT2}) is $L_{111}= y^{2}$, $L_{112}= -\frac{2}{3}xy$, $L_{221}= \frac{x^{2}}{3}$, and $L_{222}=0$.

Replacing these quantities in the integrability condition (\ref{eq.intc.8}), we find the second order integrable  potential (see Class II potentials in Ref. \ref{MitsTsam sym})
\begin{equation}
V= \frac{k}{r} +\frac{F\left( \frac{y}{x} \right)}{r^{2}} \label{eq.intc.9.6}
\end{equation}
where $k$ is an arbitrary constant, $r= \sqrt{x^{2} +y^{2}}$ and $F$ is an arbitrary smooth function of its argument.

Replacing (\ref{eq.intc.9.6}) in the remaining PDEs (\ref{eq.intc.0.1}) - (\ref{eq.intc.0.4}), we obtain the following system of equations:
\begin{eqnarray}
B_{1} \left( uF' +\frac{2F}{1+u^{2}} +\frac{k}{1+u^{2}}r \right) -B_{2} \left( F' -\frac{2uF}{1+u^{2}} -\frac{ku}{1+u^{2}}r \right) &=& 0 \label{eq.intc.9.7} \\
xB_{1,x} +\frac{u(2+3u^{2})F'}{1+u^{2}} +\frac{2u^{2}F}{(1+u^{2})^{2}} +\frac{ku^{2}}{(1+u^{2})^{2}}r &=& 0 \label{eq.intc.9.8} \\
xB_{2,y} + \frac{uF'}{1+u^{2}} +\frac{2F}{(1+u^{2})^{2}} +\frac{k}{(1+u^{2})^{2}}r &=& 0 \label{eq.intc.9.9} \\
xB_{1,y} +xB_{2,x} -\frac{2(1 +2u^{2})F'}{1+u^{2}} -\frac{4uF}{(1+u^{2})^{2}} -\frac{2ku}{(1+u^{2})^{2}}r &=& 0 \label{eq.intc.9.10}
\end{eqnarray}
where $u\equiv \frac{y}{x}$, $F'\equiv \frac{dF(u)}{du}$, $\frac{x}{r}= \frac{1}{\sqrt{1 +u^{2}}}$, and $\frac{y}{r}= \frac{u}{\sqrt{1 +u^{2}}}$.

From the structure of the above system of equations, we deduce that the components $B_{a}$ must be linear functions of $r$ with coefficients depending on $u$, that is,
\begin{equation}
B_{1}= A_{1}(u) +A_{2}(u)r, \enskip B_{2}= A_{3}(u) +A_{4}(u)r \label{eq.intc.9.10.1}
\end{equation}
where $A_{1}, A_{2}, A_{3}$, and $A_{4}$ are arbitrary smooth functions of $u$.

Then, equation (\ref{eq.intc.9.8}) gives:
\[
\underbrace{-uA_{1}' +\frac{u(2+3u^{2})F'}{1+u^{2}} +\frac{2u^{2}F}{(1+u^{2})^{2}}}_{=0} +\underbrace{\left( \frac{ku^{2}}{(1+u^{2})^{2}} +\frac{A_{2}}{1+u^{2}} -uA_{2}' \right)}_{=0} r =0 \implies
\]
\[
A_{1}'= \left( \frac{2 +3u^{2}}{1 +u^{2}} F \right)', \enskip uA_{2}' -\frac{A_{2}}{1+u^{2}} -\frac{ku^{2}}{(1+u^{2})^{2}} =0.
\]
Solving these relations, we find:
\begin{equation}
A_{1}= \frac{2 +3u^{2}}{1 +u^{2}} F +c_{1}, \enskip A_{2}= \frac{ku^{2}}{1+u^{2}} +\frac{c_{2}u}{\sqrt{1+u^{2}}} \label{eq.intc.9.11}
\end{equation}
where $c_{1}$ and $c_{2}$ are arbitrary constants.

Similarly, equation (\ref{eq.intc.9.9}) implies that
\begin{equation}
A_{3}= -uf' -f, \enskip A_{4}= -\frac{ku}{1+u^{2}} +\frac{c_{3}}{\sqrt{1+u^{2}}} \label{eq.intc.9.12}
\end{equation}
where $c_{3}$ is an arbitrary constant and $f$ is a smooth function of $u$ such that
\begin{equation}
F= (1+u^{2})f'. \label{eq.intc.9.13}
\end{equation}

Replacing (\ref{eq.intc.9.11}), (\ref{eq.intc.9.12}), and (\ref{eq.intc.9.13}) in (\ref{eq.intc.9.10.1}), we find the vector
\begin{equation}
B_{a}=
\left(
  \begin{array}{c}
    c_{1} +(2 +3u^{2})f' + \frac{ku^{2}r}{1+u^{2}} +\frac{c_{2}ur}{\sqrt{1+u^{2}}} \\
    -f -uf' -\frac{kur}{1+u^{2}} +\frac{c_{3}r}{\sqrt{1+u^{2}}} \\
  \end{array}
\right). \label{eq.intc.9.13.1}
\end{equation}

Substituting (\ref{eq.intc.9.13.1}) and (\ref{eq.intc.9.13}) in (\ref{eq.intc.9.7}), we get a quadratic polynomial equation in $r$ of the form $f_{1}(u) +f_{2}(u)r +f_{3}(u)r^{2} =0$. Solving this equation, we find $c_{3}= -c_{2}$ and for the function $f$ the additional conditions:
\begin{align}
& 3u(1+u^{2})^{2}f'f'' +(1+u^{2})ff'' +c_{1}u(1+u^{2})f'' +2(2+3u^{2})(1+u^{2})f'^{2} +2c_{1}(1+u^{2})f' =0  \label{eq.intc.9.14} \\
& \left(ku +c_{2}\sqrt{1+u^{2}} \right) (1+u^{2}) f'' +2\left[ k(1+u^{2}) +c_{2}u\sqrt{1+u^{2}} \right]f' + \frac{k(c_{1} -uf)}{1+u^{2}} =0. \label{eq.intc.9.14.1}
\end{align}
The remaining equation (\ref{eq.intc.9.10}) is satisfied identically. We note that $c_{3}= -c_{2} \implies A_{2}= -uA_{4}$.

From (\ref{eq.intc.9.13}), the potential (\ref{eq.intc.9.6}) becomes
\begin{equation}
V= \frac{k}{r} +\frac{(1+u^{2})f'}{r^{2}}. \label{eq.intc.9.15}
\end{equation}
The associated CFI (\ref{eq.intc.0}) is
\begin{equation}
J^{(3,2)}_{0}= \dot{x}L^{2} -\left( \frac{ku}{\sqrt{1+u^{2}}} +c_{2} \right) L+ \left[ \left(2+3u^{2}\right) f' +c_{1} \right] \dot{x} -\left( uf' +f \right)\dot{y} \label{eq.intc.9.16}
\end{equation}
where $f$ is a function of $u$ satisfying the second order non-linear ODEs (\ref{eq.intc.9.14}) and (\ref{eq.intc.9.14.1}).

We note that the potential (\ref{eq.intc.9.15}) is superintegrable because in addition it admits the Ermakov QFI$^{\text{\ref{MitsTsam sym}}}$
\begin{equation}
I= \frac{1}{2}L^{2} +(1+u^{2})f'. \label{eq.intc.9.16.1}
\end{equation}

Using polar coordinates $(r,\theta)$, we have $u=\tan\theta$, $\frac{x}{r}= \cos\theta$, and $\frac{y}{r}= \sin\theta$. Then, condition (\ref{eq.intc.9.13}) gives $F= \frac{df}{d\theta}$ and the integrable potential (\ref{eq.intc.9.15}) becomes
\begin{equation}
V= \frac{k}{r} +\frac{df}{d\theta}r^{-2}. \label{eq.intc.9.17}
\end{equation}
The associated CFI (\ref{eq.intc.9.16}) becomes
\begin{eqnarray}
J^{(3,2)}_{0}&=& p_{\theta}^{2} \left( \cos\theta p_{r} -\sin\theta \frac{p_{\theta}}{r} \right) + \left[ \left( 2\frac{df}{d\theta} +c_{1} \right) \cos\theta -f \sin\theta \right] p_{r} - \notag \\
&& - \left[ \left( \frac{3}{r}\frac{df}{d\theta} +\frac{c_{1}}{r} +k\right) \sin\theta +\frac{f}{r} \cos\theta +c_{2} \right] p_{\theta} \label{eq.intc.9.18}
\end{eqnarray}
while conditions (\ref{eq.intc.9.14}) and (\ref{eq.intc.9.14.1}) are written as follows:
\begin{eqnarray}
\sin\theta \left[ \left( 3\frac{df}{d\theta} +c_{1} \right) \frac{d^{2}f}{d\theta^{2}} -2f\frac{df}{d\theta} \right] +\cos\theta \left[ f\frac{d^{2}f}{d\theta^{2}} +4 \left( \frac{df}{d\theta} \right)^{2} +2c_{1}\frac{df}{d\theta} \right] &=&0 \label{eq.intc.9.19} \\
c_{2}\frac{d^{2}f}{d\theta^{2}} +k\sin\theta \left( \frac{d^{2}f}{d\theta^{2}} -f \right) +k\cos\theta \left( 2\frac{df}{d\theta} +c_{1} \right) &=& 0. \label{eq.intc.9.19.1}
\end{eqnarray}
The Ermakov QFI is $I= \frac{1}{2}p_{\theta}^{2} +\frac{df}{d\theta}$.

We consider two subcases:

- Subcase $k\neq0$.

In this case, the linear second order ODE (\ref{eq.intc.9.19.1}) is solved and gives
\begin{equation}
f(\theta)= \frac{kc_{1}\cos\theta +c_{4}\theta +c_{5}}{c_{2} +k\sin\theta} \label{eq.intc.9.20}
\end{equation}
where $c_{4}$ and $c_{5}$ are arbitrary constants.

Replacing (\ref{eq.intc.9.20}) in the remaining condition  (\ref{eq.intc.9.19}), we find that $c_{2}=c_{4}=0$ and solution (\ref{eq.intc.9.20}) becomes
\begin{equation}
f(\theta)= c_{1}\cot\theta +\frac{c_{5}}{k\sin\theta}. \label{eq.intc.9.21}
\end{equation}
Then, the superintegrable potential (\ref{eq.intc.9.17}) is written as
\begin{equation}
V= \frac{k}{r} +\frac{k_{1}}{r^{2}\sin^{2}\theta} +\frac{k_{2}\cos\theta}{r^{2}\sin^{2}\theta} = \frac{k}{r} +\frac{k_{3}}{r^{2}(1-\cos\theta)} +\frac{k_{4}}{r^{2}(1+\cos\theta)} = \frac{k}{r} +\frac{k_{3}}{r(r-x)} +\frac{k_{4}}{r(r+x)} \label{eq.intc.9.22}
\end{equation}
where $k_{1}= k_{3} +k_{4}=-c_{1}$ and $k_{2}= k_{3} -k_{4} = -\frac{c_{5}}{k}$. This is the well-known second order superintegrable potential (\ref{eq.pcf2.53}) found earlier via Holt's method.

- Subcase $k=0$.

The Kepler term vanishes and the superintegrable potential (\ref{eq.intc.9.17}) becomes (see eq. (5.1) in Theorem 2 of Ref. \ref{Tremblay 2010})
\begin{equation}
V= \frac{df}{d\theta}r^{-2}. \label{eq.intc.9.23}
\end{equation}

Condition (\ref{eq.intc.9.19.1}) gives $c_{2} \frac{d^{2}f}{d\theta^{2}}=0$, which implies that $c_{2}=0$ or $f= b_{1}\theta +b_{2}$, where $b_{1}$ and $b_{2}$ are arbitrary constants. However, due to the remaining condition (\ref{eq.intc.9.19}), only the case $c_{2}=0$ gives non-trivial results. The other case implies that $b_{1}=0 \implies \frac{df}{d\theta}=0 \implies V=0$.

Therefore, the superintegrable potential (\ref{eq.intc.9.23}) admits the CFI
\begin{eqnarray}
J^{(3,2)}_{0}&=& p_{\theta}^{2} \left( \cos\theta p_{r} -\sin\theta \frac{p_{\theta}}{r} \right) + \left[ \left( 2\frac{df}{d\theta} +c_{1} \right) \cos\theta -f \sin\theta \right] p_{r} - \notag \\
&& - \left[ \left( 3\frac{df}{d\theta} +c_{1} \right) \sin\theta +f \cos\theta \right] \frac{p_{\theta}}{r} \label{eq.intc.9.24}
\end{eqnarray}
where $f$ is an arbitrary smooth function such that
\begin{equation}
\sin\theta \left[ \left( 3\frac{df}{d\theta} +c_{1} \right) \frac{d^{2}f}{d\theta^{2}} -2f\frac{df}{d\theta} \right] +\cos\theta \left[ f\frac{d^{2}f}{d\theta^{2}} +4 \left( \frac{df}{d\theta} \right)^{2} +2c_{1}\frac{df}{d\theta} \right] =0. \label{eq.intc.9.25}
\end{equation}
The second superintegrable potential in Table III of Ref. \ref{Karlovini 2000} is derived as a subcase from the condition (\ref{eq.intc.9.25}) for $c_{1}=0$. The condition given in Ref. \ref{Karlovini 2000} is associated with the given CFI only when $C=0$. Moreover, the potential\footnote{A.V. Tsiganov, \emph{`The Drach superintegrable potentials'}, J. Phys. A: Math. Gen. \textbf{33}, 7407 (2000). \label{Tsiganov 2000}} (4.11) of Ref. \ref{Tsiganov 2000} is derived also from (\ref{eq.intc.9.25}) for $c_{1}=0$. The CFI given in Ref. \ref{Tsiganov 2000} has a minor misprint (the plus sign of the third term must be changed to minus).

\subsubsection{Parameter $a_{1}$: The components $L_{abc}$ given by (\ref{eq.pcf2.KT2}) depend on $x^{3}, xy^{2}, x^{2}y, y^{3}$}

\label{sec.int.4}

In this case, the only non-vanishing parameter is the $a_{1}$; without loss of generality, we take $a_{1}=-1$.

The associated CFI (\ref{eq.intc.0}) has an angular momentum leading part (see sec. 3.3.3 in Ref. \ref{Hietarinta 1987}), that is,
\begin{equation}
J^{(3,2)}_{0}= L^{3} +B_{1}\dot{x} +B_{2}\dot{y}. \label{eq.intc.10}
\end{equation}

The KT (\ref{eq.pcf2.KT2}) is $L_{111}=-y^{3}$, $L_{112}= xy^{2}$, $L_{221}=-x^{2}y$, and $L_{222}=x^{3}$.

Replacing these quantities in the integrability condition (\ref{eq.intc.8}), we find potentials of the general form
\begin{equation}
V= F_{1}(r) +\frac{F_{2}\left( \frac{y}{x} \right)}{r^{2}} +\frac{F_{3}\left( \frac{y}{x} \right)}{r^{3}} \label{eq.intc.10.1}
\end{equation}
where $r= \sqrt{x^{2} +y^{2}}$ and $F_{1}, F_{2}, F_{3}$ are arbitrary smooth functions of their arguments.

Replacing (\ref{eq.intc.10.1}) in the remaining PDEs (\ref{eq.intc.0.1}) - (\ref{eq.intc.0.4}), we get the following system of equations:
\begin{eqnarray}
0&=& B_{1} \left[ \frac{dF_{1}}{dr} -u(1+u^{2})\frac{F_{2}'}{r^{3}} -\frac{2F_{2}}{r^{3}} -u(1+u^{2})\frac{F_{3}'}{r^{4}} -\frac{3F_{3}}{r^{4}} \right] + \notag \\
&& + B_{2} \left[ u\frac{dF_{1}}{dr} +(1+u^{2})\frac{F_{2}'}{r^{3}} -\frac{2uF_{2}}{r^{3}} +(1+u^{2})\frac{F_{3}'}{r^{4}} -\frac{3uF_{3}}{r^{4}} \right] \label{eq.intc.11.1} \\
B_{1,x} &=& 3u^{2} \left( F_{2}' +\frac{F_{3}'}{r} \right)  \label{eq.intc.11.2} \\
B_{2,y} &=& 3 \left( F_{2}' +\frac{F_{3}'}{r} \right) \label{eq.intc.11.3} \\
B_{1,y} +B_{2,x} &=& -6u \left( F_{2}' +\frac{F_{3}'}{r} \right) \label{eq.intc.11.4}
\end{eqnarray}
where $u\equiv \frac{y}{x}$ and $F'\equiv \frac{dF(u)}{du}$. We note that $\frac{x}{r}= \frac{1}{\sqrt{1+u^{2}}}$ and $\frac{y}{r}= \frac{u}{\sqrt{1+u^{2}}}$.

From the structure of the above system of PDEs, we deduce that the quantities $B_{1}$ and $B_{2}$ must be linear functions of $r$ with coefficients depending on $u$, that is,
\begin{equation}
B_{1}= A_{1}(u) +A_{2}(u)r, \enskip B_{2}= A_{3}(u) +A_{4}(u)r \label{eq.intc.11.5}
\end{equation}
where $A_{1}, A_{2}, A_{3}$, and $A_{4}$ are arbitrary smooth functions of $u$.

Replacing (\ref{eq.intc.11.5}) in (\ref{eq.intc.11.2}), we obtain:
\[
-\frac{u}{x}A_{1}' -\frac{ur}{x}A_{2}' +\frac{x}{r}A_{2} = 3u^{2} \left( F_{2}' +\frac{F_{3}'}{r} \right) \implies
\]
\[
u\left( \sqrt{1+u^{2}} A_{1}' +3uF_{3}' \right) \frac{1}{r} +u\sqrt{1+u^{2}}A_{2}' -\frac{A_{2}}{\sqrt{1+u^{2}}} +3u^{2}F_{2}' =0 \implies
\]
\begin{equation}
A_{1}'= -\frac{3uF_{3}'}{\sqrt{1+u^{2}}}, \enskip A_{2}= \frac{-3uF_{2} +c_{1}u}{\sqrt{1+u^{2}}} \label{eq.intc.11.6}
\end{equation}
where $c_{1}$ is an arbitrary constant.

Similarly, equation (\ref{eq.intc.11.3}) implies that
\begin{equation}
A_{3}'= \frac{3F_{3}'}{\sqrt{1 +u^{2}}}, \enskip A_{4}= \frac{3F_{2} +c_{2}}{\sqrt{1 +u^{2}}} \label{eq.intc.12}
\end{equation}
where $c_{2}$ is an arbitrary constant.

Replacing (\ref{eq.intc.11.6}) and (\ref{eq.intc.12}) in (\ref{eq.intc.11.5}), we find
\begin{equation}
B_{a}=
\left(
  \begin{array}{c}
    f -uf' +\frac{-3uF_{2} +c_{1}u}{\sqrt{1+u^{2}}}r \\
    f' +\frac{3F_{2} +c_{2}}{\sqrt{1 +u^{2}}}r \\
  \end{array}
\right) \label{eq.intc.13}
\end{equation}
where $f$ is an arbitrary function of $u$ such that
\begin{equation}
3F_{3}'= \sqrt{1 +u^{2}} f''. \label{eq.intc.14}
\end{equation}

Replacing the vector (\ref{eq.intc.13}) in (\ref{eq.intc.11.1}), we find that $c_{2}=-c_{1}$, $F_{1}= \frac{k}{r}$ and the additional conditions:
\begin{eqnarray}
\frac{kf}{\sqrt{1+u^{2}}} &=& 3(1+u^{2})F_{2}F_{2}' -c_{1}(1+u^{2})F_{2}' \label{eq.intc.15.1} \\
3(1+u^{2})^{3/2}F_{2}F_{3}'&=& u(1+u^{2})fF_{2}' -(1+u^{2})^{2}f'F_{2}' +2fF_{2} +c_{1}(1+u^{2})^{3/2}F_{3}' \label{eq.intc.15.2} \\
3fF_{3}&=& \left( 1 +u^{2} \right)^{2} f'F_{3}' -u \left( 1 +u^{2} \right) fF_{3}' \label{eq.intc.15.3}
\end{eqnarray}
where $k$ is an arbitrary constant. The remaining PDE  (\ref{eq.intc.11.4}) is satisfied identically.

Therefore, the family of potentials (\ref{eq.intc.10.1}) gives the integrable potential (see eq. (3.3.42) in Ref. \ref{Hietarinta 1987})
\begin{equation}
V= \frac{k}{r} +\frac{F_{2}\left( \frac{y}{x} \right)}{r^{2}} +\frac{F_{3}\left( \frac{y}{x} \right)}{r^{3}} \label{eq.intc.15.4}
\end{equation}
and the associated CFI (\ref{eq.intc.10}) is
\begin{equation}
J^{(3,2)}_{0}= L^{3} +\left( f -u f' +\frac{-3uF_{2} +c_{1}u}{\sqrt{1+u^{2}}}r \right)\dot{x} +\left(f' +\frac{3F_{2} -c_{1}}{\sqrt{1 +u^{2}}}r \right)\dot{y} \label{eq.intc.16}
\end{equation}
where $f$ is a function of $u$ satisfying conditions (\ref{eq.intc.14}) - (\ref{eq.intc.15.3}).

\underline{NOTE:} If $k=c_{1}=F_{2}=0$, then the potential (\ref{eq.intc.15.4}) becomes $V= \frac{F_{3}\left( \frac{y}{x} \right)}{r^{3}}$ which is the potential found in example (6) in Ref. \ref{Thompson 1984}. In this case, conditions (\ref{eq.intc.15.1}) and (\ref{eq.intc.15.2}) vanish, and $F_{3}'$ is eliminated from the remaining conditions (\ref{eq.intc.14}) and (\ref{eq.intc.15.3}) as follows:
\begin{eqnarray}
F_{3}&=& \frac{(1+u^{2})^{3/2}f''}{9} \left[ (1+u^{2}) (\ln f)' -u \right] \label{eq.intc.17.1} \\
0&=& \left[ (1+u^{2})(\ln f)' -u \right]f''' +\left[ \frac{f''}{f} -\left( \frac{f'}{f} \right)^{2} \right] (1+u^{2}) f'' +5u (\ln f)'f'' -4f''. \label{eq.intc.17.2}
\end{eqnarray}
The function $F_{3}$ that defines the integrable potential is expressed in terms of an arbitrary smooth function $f$ of the same argument which is the solution of the third order non-linear ODE (\ref{eq.intc.17.2}).

The above results are simplified significantly if we use polar coordinates $(r, \theta)$, where $r= \sqrt{x^{2} +y^{2}}$ and $\tan\theta=u$. We recall that the conjugate momenta $p_{r}= \dot{r}$ and $p_{\theta}= L= r^{2}\dot{\theta}$.

The integrable potential (\ref{eq.intc.15.4}) becomes
\begin{equation}
V= \frac{k}{r} +\frac{F_{2}(\theta)}{r^{2}} +\frac{F_{3}(\theta)}{r^{3}} . \label{eq.intc.18.1}
\end{equation}

For an arbitrary function $h(\theta)$, we have the following identities:
\[
1 +u^{2}= \frac{1}{\cos^{2}\theta}, \enskip h'= \cos^{2}\theta \frac{dh}{d\theta}, \enskip h''= \cos^{4}\theta \frac{d^{2}h}{d\theta^{2}} -2\cos^{3}\theta \sin\theta \frac{dh}{d\theta}.
\]
If we introduce the transformation $N=f \cos\theta$, conditions (\ref{eq.intc.14}) - (\ref{eq.intc.15.3}) become (see eq. (3.3.43) in Ref. \ref{Hietarinta 1987}):
\begin{eqnarray}
\frac{d^{2}N}{d\theta^{2}} +N &=& 3\frac{dF_{3}}{d\theta} \label{eq.intc.18.2} \\
kN&=& \left( 3F_{2} -c_{1} \right)\frac{dF_{2}}{d\theta} \label{eq.intc.18.3} \\
2NF_{2} &=& \left( 3F_{2} -c_{1} \right) \frac{dF_{3}}{d\theta} +\frac{dN}{d\theta} \frac{dF_{2}}{d\theta} \label{eq.intc.18.4} \\
3NF_{3} &=& \frac{dN}{d\theta} \frac{dF_{3}}{d\theta}. \label{eq.intc.18.5}
\end{eqnarray}
We note that condition $b'=3g'$ in eq. (3.3.43) of Ref. \ref{Hietarinta 1987} can be solved and gives $b=3g +const$. To compare with Ref. \ref{Hietarinta 1987}, we set $w=\theta$, $k=c$, $F_{2}=g$, $F_{3}=h$, $N=a$, and $b=3g -c_{1}$.

The CFI (\ref{eq.intc.16}) is written as\footnote{We note that $x=r\cos\theta$, $y=r\sin\theta$, $\dot{x}= \cos\theta p_{r} -\sin\theta \frac{p_{\theta}}{r}$, and $\dot{y}= \sin\theta p_{r} +\cos\theta \frac{p_{\theta}}{r}$.}
\begin{equation}
J^{(3,2)}_{0}= p_{\theta}^{3} +Np_{r} +\left( \frac{1}{r}\frac{dN}{d\theta} +3F_{2} -c_{1}\right) p_{\theta}. \label{eq.intc.18.6}
\end{equation}

The potential found in Table IV of Ref. \ref{Karlovini 2000} is a special case of (\ref{eq.intc.18.1}) for $k=c_{1}=F_{2}=0$, $N= 3\frac{dg}{d\theta}$ and $F_{3}= g  +\frac{d^{2}g}{d\theta^{2}}$, where $g(\theta)$ is an arbitrary smooth function. In this case, conditions (\ref{eq.intc.18.3}) and (\ref{eq.intc.18.4}) vanish, condition (\ref{eq.intc.18.2}) is satisfied identically, and condition (\ref{eq.intc.18.5}) becomes
\begin{equation}
\frac{d^{2}g}{d\theta^{2}} \frac{d^{3}g}{d\theta^{3}} -2 \frac{dg}{d\theta} \frac{d^{2}g}{d\theta^{2}} -3g \frac{dg}{d\theta} =0 \label{eq.intc.18.7}
\end{equation}
which is the condition found in Ref. \ref{Karlovini 2000}.

Moreover, when $k=0$ and $F_{2}=k_{1}=const$, conditions (\ref{eq.intc.18.2}) - (\ref{eq.intc.18.5}) give:
\[
\frac{d^{2}N}{d\theta^{2}}= \frac{3k_{1}+c_{1}}{3k_{1} -c_{1}}N, \enskip \frac{dF_{3}}{d\theta}= \frac{2k_{1}}{3k_{1} -c_{1}}N, \enskip 3NF_{3}= \frac{dN}{d\theta} \frac{dF_{3}}{d\theta}
\]
where $c_{1} \neq 3k_{1}$ for non-trivial results. Solving this system of ODEs, we find $c_{1}= \frac{3}{2}k_{1}$ and we retrieve -as expected- the integrable potential (\ref{eq.pcf2.54c}) and its CFI (\ref{eq.pcf2.55}), which we found earlier using Holt's method.

\subsubsection{Mixed choice of the ten parameters $a_{1}, ..., a_{10}$: The components $L_{abc}$ given by (\ref{eq.pcf2.KT2}) depend on products of $x, y$ of mixed degree}

\label{sec.int.5}

1) Case $a_{1}= -a_{2}$ (the remaining parameters vanish); without loss of generality, we set $a_{1}=-1 \implies a_{2}=1$.

The KT (\ref{eq.pcf2.KT2}) is $L_{111}= -y^{3} +3y^{2}$, $L_{112}= xy^{2} -2xy$, $L_{221}= -x^{2}y +x^{2}$, and $L_{222}= x^{3}$.

The associated CFI (\ref{eq.intc.0}) takes the form
\begin{equation}
J^{(3,2)}_{0}= L^{3} +3\dot{x}L^{2} +B_{1}\dot{x} +B_{2}\dot{y}. \label{eq.intc.19}
\end{equation}

Replacing the above KT components in the integrability condition (\ref{eq.intc.8}), we find again potentials of the general form (\ref{eq.intc.9.6}). Then, from the remaining conditions (\ref{eq.intc.0.1}) - (\ref{eq.intc.0.4}), using the techniques of sections \ref{sec.int.3} and \ref{sec.int.4}, we find
\begin{equation}
B_{a}=
\left(
  \begin{array}{c}
    c_{1} +\left( 2+ 3u^{2} \right)f' +\frac{3ku^{2}r}{1+u^{2}} +\frac{u\left( c_{2} -3F \right)r}{\sqrt{1 +u^{2}}} \\
    -f -uf' -\frac{3kur}{1+u^{2}} -\frac{\left(c_{2} -3F\right)r}{\sqrt{1+u^{2}}} \\
  \end{array}
\right) \label{eq.intc.20}
\end{equation}
where $f$ is an arbitrary smooth functions such that
\begin{eqnarray}
0 &=& 3F -(1+u^{2})f' \label{eq.intc.21.1} \\
0&=& 3u(1+u^{2})^{2}f'f'' +(1+u^{2})ff'' +c_{1}u(1+u^{2})f'' +2(2+3u^{2})(1+u^{2})f'^{2} +2c_{1}(1+u^{2})f' \label{eq.intc.21.2} \\
0&=& ku(1+u^{2})f'' +\frac{c_{2}}{3} (1+u^{2})^{3/2}f'' -(1+u^{2})^{5/2} \frac{f'f''}{3} -\frac{2u}{3}(1+u^{2})^{3/2}f'^{2} + \notag \\
&& +2k(1+u^{2})f' +\frac{2c_{2}}{3}u\sqrt{1+u^{2}}f' +\frac{k(c_{1} -uf)}{1 +u^{2}}. \label{eq.intc.21.3}
\end{eqnarray}

If we use polar coordinates $(r, \theta)$ and we apply the scaling transformations $f \rightleftarrows 3f$, $c_{1} \rightleftarrows 3c_{1}$ and $c_{2} \rightleftarrows 3c_{2}$, then we find the new superintegrable potential (superintegrable because also admits the Ermakov invariant $I= \frac{1}{2}p_{\theta}^{2} +\frac{df}{d\theta}$)
\begin{equation}
V= \frac{k}{r} +\frac{df}{d\theta}r^{-2} \label{eq.intc.22}
\end{equation}
where $k\neq0$ and the smooth function $f(\theta)$ satisfies the condition (\ref{eq.intc.9.25}) and the condition
\begin{equation}
\left( c_{2} -\frac{df}{d\theta} \right) \frac{d^{2}f}{d \theta^{2}} +k \left( c_{1} +2\frac{df}{d\theta} \right)\cos\theta +k \left( \frac{d^{2}f}{d\theta^{2}} -f \right) \sin\theta= 0. \label{eq.intc.23}
\end{equation}
The associated CFI (\ref{eq.intc.19}) becomes
\begin{eqnarray}
J^{(3,2)}_{0}&=& \frac{1}{3}p_{\theta}^{3} +\left( \cos\theta p_{r} -\sin\theta \frac{p_{\theta}}{r} \right) p_{\theta}^{2} + \left[ \left( 2\frac{df}{d\theta} +c_{1} \right) \cos\theta -f\sin\theta \right] p_{r} - \notag \\
&& -\left[ \left( 3\frac{df}{d\theta} +c_{1} \right) \sin\theta +f\cos\theta \right] \frac{p_{\theta}}{r} - \left( c_{2} -\frac{df}{d\theta} +k\sin\theta \right)p_{\theta}. \label{eq.intc.24}
\end{eqnarray}
\bigskip

The results obtained above for integrable and superintegrable potentials admitting CFIs of the type $J^{(3,2)}_{0}$ are summarized, respectively, in Tables \ref{Table.cfi1} and \ref{Table.cfi2}. In these Tables, we also indicate the corresponding results of Refs. \ref{Hietarinta 1987} and \ref{Karlovini 2000} in order to show what is new and/or more general. In the latter case, we state the values of the parameters for which the corresponding potentials of Refs. \ref{Hietarinta 1987} and \ref{Karlovini 2000} are obtained. Moreover, when a potential is not included both in Refs. \ref{Hietarinta 1987} and \ref{Karlovini 2000}, it is referred to as New and in the case that it has been found by other authors we give the corresponding reference.

\newpage

\begin{longtable}{|l|c|c|}
\hline
\multicolumn{3}{|c|}{{\large{Integrable potentials}}} \\ \hline
{\large Potentials and FIs} & {\large Ref. \ref{Hietarinta 1987}} & {\large Ref. \ref{Karlovini 2000} } \\ \hline
\makecell[l]{$V_{1}= F_{1}(y+\sqrt{3}x) +F_{2}(y-\sqrt{3}x) +F_{3}(-2y)$ \\ \\ $J_{1}= \dot{x}^{3} -3\dot{x}\dot{y}^{2} + 3\left[ F_{1}(y+\sqrt{3}x) +F_{2}(y-\sqrt{3}x) -2F_{3}(-2y) \right]\dot{x}-$ \\ \qquad \enskip $-3\sqrt{3} \left[ F_{1}(y+\sqrt{3}x) -F_{2}(y-\sqrt{3}x) \right]\dot{y}$ \\ where $\frac{dF_{1}}{dw_{1}}(F_{2}-F_{3}) +\frac{dF_{2}}{dw_{2}}(F_{3}-F_{1}) +\frac{dF_{3}}{dw_{3}}(F_{1}-F_{2}) =0$, \\ $w_{1}\equiv y+\sqrt{3}x$, $w_{2}\equiv y-\sqrt{3}x$ and $w_{3}\equiv -2y$} & (3.3.34) & Table I \\ \hline
\makecell[l]{$V_{2}= c_{+} e^{k(y+\sqrt{3}x)} +c_{-} e^{k(y-\sqrt{3}x)}+ c_{0} e^{-2ky}$ \\ \\ $J_{2}=(\dot{x}^{2} -3\dot{y}^{2})\dot{x} + 3\left[c_{+} e^{k(y+\sqrt{3}x)} +c_{-} e^{k(y-\sqrt{3}x)}-2c_{0} e^{-2ky}\right] \dot{x}-$ \\ \qquad \enskip $-3\sqrt{3} \left[ c_{+} e^{k(y+\sqrt{3}x)} -c_{-} e^{k(y-\sqrt{3}x)} \right]\dot{y}$} & \makecell[c]{(3.3.37) \\ $k=-\frac{1}{2}$ \\ $c_{0}=1$ \\ $c_{+}=1$ \\ $c_{-}=1$} & Table I \\ \hline
\makecell[l]{$V_{3}= \left( \frac{4}{3}c_{1}x^{2} + c_{2}x + c_{3} \right) y^{-2/3} + c_{1}y^{4/3}$ \\ \\ $J_{3} = \dot{x}^{3} +\frac{3}{2}\dot{x}\dot{y}^{2} + \left[ \left( 4c_{1}x^{2} + 3c_{2}x + 3c_{3} \right) y^{-2/3} - 6c_{1}y^{4/3} \right] \dot{x}+$ \\ \qquad \enskip $+\left( 12c_{1}x + \frac{9}{2}c_{2} \right) y^{1/3}\dot{y}$} & \makecell[c]{(3.3.27) \\ $c_{1}=\frac{3}{4}$ \\ $c_{2}=0$} & Table I \\
\hline
\makecell[l]{$V_{4}= k(xy)^{-2/3}$ \\ \\ $J_{4}= (\dot{x}^{2}-2H_{4})x\dot{x} -(\dot{y}^{2}-2H_{4})y\dot{y}$, where $H_{4}= \frac{1}{2}(\dot{x}^{2}+\dot{y}^{2}) +V_{4}$} & New & \makecell[c]{New \\ eq. (10) \\ in Ref. \ref{Inozemtsev 1983}} \\ \hline
\makecell[l]{$V_{5}= k(x^{2} -y^{2})^{-2/3}$ \\ \\ $J_{5}= \left( \dot{y}^{2} -\dot{x}^{2} \right)L +4(y\dot{x} +x\dot{y})V_{5}$} & \makecell[c]{(3.3.24) \\ $k=1$} & \makecell[c]{Table II \\ $k=1$} \\ \hline
\makecell[l]{$V_{6}= \frac{k_{1}}{r^{2}} + \frac{k_{2} e^{\sqrt{3}\theta} +k_{3} e^{-\sqrt{3} \theta}}{r^{3}}$ \\ \\ $J_{6}= p_{\theta}^{3} +\frac{3}{4} \left[ 2k_{1} +\frac{3}{r}\left( k_{2}e^{\sqrt{3}\theta} +k_{3}e^{-\sqrt{3}\theta} \right) \right]p_{\theta} +$ \\ \qquad \enskip $+\frac{3\sqrt{3}}{4}\left( k_{2}e^{\sqrt{3}\theta} -k_{3}e^{-\sqrt{3}\theta} \right)p_{r}$} & (3.3.44) & Table IV \\ \hline
\makecell[l]{$V_{7}= \frac{k_{1}}{(a_{2}y -a_{5}x)^{2}} +\frac{k_{2}}{r} +\frac{k_{3}(a_{2}x +a_{5}y)}{r(a_{2}y -a_{5}x)^{2}}$ \\ \\ $J_{7}=(a_{2}\dot{x} +a_{5}\dot{y})L^{2} + \frac{2k_{1}r^{2}}{(a_{2}y -a_{5}x)^{2}} (a_{2}\dot{x} +a_{5}\dot{y}) -\frac{k_{2}(a_{2}y -a_{5}x)}{r}L +$ \\ \qquad \enskip $+ \frac{k_{3}r}{a_{2}y -a_{5}x}(a_{2}\dot{y} -a_{5}\dot{x}) - \frac{k_{3}(a_{2}x +a_{5}y)}{r(a_{2}y -a_{5}x)}L +\frac{2k_{3}(a_{2}x+a_{5}y)r}{(a_{2}y -a_{5}x)^{2}} (a_{2}\dot{x} +a_{5}\dot{y})$} & New & New \\ \hline
\makecell[l]{$V_{8}= \frac{k}{r} +\frac{F_{2}(\theta)}{r^{2}} +\frac{F_{3}(\theta)}{r^{3}}$ \\ \\ $J_{8}= p_{\theta}^{3} +N(\theta)p_{r} +\left( \frac{1}{r}\frac{dN}{d\theta} +3F_{2} -c_{1} \right) p_{\theta}$ \\ where $\frac{d^{2}N}{d\theta^{2}} +N= 3\frac{dF_{3}}{d\theta}$, $kN= \left( 3F_{2} -c_{1} \right)\frac{dF_{2}}{d\theta}$, \\ $2NF_{2}= \left( 3F_{2} -c_{1} \right) \frac{dF_{3}}{d\theta} +\frac{dN}{d\theta} \frac{dF_{2}}{d\theta}$, and $3NF_{3}= \frac{dN}{d\theta} \frac{dF_{3}}{d\theta}$} & (3.3.42) & \makecell[l]{Table IV \\ $k=c_{1}=0$ \\ $N=3\frac{dg}{d\theta}$ \\ $F_{2}=0$ \\ $F_{3}=g +\frac{d^{2}g}{d\theta^{2}}$} \\ \hline
\caption{\label{Table.cfi1} Integrable potentials $V(x,y)$ that admit CFIs of the type $J^{(3,2)}_{0}$.}
\end{longtable}

\newpage

\begin{longtable}{|l|c|c|}
\hline
\multicolumn{3}{|c|}{{\large{Superintegrable potentials}}} \\ \hline
{\large Potentials and FIs} & {\large Ref. \ref{Hietarinta 1987}} & {\large Ref. \ref{Karlovini 2000} } \\ \hline
\makecell[l]{$V_{s1}= c_{1}(x^{2} +4y^{2}) +\frac{c_{2}}{x^{2}} +c_{3}y$ \\ \\ $J_{s11} =\dot{x}^{2}\dot{y} +x (8c_{1}y +c_{3})\dot{x} -2 \left(c_{1}x^{2} -\frac{c_{2}}{x^{2}}\right)\dot{y}$ \\ $J_{s12}=  \frac{1}{2}\dot{x}^{2} + c_{1}x^{2} +\frac{c_{2}}{x^{2}}$ \\ $J_{s13}= \frac{1}{2}\dot{y}^{2} +4c_{1}y^{2} +c_{3}y$ \\
$J_{s14}=\dot{x}L +2c_{1}x^{2}y -\frac{2c_{2}y}{x^{2}} +\frac{c_{3}}{2}x^{2}$} & \makecell[c]{(3.3.29) \\ $c_{1}=1$ \\ $c_{3}=0$} & \makecell[c]{Table I \\ $x\leftrightarrow y$} \\ \hline
\makecell[l]{$V_{s2}=F(x+y)$ \\ (superintegrable by using only QFIs) \\ $J_{s21}= \dot{x}-\dot{y}$, $J_{s22}=t(\dot{x}-\dot{y})-x+y$, $J_{s23}=\dot{x}\dot{y} +F(x+y)$ \\ $J_{s24}= J_{s21}^{3} +3J_{s21}J_{s23}= \dot{x}^{3} -\dot{y}^{3} +3F(x+y)(\dot{x} -\dot{y})$} & New & New \\ \hline
\makecell[l]{$V_{s3}= c_{1}\sqrt{x} +c_{2}y$ \\ \\ $J_{s31}= \frac{1}{2}\dot{x}^{2} +c_{1}\sqrt{x}$, $J_{s32}= \frac{1}{2}\dot{y}^{2} +c_{2}y$, $J_{s33}= c_{2}\dot{x}^{3} +3c_{1}c_{2}\sqrt{x}\dot{x} -\frac{3}{2}c_{1}^{2}\dot{y}$} & \makecell[c]{not \\ included} & \makecell[c]{Table I \\ $c_{1}=1$} \\ \hline
\makecell[l]{$V_{s4}= k \left( \sqrt{x} \pm \sqrt{y} \right)$ \\ \\ $J_{s41}= \frac{1}{2}\dot{x}^{2} +k\sqrt{x}$, $J_{s42}= \frac{1}{2}\dot{y}^{2} \pm k\sqrt{y}$, $J_{s43}= \dot{x}^{3} -\dot{y}^{3} + 3k\left( \sqrt{x}\dot{x} \mp \sqrt{y}\dot{y} \right)$} & \makecell[c]{not \\ included} & \makecell[c]{Table I \\ $k=1$} \\ \hline
\makecell[l]{$V_{s5}= c_{1}(x^{2}+y^{2}) +\frac{c_{2}}{x^{2}} +\frac{c_{3}}{y^{2}}$ \\ \\ $J_{s51}= \frac{1}{2}\dot{x}^{2} +c_{1}x^{2} +\frac{c_{2}}{x^{2}}$, $J_{s52}= \frac{1}{2}\dot{y}^{2} +c_{1}y^{2} +\frac{c_{3}}{y^{2}}$, $J_{s53}= L^{2} +2c_{2}\frac{y^{2}}{x^{2}} +2c_{3}\frac{x^{2}}{y^{2}}$ \\ $J_{s54}= e^{\lambda t}\left[ -\dot{x}^{2}+\lambda x\dot{x} +2c_{1}x^{2} -\frac{2c_{2}}{x^{2}} \right]$, $J_{s55}= e^{\lambda t}\left[ -\dot{y}^{2}+\lambda y\dot{y} +2c_{1}y^{2} -\frac{2c_{3}}{y^{2}} \right]$ \\ $J_{s56}= (\dot{x}\dot{y} +2c_{1}xy)\frac{L}{2} -c_{2}\frac{y}{x^{2}}\dot{y} +c_{3}\frac{x}{y^{2}} \dot{x}$, where $\lambda^{2}=-8c_{1}$} & (3.2.34) & Table II \\ \hline
\makecell[l]{$V_{s6}= c_{0}\left(9x^{2} +y^{2}\right)$ \\ \\ $J_{s61}= \frac{1}{2}\dot{x}^{2} +9c_{0}x^{2}$, $J_{s62}= \frac{1}{2}\dot{y}^{2} +c_{0}y^{2}$, $J_{s63}= \dot{y}^{2}L +\frac{2c_{0}}{3}y^{3}\dot{x} -6c_{0}xy^{2}\dot{y}$} & \makecell[c]{(3.3.31) \\ $c_{0}= \frac{1}{18}$} & \makecell[c]{Table II \\ $c_{0}=1$} \\ \hline
\makecell[l]{$V_{s7}= c_{1}\left( x^{2}+y^{2} \right) + \frac{c_{2}xy +c_{3} \left( x^{2}+y^{2} \right)}{\left( x^{2}-y^{2} \right)^{2}}$ \\ \\ $J_{s71}= \dot{x}\dot{y} + c_{1} \left( x^{2}+y^{2} \right) +\frac{2c_{3} -c_{2}}{4\left( y +x \right)^{2}} -\frac{2c_{3} +c_{2}}{4 \left( y -x \right)^{2}}$ \\ $J_{s72}= \left( \dot{y}^{2} -\dot{x}^{2} \right) L -2c_{1}\left( x^{2}-y^{2} \right)L +\frac{x \left( x^{2} +3y^{2} \right) \left( c_{2}\dot{x} +2c_{3}\dot{y} \right)}{\left( x^{2}-y^{2}\right)^{2}}+$ \\ \qquad \quad $ +\frac{y(3x^{2} +y^{2})(2c_{3}\dot{x} +c_{2} \dot{y})}{(x^{2} -y^{2})^{2}}$} & \makecell[c]{not \\ included} & Table II \\ \hline
\makecell[l]{$V_{s8}= \frac{k_{1}}{y^{2}} +\frac{k_{2}}{r} +\frac{k_{3}x}{ry^{2}}$ \\ \\ $J_{s81}= \frac{L^{2}}{2} +k_{1}\frac{x^{2}}{y^{2}} + k_{3}\frac{rx}{y^{2}}$,
$J_{s82}= \dot{y}L + 2k_{1} \frac{x}{y^{2}} + k_{2}\frac{x}{r} +k_{3}\frac{2x^{2}+y^{2}}{ry^{2}}$ \\
$J_{s83}= \dot{x}L^{2} -\frac{k_{2}y}{r}L +\frac{2k_{1}r^{2}}{y^{2}}\dot{x} + \frac{k_{3} x(2x^{2} +3y^{2})}{ry^{2}}\dot{x} +\frac{k_{3}y}{r} \dot{y}$} & (3.2.36) & Table III \\ \hline
\makecell[l]{$V_{s9}= \frac{c_{1}}{r} +\frac{c_{2}\sqrt{r+x}}{r} +\frac{c_{3} \sqrt{r-x}}{r}$ \\ \\ $J_{s91}=\dot{y}L + \frac{c_{1}x}{r} + \frac{c_{3}(r+x)\sqrt{r-x} -c_{2}(r-x)\sqrt{r+x}}{r}$ \\ $J_{s92}= -\dot{x}L + G(x,y)$ \\ $J_{s93}= H_{9}L + \frac{1}{2} \left( c_{2}\sqrt{r-x} +c_{3}\sqrt{r+x} \right)\dot{x} -\frac{1}{2} \left( c_{2}\sqrt{r+x} -c_{3}\sqrt{r-x} \right) \dot{y}$ \\ where $H_{9}= \frac{1}{2}(\dot{x}^{2} +\dot{y}^{2}) +V_{s9}$, $G_{,y}= -xV_{,x}$, and $G_{,x}= 2yV_{,x} -xV_{,y}$} & \makecell[c]{not \\ included} & Table V \\ \hline
\makecell[l]{$V_{s10}= c_{0}(x^{2} +9y^{2})+c_{1}y$ \\ \\ $J_{s10a}= \frac{1}{2}\dot{x}^{2} +c_{0}x^{2}$, $J_{s10b}= \frac{1}{2}\dot{y}^{2} +9c_{0}y^{2} +c_{1}y$ \\ $J_{s10c}=
\dot{x}^{2}L -\frac{c_{1}}{18c_{0}} \dot{x}^{3}+\frac{c_{1}}{3}x^{2}\dot{x} +6c_{0}x^{2}y\dot{x} - \frac{2c_{0}}{3}x^{3}\dot{y}$} & New & New \\ \hline
\makecell[l]{$V_{s11}= cy +F(x)$ \\ where $F(x)$ is such that
$\left( F +\frac{k_{3}}{2} \right) \left( F +\frac{c}{k_{1}}x +\frac{k_{2}}{k_{1}} -k_{3} \right)^{2} =k_{4}$ \\ \\ $J_{s111}= \frac{1}{2}\dot{x}^{2} +F(x)$, \enskip $J_{s112}= \frac{1}{2}\dot{y}^{2} +cy$ \\ $J_{s113}= k_{1}\dot{x}^{3} +\dot{x}^{2}\dot{y} +\left( 3k_{1}F +cx +k_{2} \right) \dot{x} +\left( 2F +k_{3}\right)\dot{y}$} & New & \makecell{New \\ eq. (C.8) \\ in Ref. \ref{Gravel 2004} \\ for $k_{2}=0$ \\ and $k_{3}=0$} \\ \hline
\makecell[l]{$V_{s12}= \sqrt{k_{1}x} \pm \sqrt{k_{2}y}$ \\ \\ $J_{s121}= \frac{1}{2}\dot{x}^{2} +\sqrt{k_{1}x}$, \enskip $J_{s122}= \frac{1}{2}\dot{y}^{2} \pm \sqrt{k_{2}y}$ \\ $J_{s123}= k_{2}\dot{x}^{3} -k_{1}\dot{y}^{3} +3k_{2} \sqrt{k_{1}x} \dot{x} \mp3k_{1} \sqrt{k_{2}y} \dot{y}$} & New & \makecell{New \\ eq. (C.5) \\ in Ref. \ref{Gravel 2004}} \\ \hline
\makecell[l]{$V_{s13}= -\sqrt{k_{1}x} \pm \sqrt{k_{2}y}$ \\ \\ $J_{s131}= \frac{1}{2}\dot{x}^{2} -\sqrt{k_{1}x}$, \enskip $J_{s132}= \frac{1}{2}\dot{y}^{2} \pm \sqrt{k_{2}y}$ \\ $J_{s133}= k_{2}\dot{x}^{3} -k_{1}\dot{y}^{3} -3k_{2} \sqrt{k_{1}x} \dot{x} \mp3k_{1} \sqrt{k_{2}y} \dot{y}$} & New & \makecell{New \\ eq. (C.5) \\ in Ref. \ref{Gravel 2004}} \\ \hline
\makecell[l]{$V_{s14}= c_{1}y^{2} +F_{1}(x)$ \\ where $F_{1}(x)$ is such that \\ $0= k_{2}x^{2} +4k_{1}^{2} +\left( 9F_{1} -c_{1}x^{2} \right) \left( F_{1} -c_{1}x^{2} \right)^{3} -4k_{1} \left( F_{1} -c_{1}x^{2} \right) \left( 3F_{1} +c_{1}x^{2} \right) +$ \\ \qquad $+4k_{3} \left( 3F_{1} -c_{1}x^{2} \right) \left( F_{1} -c_{1}x^{2} \right)^{2} +4k_{3}^{2} \left( F_{1} -c_{1}x^{2} \right)^{2} -\frac{8k_{1}k_{3}}{3} \left( 3F_{1} -c_{1}x^{2} \right)$ \\ \\ $J_{s141}= \frac{1}{2}\dot{x}^{2} +F_{1}(x)$, \enskip $J_{s142}= \frac{1}{2}\dot{y}^{2} +c_{1}y^{2}$ \\ $I_{s143}= L\dot{x}^{2} -\left( 3yF_{1} -c_{1}x^{2}y +k_{3}y \right) \dot{x} +\frac{1}{2c_{1}} F_{1}' \left( 3F_{1} -c_{1}x^{2} +k_{3} \right)\dot{y}$} & New & \makecell{New \\ eq. (C.6) \\ in Ref. \ref{Gravel 2004} \\ for $k_{3}=0$} \\ \hline
\makecell[l]{$V_{s15}= \frac{df}{d\theta}r^{-2}$ \\ where  $\sin\theta \left[ \left( 3\frac{df}{d\theta} +c_{1} \right) \frac{d^{2}f}{d\theta^{2}} -2f\frac{df}{d\theta} \right] +\cos\theta \left[ f\frac{d^{2}f}{d\theta^{2}} +4 \left( \frac{df}{d\theta} \right)^{2} +2c_{1}\frac{df}{d\theta} \right] =0$ \\ \\ $J_{s151}= \frac{1}{2}p_{\theta}^{2} +\frac{df}{d\theta}$ \\
$J_{s152}= p_{\theta}^{2} \left( \cos\theta p_{r} -\sin\theta \frac{p_{\theta}}{r} \right) + \left[ \left( 2\frac{df}{d\theta} +c_{1} \right) \cos\theta -f \sin\theta \right] p_{r}-$ \\ \qquad \quad \enskip $- \left[ \left( 3\frac{df}{d\theta} +c_{1} \right) \sin\theta +f\cos\theta \right] \frac{p_{\theta}}{r}$} & \makecell[c]{not \\ included} & \makecell[c]{Table III \\ $c_{1}=0$} \\ \hline
\makecell[l]{$V_{s16}= \frac{k}{r} +\frac{df}{d\theta}r^{-2}$ \\
where $k\neq0$ \\
$\sin\theta \left[ \left( 3\frac{df}{d\theta} +c_{1} \right) \frac{d^{2}f}{d\theta^{2}} -2f\frac{df}{d\theta} \right] +\cos\theta \left[ f\frac{d^{2}f}{d\theta^{2}} +4 \left( \frac{df}{d\theta} \right)^{2} +2c_{1}\frac{df}{d\theta} \right] =0$ and \\ $\left( c_{2} -\frac{df}{d\theta} \right) \frac{d^{2}f}{d \theta^{2}} +k \left( c_{1} +2\frac{df}{d\theta} \right)\cos\theta +k \left( \frac{d^{2}f}{d\theta^{2}} -f \right) \sin\theta= 0$ \\ \\ $J_{s161}= \frac{1}{2}p_{\theta}^{2} +\frac{df}{d\theta}$ \\ $J_{s162}= \frac{1}{3}p_{\theta}^{3} +\left( \cos\theta p_{r} -\sin\theta \frac{p_{\theta}}{r} \right) p_{\theta}^{2} + \left[ \left( 2\frac{df}{d\theta} +c_{1} \right) \cos\theta -f\sin\theta \right] p_{r} -$ \\ \qquad \quad \enskip
$-\left[ \left( 3\frac{df}{d\theta} +c_{1} \right) \sin\theta +f\cos\theta \right] \frac{p_{\theta}}{r} - \left( c_{2} -\frac{df}{d\theta} +k\sin\theta \right)p_{\theta}$} & New & New \\ \hline
\caption{\label{Table.cfi2} Superintegrable potentials $V(x,y)$ that admit CFIs of the type $J^{(3,2)}_{0}$.}
\end{longtable}

\textbf{Note 1:} In Table \ref{Table.cfi2} we include also the additional LFIs/QFIs which make a potential that admits a CFI of the type $J^{(3,2)}_{0}$ superintegrable. In some cases, the knowledge of these (autonomous or time-dependent) LFIs/QFIs is sufficient for showing that the potential is superintegrable (see e.g. potentials $V_{s1}$, $V_{s2}$, $V_{s5}$, and $V_{s8}$). In these cases, the additional CFI can be expressed in terms of the LFIs/QFIs. \newline

\textbf{Note 2:} The potential (see eq. (3.2.36) of Ref. \ref{Hietarinta 1987} and Table III of Ref. \ref{Karlovini 2000})
\begin{equation*}
V_{s8(2)}=\frac{A}{r} + \frac{B}{r(r+x)} +\frac{C}{r(r-x)} \label{eq.pcf2.53}
\end{equation*}
where $A, B, C$ are arbitrary constants is of the type $V_{s8}$ for $k_{1}=B+C, k_{2}=A$, and $k_{3}=C-B$. Therefore, there is no need to be discussed separately as done in Ref. \ref{Karlovini 2000}. \newline

\textbf{Note 3:} In section 3.4 of Ref. \ref{Fokas 1980}, the authors attempted to determine all potentials of the form $V=F(x^{2} +\nu y^{2})$, where $\nu$ is an arbitrary constant and $F$ an arbitrary smooth function, that admit autonomous CFIs. They have found the following three potentials (see eqs. (3.15a), (3.15b) and (3.19) of Ref. \ref{Fokas 1980}):
\[
V_{(1a)}= \frac{1}{2}x^{2} +\frac{9}{2}y^{2}, \enskip V_{(1b)}= \frac{1}{2}x^{2} +\frac{1}{18}y^{2}, \enskip V_{(1c)}= (x^{2} -y^{2})^{-2/3}.
\]
These potentials are subcases of two families of potentials ($V_{5}$ and $V_{s6}$ - see section \ref{sec.Holt2}, cases 3) and 5) ) admitting non-trivial CFIs, which are collected in Tables \ref{Table.cfi1} and \ref{Table.cfi2}.

Indeed, the potential $V_{(1a)}$ can be derived from the potential $\bar{V}_{s6}\equiv V_{s6}(x \leftrightarrow y)= c_{0}(x^{2} +9y^{2})$ for $c_{0}=\frac{1}{2}$. In this case, the associated CFI is
\[
\bar{J}_{s63}\equiv J_{s63}(x\leftrightarrow y, c_{0}=1/2)= (x\dot{y} -y\dot{x})\dot{x}^{2} +3x^{2}y\dot{x} -\frac{1}{3}x^{3}\dot{y}.
\]

The potential $V_{(1b)}$ is a subcase of the superintegrable potential $V_{s6}=c_{0}(9x^{2} +y^{2})$ for $c_{0} =\frac{1}{18}$. In this case, the associated CFI is
\[
J_{s63}(c_{0}=1/18)= (x\dot{y} -y\dot{x})^{2} \dot{y}^{2} +\frac{1}{27}y^{3}\dot{x} -\frac{1}{3}xy^{2}\dot{y}.
\]
There is a misprint in the FI (3.15b) of Ref. \ref{Fokas 1980} where the $p_{1}=\dot{x}$ in the last term must be $p_{2}=\dot{y}$.

Finally, the potential $V_{(1c)}$ is a subcase of the integrable potential $V_{5}= k(x^{2}-y^{2})^{-2/3}$ for $k=1$. \newline

\textbf{Note 4:} For $k_{3}=1$ the integrable potential $V_{7}$ in Table \ref{Table.cfi1} reduces to the potential given in eq. (4.8) of Ref. \ref{Tsiganov 2000}. The author in Ref. \ref{Tsiganov 2000} claims that $V_{7}(k_{3}=1)$ is superintegrable due to three independent FIs which are the Hamiltonian, an additional QFI (which is not given), and a CFI of the type $J^{(3,2)}_{0}$. To our knowledge, this statement is not true, because the only QFI allowed by the potential $V_{7}$ is the Hamiltonian; therefore, $V_{7}$ is a new purely third order integrable potential due to the additional CFI $J_{7}$ (see Table \ref{Table.cfi1}).

Moreover, for $a_{5}=0$, the superintegrable potential $V_{s8} = \frac{k_{1}}{y^{2}} +\frac{k_{2}}{r} +\frac{k_{3}x}{ry^{2}} = V_{7}(a_{5}=0)$ is obtained and the CFI $J_{s83}= J_{7}(a_{5}=0)$. \newline

\textbf{Note 5:} The potential $V_{4}=k(xy)^{-2/3}$ has been found by Inozemtsev (see eq. (10) of Ref. \ref{Inozemtsev 1983}); however, it is not mentioned in Refs. \ref{Hietarinta 1987} and \ref{Karlovini 2000}. There is a comment made in Ref. \ref{Hietarinta 1987} -below eq. (3.3.25)- concerning Ref. \ref{Inozemtsev 1983} which is not correct because Inozemtsev by using Holt's method found the potential $V_{4}=k(xy)^{-2/3}$ and not the potential $V_{5}=k (x^{2} -y^{2})^{-2/3}$ which was found in Ref. \ref{Fokas 1980} for $k=1$. In the case 5) of section \ref{sec.Holt2}, we have found the potential $V_{5}$ using Holt's method. \newline

\textbf{Note 6:} The superintegrable potential (see section \ref{sec.Holt5}) $V_{s10}= c_{0}(x^{2} +9y^{2}) +c_{1}y$ is a new result which generalizes the potentials (3.15a) and (3.15b) of Ref. \ref{Fokas 1980}. It holds that $V_{s10}(c_{1}=0, x\leftrightarrow y)= V_{s6}$. \newline

\textbf{Note 7:} In Theorem 2 of Ref. \ref{Tremblay 2010}, it is found that there are only four integrable/superintegrable potentials separating in polar coordinates $(r,\theta)$ which admit a CFI. All these potentials are included in Tables \ref{Table.cfi1} and \ref{Table.cfi2}, either as they are or as special cases. Indeed, we have the following: \newline
a) Potentials (4.15) and (4.20) of Ref. \ref{Tremblay 2010} are the well-known second order superintegrable potentials $V_{s8}(x \leftrightarrow y)$ and $V_{s5}$ (see also Note 1 above), respectively. Indeed, for the potential (4.15) of Ref. \ref{Tremblay 2010}, we have:
\[
V= \frac{k_{2}}{r} +\frac{k_{1}+k_{3} \sin\theta}{r^{2}\cos^{2}\theta}= \frac{k_{2}}{r} +\frac{k_{1}+\frac{k_{3}y}{r}}{x^{2}}= \frac{k_{2}}{r} +\frac{k_{1}}{x^{2}} +\frac{k_{3}y}{rx^{2}}= V_{s8}(x\leftrightarrow y).
\]
b) The potential (4.21) of Ref. \ref{Tremblay 2010} is a subcase of the integrable potential $V_{1}$. A detailed derivation is given in section \ref{sec.int.1}, Case 2). \newline
c) The potential (5.1) of Ref. \ref{Tremblay 2010} is the superintegrable potential $V_{s15}$. The associated condition (\ref{eq.intc.9.25}) coincides with eq. (4.27) of Ref. \ref{Tremblay 2010} if we make the transformation $f \leftrightarrow f -c_{0}$ where $c_{0}$ is an arbitrary constant; alternatively, we may add the constant $c_{0}$ in the function $A_{3}(u)$ of equation (\ref{eq.intc.9.12}).  Comparing with the notation of Ref. \ref{Tremblay 2010}, we have $f=T$, $c_{1}=\beta_{1}$, and $c_{0}=\beta_{2}$.

We note that $V_{s16}$ is a new superintegrable separable potential in polar coordinates which is not included in Ref. \ref{Tremblay 2010}.\newline

\textbf{Note 8 (separability):} According to the review paper\footnote{W. Miller, S. Post and P. Winternitz, \emph{`Classical and quantum superintegrability with applications'}, J. Phys. A: Math. Theor. \textbf{46}, 423001 (2013). \label{Miller review}} cited in Ref. \ref{Miller review}, we have the following classification (see sec. 5.3 of Ref. \ref{Miller review}) on the separability of a dynamical system in $E^{2}$: \newline
- Potentials separating in Cartesian coordinates $(x,y)$ are of the form (see sec. 5.3.1 of Ref. \ref{Miller review})
\[
V(x,y)= F_{1}(x) +F_{2}(y).
\]
- Potentials separating in polar coordinates $(r,\theta)$, where $x=r\cos\theta$ and $y=r\sin\theta$, are of the form (see sec. 5.3.2 of Ref. \ref{Miller review})
\[
V(r,\theta)= F_{1}(r) +\frac{F_{2}(\theta)}{r^{2}}.
\]
- Potentials separating in parabolic coordinates $(\xi, \eta)$, where $x= \frac{\xi^{2} -\eta^{2}}{2}$ and $y=\xi \eta$, are of the form (see sec. 5.3.3 of Ref. \ref{Miller review})
\[
V(\xi,\eta)= \frac{F_{1}(\xi) +F_{2}(\eta)}{\xi^{2} +\eta^{2}}.
\]

It is well-known that a separable Netwonian potential in $E^{2}$ admits always a QFI, that is, there is a one-to-one correspondence between the separation of variables and the QFIs. Therefore, 2d separable potentials allowing an additional CFI become third order superintegrable potentials. It has been proved\footnote{I. Popper, S. Post and P. Winternitz, \emph{`Third-order superintegrable systems separable in parabolic coordinates'}, J. Math. Phys. \textbf{53}, 062105 (2012). \label{Popper 2012}} that such potentials separating in parabolic coordinates are in fact second order superintegrable, because their additional CFI reduces to two QFIs (i.e. there exist two different coordinate systems in which the potential can be separated) whose Poisson bracket gives the CFI. Therefore, purely third order superintegrable potentials with a separation of variables exist only in the case of Cartesian and polar coordinates (see also the last paragraph of sec. 5.3.3 in \ref{Miller review}).

All the potentials of Table \ref{Table.cfi1} are non-separable (i.e. they do not admit QFIs other than the Hamiltonian) third order integrable potentials. However, all the potentials of Table \ref{Table.cfi2} are separable either in one or two sets of coordinates. The latter are the potentials mentioned in Note 1.

\section{The CFI $J^{(3,1)}_{1}$ where $\ell=1$}

\label{sec.pot.cfi1}

In order to simplify the notation, we set $L_{(0)ab}=C_{ab}$, $L_{(2)ab}= D_{ab}$ and $L_{(1)a}=L_{a}$. The CFI (\ref{eq.cfi1}) for $\ell=1$ becomes
\begin{equation}
J^{(3,1)}_{1}= -tC_{(ab;c)} \dot{q}^{a} \dot{q}^{b}\dot{q}^{c} + \left( t^{2}D_{ab} +C_{ab} \right) \dot{q}^{a} \dot{q}^{b} + +tL_{a} \dot{q}^{a}+ \frac{t^{2}}{2}L_{a}V^{,a} + G(q) \label{eq.pcf1}
\end{equation}
where the symmetric tensor (not a KT!) $C_{ab}$ is given by (\ref{eq.KT2}), the generated third order KT $C_{(ab;c)}$ is given by (\ref{eq.KT3}), $D_{ab}$ is a second order KT given by (\ref{FL.14b}), and the vector $L_{a}$ satisfies the conditions:
\begin{eqnarray}
L_{(a;b)}&=& -3 C_{(ab;c)} V^{,c} -2D_{ab} \label{eq.pcf1.1} \\
\left( L_{b}V^{,b} \right)_{,a} &=& 4 D_{ab}V^{,b} \label{eq.pcf1.2} \\
G_{,a}&=& 2C_{ab}V^{,b} -L_{a}. \label{eq.pcf1.3}
\end{eqnarray}
These conditions must be also supplemented by the integrability conditions $\left( L_{b}V^{,b} \right)_{,[xy]} =0$ and $G_{,[xy]}=0$.

From the results of section \ref{sec.E2.geometry}, we have for the second order KT $D_{ab}$ and the symmetric tensor $C_{ab}$:
\begin{equation}
D_{ab}=\left(
\begin{array}{cc}
\gamma y^{2}+2\alpha y+A & -\gamma xy-\alpha x-\beta y+C \\
-\gamma xy-\alpha x-\beta y+C & \gamma x^{2}+2\beta x+B%
\end{array}
\right) \label{eq.pcf1.4}
\end{equation}
\begin{eqnarray}
C_{11}&=& 3b_{2}xy^{2} +3b_{5}y^{3} +3b_{3}xy +3(b_{10}+b_{8})y^{2} +b_{4}x +3b_{15}y +b_{12} \notag \\
C_{12}&=& -3b_{2}x^{2}y -3b_{5}xy^{2} -\frac{3}{2}b_{3}x^{2} -\frac{3}{2}(2b_{10}+b_{8})xy -\frac{3}{2}b_{6}y^2 +\frac{3}{2}(b_{9} -b_{15})x -\frac{3}{2}b_{11}y +b_{13} \notag \\
C_{22}&=& 3b_{2}x^{3} +3b_{5}x^{2}y +3b_{10}x^{2} +3b_{6}xy +3(b_{1} +b_{11})x +b_{7}y +b_{14} \label{eq.pcf1.5}
\end{eqnarray}
and for the third order KT\footnote{We note that if the symmetric tensor $C_{ab}$ is a second order KT (i.e. $b_{1}=b_{2}= ...=b_{9}=0$), then $C_{(ab;c)}=0$ and the CFI $J^{(3,1)}_{0}$ reduces to a QFI studied in Ref. \ref{MitsTsam sym}. Therefore, in order to have new results, at least one of the parameters $b_{1}, b_{2}, ..., b_{9}$ must be non-zero.} $C_{(ab;c)}$:
\begin{eqnarray}
C_{(11;1)}&=& 3b_{2}y^{2} +3b_{3}y +b_{4} \notag \\
C_{(11;2)}&=& -2b_{2}xy +b_{5}y^{2} -b_{3}x +b_{8}y +b_{9} \notag \\
C_{(22;1)}&=& b_{2}x^{2}-2b_{5}xy -b_{8}x -b_{6}y +b_{1} \label{eq.pcf1.6} \\
C_{(22;2)}&=& 3b_{5}x^{2} +3b_{6}x +b_{7}. \notag
\end{eqnarray}

Substituting the above quantities in the conditions (\ref{eq.pcf1.1}) - (\ref{eq.pcf1.3}) and taking into account the associated integrability conditions, we find the following system of PDEs:
\begin{eqnarray}
0 &=& L_{1,x} +3\left( 3b_{2}y^{2} +3b_{3}y +b_{4} \right) V_{,x} +3\left( b_{5}y^{2} -2b_{2}xy -b_{3}x +b_{8}y +b_{9} \right)V_{,y} +2\gamma y^{2} +4\alpha y +2A \label{eq.pcf1a} \\
0 &=& L_{1,y} + L_{2,x} +6\left( b_{5}y^{2} +b_{8}y -2b_{2}xy -b_{3}x +b_{9} \right) V_{,x} -6\left( 2b_{5}xy -b_{2}x^{2} +b_{8}x +b_{6}y -b_{1} \right) V_{,y} - \notag \\
&& -4\left(\gamma xy +\alpha x +\beta y -C \right) \label{eq.pcf1b} \\
0 &=& L_{2,y} +3\left( b_{2}x^{2} -2b_{5}xy -b_{8}x -b_{6}y +b_{1} \right)V_{,x} +3\left( 3b_{5}x^{2} +3b_{6}x +b_{7} \right) V_{,y} +2\gamma x^{2} +4\beta x +2B \label{eq.pcf1c} \\
0&=& L_{1}V_{,xx} +L_{2}V_{,xy} +L_{1,x}V_{,x} +L_{2,x}V_{,y} -4\left(\gamma y^{2} +2\alpha y +A\right)V_{,x} + 4\left( \gamma xy +\alpha x +\beta y -C \right)V_{,y} \label{eq.pcf1d} \\
0&=& L_{1}V_{,xy} +L_{2}V_{,yy} +L_{1,y}V_{,x} +L_{2,y}V_{,y} + 4\left( \gamma xy +\alpha x +\beta y -C \right)V_{,x} -4\left(\gamma x^{2} +2\beta x +B\right)V_{,y} \label{eq.pcf1e} \\
0&=& G_{,x} +L_{1} -2\left( 3b_{2}xy^{2} +3b_{5}y^{3} +3b_{3}xy +3(b_{10}+b_{8})y^{2} +b_{4}x +3b_{15}y +b_{12} \right) V_{,x} - \notag \\
&& -2\left( -3b_{2}x^{2}y -3b_{5}xy^{2} -\frac{3}{2}b_{3}x^{2} -\frac{3}{2}(2b_{10}+b_{8})xy -\frac{3}{2}b_{6}y^2 +\frac{3}{2}(b_{9} -b_{15})x -\frac{3}{2}b_{11}y +b_{13} \right) V_{,y} \label{eq.pcf1f} \\
0&=& G_{,y} +L_{2} -2\left( 3b_{2}x^{3} +3b_{5}x^{2}y +3b_{10}x^{2} +3b_{6}xy +3(b_{1} +b_{11})x +b_{7}y +b_{14} \right) V_{,y} - \notag \\
&& -2\left( -3b_{2}x^{2}y -3b_{5}xy^{2} -\frac{3}{2}b_{3}x^{2} -\frac{3}{2}(2b_{10}+b_{8})xy -\frac{3}{2}b_{6}y^2 +\frac{3}{2}(b_{9} -b_{15})x -\frac{3}{2}b_{11}y +b_{13} \right) V_{,x} \label{eq.pcf1g} \\
0 &=& \left( \gamma xy +\alpha x +\beta y -C \right) (V_{,xx} - V_{,yy}) +\left[\gamma (y^{2} -x^{2}) -2\beta x +2\alpha y +A -B \right]V_{,xy} - \notag \\
&& -3(\gamma x +\beta) V_{,y} +3(\gamma y +\alpha) V_{,x} \label{eq.pcf1h} \\
0&=& \left[ -3b_{2}x^{2}y -3b_{5}xy^{2} -\frac{3}{2}b_{3}x^{2} -\frac{3}{2}(2b_{10}+b_{8})xy -\frac{3}{2}b_{6}y^2 +\frac{3}{2}(b_{9} -b_{15})x -\frac{3}{2}b_{11}y +b_{13} \right] \left( V_{,xx} -V_{,yy} \right) + \notag \\
&& + \left[ 3(b_{2}x+b_{5}y)(x^{2} -y^{2}) +3b_{10}x^{2} -3(b_{10}+b_{8})y^{2} +3(b_{6}-b_{3})xy +(3b_{1} +3b_{11} -b_{4})x + \right. \notag \\
&& \left.+ (b_{7} -3b_{15})y +b_{14}-b_{12} \right] V_{,xy} +\left( -12b_{5}y^{2} -12b_{2}xy -6b_{3}x -9b_{10}y -\frac{15}{2}b_{8}y +\frac{3}{2}b_{9} -\frac{9}{2}b_{15} \right) V_{,x} +\notag \\
&& + \left( 12b_{2}x^{2} +12b_{5}xy +\frac{3}{2}b_{8}x +9b_{10}x +6b_{6}y +3b_{1} +\frac{9}{2}b_{11} \right) V_{,y} +\frac{1}{2} \left( L_{1,y} -L_{2,x} \right) \label{eq.pcf1i}
\end{eqnarray}
where (\ref{eq.pcf1h}) is the well-known Bertrand-Darboux equation (arises from the condition $G_{,[xy]}=0$). Therefore, we have to solve an overdetermined system of nine PDEs with four unknown functions $L_{1}(x,y)$, $L_{2}(x,y)$, $V(x,y)$ and $G(x,y)$. Since a general solution is not possible to be found, we consider several special cases.

\subsection{Case $D_{ab}=0$ and $L^{a}=Z (V^{,y}, -V^{,x})$}

\label{sec.pot.cfi1.1}

In this case, the CFI (\ref{eq.pcf1}) becomes
\begin{eqnarray}
J^{(3,1)}_{1}(D_{ab}=0)&=& -tC_{(ab;c)} \dot{q}^{a} \dot{q}^{b}\dot{q}^{c} + C_{ab}\dot{q}^{a} \dot{q}^{b} +tL_{a} \dot{q}^{a} + G(q) \notag \\
&=& t\underbrace{\left( -C_{(ab;c)} \dot{q}^{a} \dot{q}^{b}\dot{q}^{c} +L_{a} \dot{q}^{a}\right)}_{\equiv J}+ C_{ab}\dot{q}^{a} \dot{q}^{b} + G(q) \label{eq.pcf1.7.0}
\end{eqnarray}
and the system of equations (\ref{eq.pcf1.1}) - (\ref{eq.pcf1.3}) reduces to:
\begin{eqnarray}
L_{(a;b)}&=& -3 C_{(ab;c)} V^{,c} \label{eq.pcf1.7} \\
G_{,a}&=& 2C_{ab}V^{,b} -L_{a} \label{eq.pcf1.8}
\end{eqnarray}
where the symmetric tensor $C_{ab}$ is given by (\ref{eq.pcf1.5}), the reducible KT $C_{(ab;c)}$ by (\ref{eq.pcf1.6}), the vector\footnote{See section \ref{sec.pot.cfi2.2Holt} for more details about the choice (\ref{eq.pcf1.9}).}
\begin{equation}
L_{a}= Z
\left(
  \begin{array}{c}
    V_{,y} \\
    -V_{,x} \\
  \end{array}
\right) \label{eq.pcf1.9}
\end{equation}
and $Z(x,y)$ is an arbitrary smooth function.

Because the coefficient of $t$ in the CFI (\ref{eq.pcf1.7.0})
\begin{equation}
J= -C_{(ab;c)} \dot{q}^{a} \dot{q}^{b}\dot{q}^{c} +L_{a} \dot{q}^{a} \label{eq.pcf1.7.1}
\end{equation}
is an autonomous CFI of the form (\ref{eq.pcf2.3}), potentials that admit the time-dependent CFI (\ref{eq.pcf1.7.0}) admit also the autonomous CFI (\ref{eq.pcf1.7.1}). Therefore, integrable potentials found in section \ref{sec.pot.cfi2.2Holt} can become superintegrable by satisfying the additional condition (\ref{eq.pcf1.8}) which produces the additional time-dependent CFI (\ref{eq.pcf1.7.0}).

Replacing the vector (\ref{eq.pcf1.9}) in (\ref{eq.pcf1.7}) and following the method of section \ref{sec.pot.cfi2.2Holt}, we find that
\begin{equation}
Z(x,y)= F(V) + Y(x,y) \label{eq.pcf1.12}
\end{equation}
where $F(V)$ is an arbitrary (smooth) function of the potential $V(x,y)$ and
\begin{eqnarray}
Y(x,y) &=& 3(b_{2}y -b_{5}x)(x^{2}+y^{2}) -\frac{3}{2} (3b_{6}-b_{3})x^{2} +\frac{3}{2}(3b_{3}-b_{6})y^{2} -3b_{8}xy - \notag \\
&& -3(b_{7}+b_{9})x +3(b_{4}+b_{1})y +c \label{eq.pcf1.13}
\end{eqnarray}
where $c$ is an arbitrary constant, and the system of equations:
\begin{align}
& \left[ F(V) + Y\right] V_{,xy} +F'V_{,x}V_{,y} +3C_{(11;1)}V_{,x} -3C_{(22;2)}V_{,y} = 0 \label{eq.pcf1.14a} \\
& \left[ F(V) + Y\right](V_{,yy}-V_{,xx}) +F'\left[ (V_{,y})^{2} - (V_{,x})^{2} \right] +3\left( C_{(11;1)} +3C_{(22;1)} \right) V_{,y} +3\left( C_{(22;2)} +3C_{(11;2)} \right)V_{,x} =0 \label{eq.pcf1.14b}
\end{align}
where $F' \equiv \frac{dF}{dV}$. Equations (\ref{eq.pcf1.14a}) and (\ref{eq.pcf1.14b}) must be supplemented also with the integrability condition (\ref{eq.pcf1i}) of the function $G(x,y)$:
\begin{eqnarray}
0&=& \left[ -3b_{2}x^{2}y -3b_{5}xy^{2} -\frac{3}{2}b_{3}x^{2} -\frac{3}{2}(2b_{10}+b_{8})xy -\frac{3}{2}b_{6}y^2 +\frac{3}{2}(b_{9} -b_{15})x -\frac{3}{2}b_{11}y +b_{13} \right] \left( V_{,xx} -V_{,yy} \right) + \notag \\
&& + \left[ 3(b_{2}x+b_{5}y)(x^{2} -y^{2}) +3b_{10}x^{2} -3(b_{10}+b_{8})y^{2} +3(b_{6}-b_{3})xy +(3b_{1} +3b_{11} -b_{4})x + (b_{7} -3b_{15})y+ \right. \notag \\
&& \left. +b_{14}-b_{12} \right] V_{,xy}
-\left[ \frac{9}{2}b_{5}(x^{2} +3y^{2}) +9b_{2}xy +\frac{9}{2}(b_{3}+b_{6})x +9(b_{8}+b_{10})y +\frac{3}{2}(b_{7} +3b_{15}) \right] V_{,x} +\notag \\
&& + \left[ \frac{9}{2}b_{2}(3x^{2} +y^{2}) +9b_{5}xy +9b_{10}x +\frac{9}{2}(b_{3} +b_{6})y +\frac{3}{2}(b_{4} +3b_{1} +3b_{11}) \right] V_{,y}+ \frac{1}{2} (F+Y)(V_{,xx} +V_{,yy}) + \notag \\ && +\frac{1}{2}F'\left[ (V_{,x})^{2} +(V_{,y})^{2} \right]. \label{eq.pcf1.14c}
\end{eqnarray}
Finally, the function $G(x,y)$ is computed by integrating the condition (\ref{eq.pcf1.8}).

We consider several cases.

\subsubsection{Case $F(V)=0$, $b_{1}=1$, $b_{11}=-1$ and the remaining parameters are fixed to zero}

\label{sec.pot.cfi1.1a}

We find $C_{ab}=
\left(
  \begin{array}{cc}
    0 & \frac{3}{2}y \\
    \frac{3}{2}y & 0 \\
  \end{array}
\right)$, $C_{(22;1)}=1$, and $Z= Y= 3y$.

The system of equations (\ref{eq.pcf1.14a}) - (\ref{eq.pcf1.14c}) becomes:
\begin{eqnarray}
V_{,xy} &=& 0 \label{eq.pcf1.15a} \\
y(V_{,yy} -V_{,xx}) +3V_{,y} &=& 0 \label{eq.pcf1.15b} \\
V_{,xx} &=& 0. \label{eq.pcf1.15c}
\end{eqnarray}
The solution of the above system of equations is the potential
\begin{equation}
V(x,y)= \frac{c_{2}}{y^{2}} +c_{3}x \label{eq.pcf1.16}
\end{equation}
where $c_{2}$ and $c_{3}$ are arbitrary constants. If we interchange $x \leftrightarrow y$, the potential (\ref{eq.pcf1.16}) is a subcase of the superintegrable potential $V_{s1}$ (see Table \ref{Table.cfi2} and case 2 of section \ref{sec.Holt2}) for $c_{1}=0$.

By integrating the condition (\ref{eq.pcf1.8}), we get $G(x,y)= 3c_{3}y^{2}$ and the associated CFI (\ref{eq.pcf1.7.0}) is
\begin{equation}
J^{(3,1)}_{1}= -t\dot{x}\dot{y}^{2} +y\dot{x}\dot{y} -\frac{2c_{2}}{y^{2}}t \dot{x} -c_{3}ty\dot{y} +c_{3}y^{2} =tJ_{s11}(c_{1}=0, x \leftrightarrow y) -x\dot{x}\dot{y} -c_{3}x^{2}. \label{eq.pcf1.17}
\end{equation}

\subsubsection{Case $F(V)=\lambda V \implies F'=\lambda\neq0$, $b_{4}\neq0$ and the remaining parameters of $C_{(ab;c)}$ are fixed to zero.}

\label{sec.pot.cfi1.1b}

We find $C_{(11;1)}=b_{4}$, $Y= 3b_{4}y$ and $Z=\lambda V -3b_{4}y$.

Substituting in equations (\ref{eq.pcf1.14a}) and (\ref{eq.pcf1.14b}), we find the potential (see case 5 of section \ref{sec.Holt1})
\begin{equation}
V(x,y)= V_{s3}= c_{1}\sqrt{x} +c_{2}y \label{eq.pcf1.20}
\end{equation}
where $c_{1}\equiv\frac{1}{\lambda}\neq0$ and $c_{2}\equiv -\frac{3b_{4}}{\lambda}=-3c_{1}b_{4}$. This is the superintegrable potential $V_{s3}$ of Table \ref{Table.cfi2}.

Replacing the potential (\ref{eq.pcf1.20}) in (\ref{eq.pcf1.14c}), we find that $b_{10}=b_{11}=b_{13}= b_{15}=0$ and $b_{12}, b_{14}$ are the only surviving parameters. Therefore, $Z(x,y)= \sqrt{x}$ and $C_{ab}=
\left(
  \begin{array}{cc}
    b_{4}x+b_{12} & 0 \\
    0 & b_{14} \\
  \end{array}
\right)$. Substituting these results in the remaining condition (\ref{eq.pcf1.8}), we get $G(x,y)= \frac{8b_{4}}{3\lambda} x\sqrt{x} +\frac{2b_{12}}{\lambda} \sqrt{x} -\frac{6b_{4}b_{14}}{\lambda}y +\frac{y}{2\lambda}$.

The associated CFI (\ref{eq.pcf1.7.0}) is
\begin{eqnarray}
J^{(3,1)}_{1}&=& -tb_{4}\dot{x}^{3} +b_{4}x\dot{x}^{2} -\frac{3b_{4}}{\lambda}t\sqrt{x}\dot{x} -\frac{t}{2\lambda} \dot{y} +\frac{8b_{4}}{3\lambda}x\sqrt{x} +\frac{y}{2\lambda} +2b_{12}J_{s31} +2b_{14}J_{s32} \notag \\
&=& \frac{1}{3c_{1}} \left( J_{s33}t -c_{2}x\dot{x}^{2} +\frac{3c_{1}^{2}}{2}y -\frac{8c_{1}c_{2}}{3}x\sqrt{x} \right) +2b_{12}J_{s31} +2b_{14}J_{s32} \label{eq.pcf1.21}
\end{eqnarray}
where $J_{s31}$ and $J_{s32}$ (see Table \ref{Table.cfi2}) are the QFIs resulting from the separability of the potential $V_{s3}$, and $J_{s33}$ (see Table \ref{Table.cfi2}) is the autonomous CFI of $V_{s3}$. The CFI (\ref{eq.pcf1.21}) contains the new time-dependent CFI
\begin{equation}
J_{s34}= tJ_{s33} -c_{2}x\dot{x}^{2} +\frac{3c_{1}^{2}}{2}y -\frac{8c_{1}c_{2}}{3}x\sqrt{x}. \label{eq.pcf1.22}
\end{equation}

\subsubsection{Case $F(V)=\lambda V^{2} \implies F'=2\lambda V$, $\lambda\neq0$, $b_{4}=-b_{7}$ and the remaining parameters are fixed to zero.}

\label{sec.pot.cfi1.1c}

We find $C_{(11;1)}=-b_{7}$, $C_{(22;2)}=b_{7}$, $Y= -3b_{7}(x+y)$ and $Z=\lambda V^{2} -3b_{7}(x+y)$.

Substituting in equations (\ref{eq.pcf1.14a}) and (\ref{eq.pcf1.14b}), we find the superintegrable potential (see case 6 of section \ref{sec.Holt1} and Table \ref{Table.cfi2})
\begin{equation}
V(x,y)= V_{s4}= k(\sqrt{x} \pm \sqrt{y}) \label{eq.pcf1.23}
\end{equation}
where $k^{2}\equiv \frac{3b_{7}}{\lambda}$. The potential (\ref{eq.pcf1.23}) satisfies equation (\ref{eq.pcf1.14c}) identically.

We compute: $Z(x,y)= \pm6b_{7}\sqrt{xy}$ and $C_{ab}=
\left(
  \begin{array}{cc}
   -b_{7}x & 0 \\
    0 & b_{7}y \\
  \end{array}
\right)$. Substituting these quantities in the remaining condition (\ref{eq.pcf1.8}), we get $G(x,y)= -\frac{8kb_{7}}{3}x\sqrt{x} \pm \frac{8kb_{7}}{3} y\sqrt{y}$.

The associated CFI (\ref{eq.pcf1.7.0}) is
\begin{eqnarray}
J^{(3,1)}_{1}&=& t\underbrace{\left( \dot{x}^{3} -\dot{y}^{3} +3k\sqrt{x} \dot{x} \mp 3k\sqrt{y}\dot{y} \right)}_{= J_{s43}} -x\dot{x}^{2} +y\dot{y}^{2} -\frac{8k}{3}x\sqrt{x} \pm \frac{8k}{3} y\sqrt{y} \label{eq.pcf1.24}
\end{eqnarray}
where $J_{s43}$ (see Table \ref{Table.cfi2}) is the autonomous CFI of $V_{s4}$.

\subsubsection{Case $F(V)=0$, $b_{8}=-\frac{1}{3}$ and the remaining parameters of $C_{(ab;c)}$ are fixed to zero.}

\label{sec.pot.cfi1.1d}

We find $C_{(11;2)}=-\frac{y}{3}$, $C_{(22;1)}=\frac{x}{3}$ and $Z=Y=xy$.

Substituting the above quantities in equations (\ref{eq.pcf1.14a}) and (\ref{eq.pcf1.14b}), we find the superintegrable potential (see case 2 of section \ref{sec.Holt2} and Table \ref{Table.cfi2})
\begin{equation}
V(x,y)= V_{s5}= c_{1}(x^{2} +y^{2}) +\frac{c_{2}}{x^{2}} + \frac{c_{3}}{y^{2}} \label{eq.pcf1.25}
\end{equation}
where $c_{1}, c_{2}$, and $c_{3}$ are arbitrary constants.

Replacing the potential (\ref{eq.pcf1.25}) in (\ref{eq.pcf1.14c}), we find that $c_{1}=0$ and $b_{11}=b_{12}= b_{14}=0$. Therefore, we have:
\[
V=V_{s5}(c_{1}=0)= \frac{c_{2}}{x^{2}} + \frac{c_{3}}{y^{2}}, \enskip C_{ab}=
\left(
  \begin{array}{cc}
    (3b_{10}-1)y^{2} +b_{12} & \left(\frac{1}{2}-3b_{10}\right) xy \\
    \left(\frac{1}{2}-3b_{10}\right) xy & -x^{2}+b_{14} \\
  \end{array}
\right).
\]
Substituting in the remaining condition (\ref{eq.pcf1.8}), we obtain $b_{10}=-\frac{1}{3}$ and the function
\[
G(x,y)= -\frac{4c_{2}y^{2}}{x^{2}} -\frac{2c_{3}x^{2}}{y^{2}} +2b_{12}\frac{c_{2}}{x^{2}} + 2b_{14}\frac{c_{3}}{y^{2}}.
\]

The associated CFI (\ref{eq.pcf1.7.0}) is
\begin{eqnarray}
J^{(3,1)}_{1}&=& t\underbrace{\left( y\dot{x}^{2}\dot{y} -x\dot{x}\dot{y}^{2} -2c_{3}\frac{x}{y^{2}}\dot{x} +2c_{2}\frac{y}{x^{2}}\dot{y} \right)}_{= J_{s56}(c_{1}=0)} -2y^{2}\dot{x}^{2} -x^{2}\dot{y}^{2} +3xy\dot{x}\dot{y} -\frac{4c_{2}y^{2}}{x^{2}} -\frac{2c_{3}x^{2}}{y^{2}} +\notag \\ && +2b_{12}J_{s51}(c_{1}=0) + 2b_{14}J_{s52}(c_{1}=0) \label{eq.pcf1.26}
\end{eqnarray}
where $J_{s56}$ (see Table \ref{Table.cfi2}) is the autonomous CFI of $V_{s5}$ and $J_{s51}, J_{s52}$ (see Table \ref{Table.cfi2}) are the QFIs arising from the separability of $V_{s5}$. The expression (\ref{eq.pcf1.26}) contains the new time-dependent CFI
\begin{equation}
J_{s57}= tJ_{s56}(c_{1}=0) -2y^{2}\dot{x}^{2} -x^{2}\dot{y}^{2} +3xy\dot{x}\dot{y} -\frac{4c_{2}y^{2}}{x^{2}} -\frac{2c_{3}x^{2}}{y^{2}}. \label{eq.pcf1.27}
\end{equation}

\subsubsection{Case $F(V)=0$, $b_{3}=b_{6}=1$ and the remaining parameters are fixed to zero.}

\label{sec.pot.cfi1.1e}

We find $C_{(11;1)}=3y$, $C_{(11;2)}=-x$, $C_{(22;1)}= -y$, $C_{(22;2)}= 3x$ and $Z=Y= 3(y^{2} -x^{2})$.

Substituting the above quantities in equations (\ref{eq.pcf1.14a}) and (\ref{eq.pcf1.14b}), we find the superintegrable potential (see case 4 of section \ref{sec.Holt2} and Table \ref{Table.cfi2})
\begin{equation}
V(x,y)= V_{s7}= c_{1}(x^{2} +y^{2}) +\frac{c_{2}xy +c_{3}(x^{2}+y^{2})}{(x^{2}-y^{2})^{2}} \label{eq.pcf1.28}
\end{equation}
where $c_{1}, c_{2}$, and $c_{3}$ are arbitrary constants.

From condition (\ref{eq.pcf1.14c}), we get $c_{1}=0$ and we have:
\[
V=V_{s7}(c_{1}=0)= \frac{c_{2}xy +c_{3}(x^{2} +y^{2})}{(x^{2}-y^{2})^{2}}, \enskip C_{ab}=
\left(
  \begin{array}{cc}
    3xy & -\frac{3}{2}(x^{2}+y^{2}) \\
    -\frac{3}{2}(x^{2}+y^{2}) xy & 3xy \\
  \end{array}
\right).
\]
Integrating the remaining condition (\ref{eq.pcf1.8}), we obtain $G(x,y)= \frac{12c_{2}x^{2}y^{2}}{(x^{2}-y^{2})^{2}} +\frac{12c_{3}xy(x^{2}+y^{2})}{(x^{2}-y^{2})^{2}}$.

The associated CFI (\ref{eq.pcf1.7.0}) is
\begin{equation}
J_{s73}= -tJ_{s72}(c_{1}=0) +xy(\dot{x}^{2} +\dot{y}^{2}) -(x^{2}+y^{2})\dot{x}\dot{y} +\frac{4c_{2}x^{2}y^{2}}{(x^{2}-y^{2})^{2}} +\frac{4c_{3}xy(x^{2}+y^{2})}{(x^{2}-y^{2})^{2}} \label{eq.pcf1.29}
\end{equation}
where $J_{s72}$ (see Table \ref{Table.cfi2}) is the autonomous CFI of $V_{s7}$.

\subsubsection{Case $F(V)=0$, $b_{2}=1$ and the remaining parameters are fixed to zero.}

\label{sec.pot.cfi1.1f}

We find $C_{(11;1)}=3y^{2}$, $C_{(11;2)}=-2xy$, $C_{(22;1)}= x^{2}$, and $Z=Y= 3yr^{2}$ where $r= \sqrt{x^{2} +y^{2}}$.

Substituting the above quantities in the system of equations (\ref{eq.pcf1.14a}) - (\ref{eq.pcf1.14c}), we find the superintegrable potential (see case 1 of section \ref{sec.Holt3} and Table \ref{Table.cfi2})
\begin{equation}
V(x,y)= V_{s8}(k_{2}=0) = \frac{k_{1}}{y^{2}} +\frac{k_{3}x}{ry^{2}} \label{eq.pcf1.30}
\end{equation}
where $k_{1}$ and $k_{3}$ are arbitrary constants. The symmetric tensor $C_{ab}=3x
\left(
  \begin{array}{cc}
    y^{2} & -xy \\
    -xy & x^{2} \\
  \end{array}
\right)$.

Integrating the remaining condition (\ref{eq.pcf1.8}), we get $G(x,y)= 3\frac{2k_{1}r^{2}x +k_{3}r(2x^{2} +y^{2})}{y^{2}}$.

The associated CFI (\ref{eq.pcf1.7.0}) is
\begin{equation}
J_{s84} = -tJ_{s83}(k_{2}=0) +x(x\dot{y} -y\dot{x})^{2} +\frac{2k_{1}r^{2}x +k_{3}r(2x^{2} +y^{2})}{y^{2}} \label{eq.pcf1.31}
\end{equation}
where $J_{s83}$ (see Table \ref{Table.cfi2}) is the autonomous CFI of $V_{s8}$.

\subsubsection{Case $F(V)=0$ and $b_{2},b_{5}$ are the only non-vanishing parameters.}

\label{sec.pot.cfi1.1g}

We rename the parameters as $b_{2}=a_{2}$ and $b_{5}=a_{5}$.

We find $C_{(11;1)}=3a_{2}y^{2}$, $C_{(11;2)}=-2a_{2}xy +a_{5}y^{2}$, $C_{(22;1)}= a_{2}x^{2} -2a_{5}xy$, $C_{(22;2)}= 3a_{5}x^{2}$, and $Z=Y= 3(a_{2}y -a_{5}x)r^{2}$ where $r= \sqrt{x^{2} +y^{2}}$.

Substituting the above quantities in the system of equations (\ref{eq.pcf1.14a}) - (\ref{eq.pcf1.14c}), we find the integrable potential (see case 2 of section \ref{sec.Holt3} and Table \ref{Table.cfi1})
\begin{equation}
V(x,y)= V_{7}(k_{2}=0) = \frac{k_{1}}{(a_{2}y-a_{5}x)^{2}} +\frac{k_{3}(a_{2}x+a_{5}y)}{r(a_{2}y-a_{5}x)^{2}}  \label{eq.pcf1.32}
\end{equation}
where $k_{1}$ and $k_{3}$ are arbitrary constants. The symmetric tensor $C_{ab}= 3(a_{2}x +a_{5}y)
\left(
  \begin{array}{cc}
    y^{2} & -xy \\
    -xy & x^{2} \\
  \end{array}
\right)$.

Integrating the remaining condition (\ref{eq.pcf1.8}), we get $G(x,y)= 3 \left[ \frac{2k_{1}r^{2}(a_{2}x +a_{5}y)}{(a_{2}y -a_{5}x)^{2}} + \frac{2k_{3}r(a_{2}x +a_{5}y)^{2}}{(a_{2}y -a_{5}x)^{2}} +k_{3}r \right]$.

The associated CFI (\ref{eq.pcf1.7.0}) is
\begin{equation}
J_{71}= -tJ_{7}(k_{2}=0) +(a_{2}x +a_{5}y)(x\dot{y} -y\dot{x})^{2} + \frac{2k_{1}r^{2}(a_{2}x +a_{5}y)}{(a_{2}y -a_{5}x)^{2}} +\frac{2k_{3}r(a_{2}x +a_{5}y)^{2}}{(a_{2}y -a_{5}x)^{2}} +k_{3}r \label{eq.pcf1.33}
\end{equation}
where $J_{7}$ (see Table \ref{Table.cfi1}) is the autonomous CFI of $V_{7}$. We note that \emph{the additional time-dependent CFI $J_{71}$ makes the integrable potential $V_{7}(k_{2}=0)$ superintegrable.}
\bigskip

We collect the results of this section in Table \ref{Table.time}.

\begin{longtable}{|l|l|}
\hline
\multicolumn{2}{|c|}{{\large{Superintegrable potentials}}} \\
\hline
{\large Potential} & {\large CFI of the type $J^{(3,1)}_{1}$} \\ \hline
$V_{s1}(c_{1}=0)= \frac{c_{2}}{x^{2}} +c_{3}y$ & $J_{s15}= tJ_{s11}(c_{1}=0) -x\dot{x}\dot{y} -c_{3}x^{2}$ \\ \hline
$V_{s3}=c_{1}\sqrt{x} +c_{2}y$ & $J_{s34}= tJ_{s33} -c_{2}x\dot{x}^{2} +\frac{3c_{1}^{2}}{2}y -\frac{8c_{1}c_{2}}{3}x\sqrt{x}$ \\ \hline
$V_{s4}=k(\sqrt{x} \pm \sqrt{y})$ & $J_{s44}= tJ_{s43} -x\dot{x}^{2} +y\dot{y}^{2} -\frac{8k}{3}x\sqrt{x} \pm \frac{8k}{3} y\sqrt{y}$ \\ \hline
$V_{s5}(c_{1}=0)= \frac{c_{2}}{x^{2}} +\frac{c_{3}}{y^{2}}$ & $J_{s57}= tJ_{s56}(c_{1}=0) -2y^{2}\dot{x}^{2} -x^{2}\dot{y}^{2} +3xy\dot{x}\dot{y} -\frac{4c_{2}y^{2}}{x^{2}} -\frac{2c_{3}x^{2}}{y^{2}}$ \\ \hline
$V_{s7}(c_{1}=0)= \frac{c_{2}xy +c_{3}(x^{2} +y^{2})}{(x^{2}-y^{2})^{2}}$ & \makecell[l]{$J_{s73}= -tJ_{s72}(c_{1}=0) +xy(\dot{x}^{2} +\dot{y}^{2}) -(x^{2}+y^{2})\dot{x}\dot{y} +\frac{4c_{2} x^{2}y^{2}}{(x^{2}-y^{2})^{2}}+$ \\ \qquad \quad $+\frac{4c_{3}xy (x^{2}+y^{2})}{(x^{2}-y^{2})^{2}}$} \\ \hline
$V_{s8}(k_{2}=0)=\frac{k_{1}}{y^{2}} +\frac{k_{3}x}{ry^{2}}$ & $J_{s84} = -tJ_{s83}(k_{2}=0) +x(x\dot{y} -y\dot{x})^{2} +\frac{2k_{1}r^{2}x +k_{3}r(2x^{2} +y^{2})}{y^{2}} $ \\ \hline
$V_{7}(k_{2}=0) = \frac{k_{1}}{(a_{2}y-a_{5}x)^{2}} +\frac{k_{3}(a_{2}x+a_{5}y)}{r(a_{2}y-a_{5}x)^{2}}$ & \makecell[l]{$J_{71}= -tJ_{7}(k_{2}=0) +(a_{2}x +a_{5}y)(x\dot{y} -y\dot{x})^{2} + \frac{2k_{1}r^{2}(a_{2}x +a_{5}y)}{(a_{2}y -a_{5}x)^{2}}+$ \\ \qquad \enskip $+\frac{2k_{3}r(a_{2}x +a_{5}y)^{2}}{(a_{2}y -a_{5}x)^{2}} +k_{3}r$} \\ \hline
\caption{\label{Table.time} Superintegrable potentials $V(x,y)$ that admit time-dependent CFIs of the type $J^{(3,1)}_{1}$.}
\end{longtable}

\textbf{Note 1:} All the time-dependent CFIs contained in Table \ref{Table.time} are new. Indeed, in Refs. \ref{Hietarinta 1987} and \ref{Karlovini 2000} only autonomous CFIs are considered. \newline

\textbf{Note 2:} The potentials $V_{s1}, V_{s3}, V_{s4}, V_{s5}, V_{s7}$, and $V_{s8}$ are found to be superintegrable already in section \ref{sec.pot.cfi2.2Holt} via Holt's method (see Table \ref{Table.cfi2}). However, the additional time-dependent CFIs of Table \ref{Table.time} are equally important because they can be used to integrate the dynamical equations. \newline

\textbf{Note 3:} Due to the time-dependent CFI $J_{71}$, it follows that the integrable potential $V_{7}(k_{2}=0)$ is superintegrable. This is a new result, not included in Table \ref{Table.cfi2}, which illustrates the importance of time-dependent FIs in establishing superintegrability. \newline

\textbf{Note 4:} The integrable and superintegrable potentials of Tables \ref{Table.cfi1} and \ref{Table.cfi2} admitting a time-dependent CFI of the type $J^{(3,1)}_{1}$ are included in Table \ref{Table.time}. The remaining potentials of Tables \ref{Table.cfi1} and \ref{Table.cfi2} do not admit CFIs of this type.

\section{The CFI $I^{(3)}_{e}$}

\label{sec.pot.cfi3}

We have the time-dependent CFI (we set $L_{a}=B_{a}$)
\begin{equation}
I^{(3)}_{e}= e^{\lambda t} \left( -L_{(ab;c)} \dot{q}^{a} \dot{q}^{b} \dot{q}^{c} + \lambda L_{ab} \dot{q}^{a} \dot{q}^{b} + \lambda B_{a} \dot{q}^{a} + B_{a}V^{,a} \right) \label{eq.pcf3}
\end{equation}
where $\lambda\neq0$, the symmetric tensor $L_{ab}$ is given by (\ref{eq.KT2}), the generated third order KT $L_{(ab;c)}$ is given by (\ref{eq.KT3}) and the vector $B_{a}$ satisfies the conditions:
\begin{eqnarray}
B_{(a;b)} &=& -\frac{3}{\lambda} L_{(ab;c)} V^{,c} -\lambda L_{ab} \label{eq.pcf3.1} \\
\left(B_{c}V^{,c}\right)_{,a}&=& 2\lambda L_{ab} V^{,b} -\lambda^{2}B_{a}. \label{eq.pcf3.2}
\end{eqnarray}

From section \ref{sec.E2.geometry}, we have:
\begin{eqnarray}
L_{11}&=& 3b_{2}xy^{2} +3b_{5}y^{3} +3b_{3}xy +3(b_{10}+b_{8})y^{2} +b_{4}x +3b_{15}y +b_{12} \notag \\
L_{12}&=& -3b_{2}x^{2}y -3b_{5}xy^{2} -\frac{3}{2}b_{3}x^{2} -\frac{3}{2}(2b_{10}+b_{8})xy -\frac{3}{2}b_{6}y^2 +\frac{3}{2}(b_{9} -b_{15})x -\frac{3}{2}b_{11}y +b_{13} \label{eq.pcf3.3} \\
L_{22}&=& 3b_{2}x^{3} +3b_{5}x^{2}y +3b_{10}x^{2} +3b_{6}xy +3(b_{1} +b_{11})x +b_{7}y +b_{14} \notag
\end{eqnarray}
and
\begin{eqnarray}
L_{(11;1)}&=& 3b_{2}y^{2} +3b_{3}y +b_{4} \notag \\
L_{(11;2)}&=& -2b_{2}xy +b_{5}y^{2} -b_{3}x +b_{8}y +b_{9} \notag \\
L_{(22;1)}&=& b_{2}x^{2}-2b_{5}xy -b_{8}x -b_{6}y +b_{1} \label{eq.pcf3.4} \\
L_{(22;2)}&=& 3b_{5}x^{2} +3b_{6}x +b_{7} \notag
\end{eqnarray}
where $b_{1}, b_{2}, ..., b_{15}$ are arbitrary constants. We note that at least one of the nine parameters $b_{1}, b_{2}, ..., b_{9}$ must not vanish in order the FI (\ref{eq.pcf3}) to be cubic (i.e. $L_{(ab;c)}\neq0$).

Substituting equations (\ref{eq.pcf3.3}) and (\ref{eq.pcf3.4}) in conditions (\ref{eq.pcf3.1}) and (\ref{eq.pcf3.2}), we obtain a system of six PDEs (including the integrability condition of (\ref{eq.pcf3.2}) ) with three unknown functions $V(x,y)$, $B_{1}(x,y)$, $B_{2}(x,y)$ and fifteen free parameters $b_{1}, b_{2}, ..., b_{15}$. This system cannot be solved in full generality; therefore, we consider special cases.

\subsection{Case $B_{a}V^{,a}=s=const$}

\label{sec.pot.cfi3.1}

We assume
\begin{equation}
B_{a}V^{,a}=s \label{eq.pcf3.5}
\end{equation}
where $s$ is an arbitrary constant.

Substituting assumption (\ref{eq.pcf3.5}) in (\ref{eq.pcf3.2}), we find that the vector
\begin{equation}
B_{a}= \frac{2}{\lambda}L_{ab}V^{,b}. \label{eq.pcf3.6}
\end{equation}
Replacing the vector (\ref{eq.pcf3.6}) in the conditions (\ref{eq.pcf3.1}) and (\ref{eq.pcf3.5}), we find the following system of equations:
\begin{eqnarray}
0&=& L_{11} (V_{,x})^{2} +2L_{12}V_{,x}V_{,y} +L_{22} (V_{,y})^{2} -\frac{s\lambda}{2} \label{eq.pcf3.7a} \\
0&=& 2L_{11}V_{,xx} +2L_{12}V_{,xy} +5L_{(11;1)}V_{,x} +\left( 3L_{(11;2)} +2L_{12,x} \right)V_{,y} +\lambda^{2}L_{11} \label{eq.pcf3.7b} \\
0&=& 2L_{22}V_{,yy} +2L_{12}V_{,xy} +5L_{(22;2)}V_{,y} +\left( 3L_{(22;1)} +2L_{12,y} \right)V_{,x} +\lambda^{2}L_{22} \label{eq.pcf3.7c} \\
0&=& L_{12}\left( V_{,xx} +V_{,yy} \right) +\left( L_{11} +L_{22} \right)V_{,xy} +\left( 3L_{(11;2)} +L_{12,x} +L_{11,y} \right) V_{,x} +\notag \\
&& +\left( 3L_{(22;1)} +L_{22,x} +L_{12,y} \right) V_{,y} +\lambda^{2}L_{12}. \label{eq.pcf3.7d}
\end{eqnarray}
Using equations (\ref{eq.pcf3.3}) and (\ref{eq.pcf3.4}), the system of equations (\ref{eq.pcf3.7a}) - (\ref{eq.pcf3.7d}) becomes:
\begin{align}
0= & L_{11} (V_{,x})^{2} +2L_{12}V_{,x}V_{,y} +L_{22} (V_{,y})^{2} -\frac{s\lambda}{2} \label{eq.pcf3.8a} \\
0= & 2L_{11}V_{,xx} +2L_{12}V_{,xy} +5L_{(11;1)}V_{,x} -3\left( b_{5}y^{2} +6b_{2}xy +3b_{3}x +2b_{10}y -2b_{9} +b_{15} \right)V_{,y} +\lambda^{2}L_{11} \label{eq.pcf3.8b} \\
0= & 2L_{22}V_{,yy} +2L_{12}V_{,xy} +5L_{(22;2)}V_{,y} -3\left[ b_{2}x^{2} +6b_{5}xy +2(b_{8}+b_{10})x +3b_{6}y -b_{1} +b_{11} \right]V_{,x} +\lambda^{2}L_{22} \label{eq.pcf3.8c} \\
0= & L_{12}\left( V_{,xx} +V_{,yy} \right) +\left( L_{11} +L_{22} \right)V_{,xy} +3\left[ 3b_{5}y^{2} -2b_{2}xy -b_{3}x +\left( b_{10} +\frac{5}{2}b_{8} \right)y +\frac{1}{2}(3b_{9} +b_{15}) \right] V_{,x} +\notag \\
& +3\left[ 3b_{2}x^{2} -2b_{5}xy -b_{6}y +\left( b_{10} -\frac{3}{2}b_{8} \right)x +2b_{1} +\frac{b_{11}}{2} \right] V_{,y} +\lambda^{2}L_{12}. \label{eq.pcf3.8d}
\end{align}
This is a system of four PDEs with one unknown function $V(x,y)$. Solving this system, we find 2d potentials that admit time-dependent CFIs of the form (\ref{eq.pcf3}). Because the system of PDEs (\ref{eq.pcf3.8a}) - (\ref{eq.pcf3.8d}) depends on sixteen free parameters $s, b_{1}, b_{2}, ..., b_{15}$, a general solution cannot be found; therefore, special solutions are considered.
\bigskip

1) Case $b_{3}= -b_{6}\equiv \frac{b}{3}$, where $b$ is a non-zero arbitrary constant, and the remaining parameters are fixed to zero.

We find $L_{11}=bxy$, $L_{12}=\frac{b}{2}(y^{2}-x^{2})$, $L_{22}= -bxy$, $L_{(11;1)}= by$, $L_{(11;2)}=-\frac{b}{3}x$, $L_{(22;1)}= \frac{b}{3}y$ and $L_{(22;2)}= -bx$.

The system of PDEs (\ref{eq.pcf3.8a}) - (\ref{eq.pcf3.8d}) becomes:
\begin{eqnarray}
0&=& xy \left[(V_{,x})^{2} -(V_{,y})^{2}\right] +(y^{2}-x^{2})V_{,x}V_{,y} \label{eq.pcf3.9a} \\
0&=& xy(2V_{,xx} +\lambda^{2})+(y^{2}-x^{2})V_{,xy} +5yV_{,x} -3xV_{,y} \label{eq.pcf3.9b} \\
0&=& xy(2V_{,yy} +\lambda^{2}) +(x^{2}-y^{2})V_{,xy} +5xV_{,y} -3yV_{,x} \label{eq.pcf3.9c} \\
0&=& (y^{2}-x^{2})\left( V_{,xx} +V_{,yy} +\lambda^{2}\right) -2xV_{,x} +2yV_{,y}. \label{eq.pcf3.9d}
\end{eqnarray}
Solving the system (\ref{eq.pcf3.9a}) - (\ref{eq.pcf3.9d}), we find the potential
\begin{equation}
V(x,y)= -\frac{\lambda^{2}}{8}r^{2} +\frac{k}{r^{2}} \label{eq.pcf3.10}
\end{equation}
where $r^{2}=x^{2}+y^{2}$ and $k$ is an arbitrary constant. Substituting the potential (\ref{eq.pcf3.10}) in (\ref{eq.pcf3.6}), we get
\begin{equation}
B_{a}=\frac{2b}{\lambda}\left(  \frac{\lambda^{2}}{8}r^{2} +\frac{k}{r^{2}} \right)
\left(
  \begin{array}{c}
   -y \\
   x \\
  \end{array}
\right). \label{eq.pcf3.11}
\end{equation}

The associated CFI (\ref{eq.pcf3}) is
\begin{equation}
I^{(3)}_{e}= p_{\theta} e^{\lambda t} \left[ \dot{x}^{2} +\dot{y}^{2} -\lambda(x\dot{x} +y\dot{y}) + \frac{\lambda^{2}}{4}r^{2} +\frac{2k}{r^{2}} \right] \label{eq.pcf3.12}
\end{equation}
where $p_{\theta}= x\dot{y} -y\dot{x}$ is the angular momentum of the system. However, it is well-known (see e.g. Ref. \ref{Hietarinta 1987} and Tables in Ref. \ref{MitsTsam sym}) that potentials of the form $V=F(r)$ are integrable because they admit the additional LFI $p_{\theta}$. Therefore, the CFI (\ref{eq.pcf3.12}) is the product of two independent FIs: the LFI $p_{\theta}$ and the time-dependent QFI
\begin{equation}
I= e^{\lambda t} \left[ \dot{x}^{2} +\dot{y}^{2} -\lambda(x\dot{x} +y\dot{y}) + \frac{\lambda^{2}}{4}r^{2} +\frac{2k}{r^{2}} \right]. \label{eq.pcf3.13}
\end{equation}
The Hamiltonian, the angular momentum and the time-dependent QFI (\ref{eq.pcf3.13}) are independent; therefore, the potential (\ref{eq.pcf3.10}) is superintegrable. This appears to be a rather new result\footnote{E.G. Kalnins, J.M. Kress, G.S. Pogosyan and W. Miller Jr., \emph{`Completeness of superintegrability in two-dimensional constant-curvature spaces'}, J. Phys. A: Math. Gen. \textbf{34}, 4705 (2001). \label{Kalnins 2001}} since it is not included in Refs. \ref{Hietarinta 1987}, \ref{Karlovini 2000}, \ref{MitsTsam sym} and \ref{Kalnins 2001}.
\bigskip

2) Case $b_{3}=b_{6}$ and the remaining parameters are fixed to zero.

We find the potential (oscillator) $V= -\frac{\lambda^{2}}{8}r^{2}$ which is a well-known superintegrable potential (using only QFIs).
\bigskip

3) Case $b_{3}=b_{4}$ and the remaining parameters are fixed to zero.

We find the potential $V=-\frac{\lambda^{2}}{8}r^{2} -\frac{\lambda^{2}}{12}y$. This is a superintegrable potential as showed in Ref. \ref{MitsTsam sym} using time-dependent QFIs. Specifically, it is a special case of the potential $V_{274}$  for $k_{1}=k_{2}=c_{1}=0$ and $c_{2}=\frac{1}{3}$ (see sec. 8, case 7d and last Table in Ref. \ref{MitsTsam sym}).
\bigskip

We note that other choices of the free parameters may produce new superintegrable potentials.
\bigskip

We collect the results of this section in Table \ref{Table.cfi3}.

\begin{longtable}{|l|c|c|}
\hline
{\large Potentials and FIs} & {\large Ref. \ref{Hietarinta 1987}} & {\large Ref. \ref{Karlovini 2000} } \\ \hline
\makecell[l]{$V_{s11}= -\frac{\lambda^{2}}{8}r^{2} +\frac{k}{r^{2}}$ \\ \\ $p_{\theta}=x\dot{y} -y\dot{x}$, \enskip $J_{s11a}=p_{\theta}J_{s11b}$ \\ $J_{s11b}= e^{\lambda t} \left[ \dot{x}^{2} +\dot{y}^{2} -\lambda(x\dot{x} +y\dot{y}) + \frac{\lambda^{2}}{4}r^{2} +\frac{2k}{r^{2}} \right]$} & New & New \\ \hline
\caption{\label{Table.cfi3} Superintegrable potentials $V(x,y)$ that admit CFIs of the type $J^{(3)}_{e}$.}
\end{longtable}

\section{Conclusions}

\label{conclusions}

The purpose of the present article is to show that the existing results (to
our knowledge), summarized mainly in the review articles of Refs. \ref{Hietarinta
1987} and \ref{Karlovini 2000}, concerning the integrable and the superintegrable potentials of 2d conservative autonomous Newtonian dynamical systems that admit autonomous CFIs can be obtained by a \emph{single algorithmic method} based on the general Theorem 1 of Ref. \ref{Mits 2021} and specialized for the CFIs in Theorem \ref{theorem.CFIs} of section \ref{sec.theorem.CFIs}.

In addition to the existing results, the application of the algorithm
provided new integrable and superintegrable potentials some of them
containing the known ones for special values of the parameters. Also, we
have found new time-dependent CFIs which do not appear in the current literature.

It is emphasized that the CFIs we have found are special solutions of general systems of PDEs; therefore, other studies using Theorem \ref{theorem.CFIs} and the methods demonstrated above (e.g. Holt's method or the integrability condition method) will be able to produce more CFIs by making different suitable choices of the free parameters.

In this respect, it would be useful to update the relevant results on this
topic, given in the previous main sources, and also add the new ones obtained above in a format that could be easily used for reference purposes. This is done in Tables \ref{Table.cfi1}, \ref{Table.cfi2}, \ref{Table.time} and \ref{Table.cfi3} where we give: \newline
a. The known integrable and superintegrable potentials together with their appropriate source. \newline
b. The new ones obtained above. \newline
c. The known ones which are obtained by special values of the parameters of more
general CFIs obtained here.

It is profound that the next step is to determine the integrable/superintegrable Newtonian dynamical systems with more degrees of freedom and, at a later stage, to exploit the known results on the KTs of curved spaces to determine integrable/superintegrable dynamical systems in General Relativity and Cosmology.

\section*{Data Availability}

The data that supports the findings of this study are available within the article.

\section*{Conflict of interest}

The authors declare no conflict of interest.

\bigskip

\bigskip

\theendnotes

\end{document}